\documentclass[12pt]{article}

\usepackage{graphicx}
\usepackage{amsmath}
\usepackage{amssymb}
\usepackage{amsfonts}
\usepackage{graphicx}
\usepackage{mathrsfs}

\newcommand{\be}{\begin{equation}}
\newcommand{\ee}{\end{equation}}
\newcommand{\ba}{\begin{eqnarray}}
\newcommand{\ea}{\end{eqnarray}}
\newcommand{\nnb}{\nonumber}

\newcommand{\tr}{\mathrm{tr}}
\newcommand{\Tr}{\mathrm{Tr}}
    
\newcommand{\pd}[2]{\frac{\partial #1}{\partial #2}}

\newcommand{\ep}{\varepsilon}                                    

\newcommand{\om}{\omega}
\newcommand{\Om}{\Omega}
\newcommand{\re}{\mathrm{Re}}
\newcommand{\im}{\mathrm{Im}}	
\newcommand{\sst}{\hbox{$SU(3)\times SU(3)$}}

\def\slas#1{\rlap{\begin{picture}(10,10)(-5,0)
\put(0,0){\line(2,1){15}}
\end{picture}} #1 }

\catcode`\@=11
\def\marginnote#1{}

\def\titlepage{\@restonecolfalse\if@twocolumn\@restonecoltrue\onecolumn
     \else \newpage \fi \thispagestyle{empty}\c@page\z@
        \def\thefootnote{\fnsymbol{footnote}} }

\def\endtitlepage{\if@restonecol\twocolumn \else  \fi
        \def\thefootnote{\arabic{footnote}} \setcounter{footnote}{0}}

\catcode`@=12 \relax

\makeindex

\begin{document}
\topmargin -1.1cm
\begin{titlepage}
\begin{flushright}
LPTENS-07/31\\
ROM2F/2007/12\\
July 2007
\end{flushright}
\vskip 2.5cm
\begin{center}
{\Large\bf Effective actions and $N=1$ vacuum conditions}\\
\vspace{3mm} {\Large \bf from $SU(3)\times SU(3)$ compactifications}\\
\vskip 1.5cm 
{\bf Davide Cassani$^{a,b}$ and Adel Bilal$^{a}$}\\
\vskip.6cm 
$^a$ Laboratoire de Physique Th\'eorique,
\'Ecole Normale Sup\'erieure - CNRS\footnote{
\hbox{unit\'e mixte du CNRS et de l'\'Ecole Normale Sup\'erieure 
associ\'ee \`a l'Universit\'e Paris 6 Pierre et Marie Curie;}

\hskip .3cm e-mail: \texttt{lastname@lpt.ens.fr}}
\\
24 rue Lhomond, 75231 Paris Cedex 05, France\\
\vskip .5cm
$^b$ Dipartimento di Fisica, Universit\`a di Roma ``Tor Vergata''\\
Via della Ricerca Scientifica 1, 00133 Roma, Italy
\end{center}

\vskip .7cm

\begin{center}
{\bf Abstract}
\end{center}
\begin{quote}
We consider compactifications of type II string theory on general $SU(3)\times SU(3)$ structure backgrounds allowing for a very large set of fluxes, possibly nongeometric ones. We study the effective 4d low energy theory which is a gauged $N=2$ supergravity, and discuss how its data are obtained from the formalism of the generalized geometry on $T\oplus T^*$. In particular we relate Hitchin's special K\"ahler metrics on the spaces of even and odd pure spinors to the metric on the supergravity moduli space of internal metric and $B$-field fluctuations. We derive the $N=1$ vacuum conditions from this $N=2$ effective action, as well as from its $N=1$ truncation. We prove a direct correspondence between these conditions and an integrated version of the pure spinor equations characterizing the $N=1$ backgrounds at the ten dimensional level.
\end{quote}
\end{titlepage}
\setcounter{footnote}{0} \setcounter{page}{0}
\newpage
\begin{small}
\tableofcontents
\end{small}

\setcounter{equation}{0}

\newpage

\section{Introduction}

\setcounter{equation}{0}
Flux compactifications \cite{FluxReviews} have opened new perspectives in the search for supersymmetric string vacua, as well as in the study of the corresponding four dimensional effective actions.

In susy-preserving string compactifications one usually starts by solving the conditions for a ten dimensional supersymmetric background with a 4d$\times$6d factorized topology. Super-symmetry translates into differential conditions for the spinors existing on the compact manifold, and this strongly constrains the geometry. Typically the outcome of this first step is a continuous family of ten-dimensional solutions, and a corresponding class of geometrically characterized 6d manifolds. A four-dimensional effective theory describing the low-energy physics of the fluctuations around this solution can then be obtained by a Kaluza-Klein reduction. The moduli parameterizing the family of 10d vacua manifest themselves in the 4d theory as massless fields: their constant configurations also parameterize the 4d vacua, which will be automatically supersymmetric. The most prominent example in which this is realized is given by Calabi-Yau compactifications without fluxes.

A second possible approach to compactifications is to work `off-shell': one can dimensionally reduce the higher dimensional theory on the most general class of manifolds which satisfy the minimal requirements allowing to define a supergravity theory in 4d. A necessary condition in this sense is the existence of globally defined spinors on the compact manifold. However, since one does not demand to start from a 10d supersymmetric solution, no differential constraints are imposed on such spinors. This off-shell option is advantageous when compactifying in the presence of fluxes, for reasons that will be apparent from the following discussion. 

In KK reductions, one expands the 10d fields in modes of appropriate wave operators on the compact manifold. An effective theory describing the massless physics in 4d is then defined by truncating the spectrum to the zero modes and integrating over the internal space. 

Switching on fluxes induces additional terms to the field equations of motion. In particular, from a 10d perspective fluxes backreact on the geometry via their contribution to the energy-momentum tensor in the Einstein equations. It follows that the Ricci-flatness condition is removed, and a Minkowski$_4\,\times\,$Calabi-Yau background is no more available. 

{F}rom an effective four dimensional viewpoint, the modification of the mass operators due to the fluxes implies that some of the previously massless 4d fields acquire masses. In this situation, a necessary condition for performing the Kaluza-Klein reduction is the existence of a hierarchy of scales in which the mass scale induced by the fluxes is well below the scale of the KK excitations which are truncated. If this is realized, it is in principle possible to restrict to a set of light degrees of freedom and define a consistent low energy effective theory. However, a direct identification of the correct modes to be kept has not been achieved yet, and a reasonable way to proceed which has been adopted in the literature \cite{GurLouisMicuWaldr, GurrMicuIIBhalfFlat, GLW1, TomasMirrorSymFl, HousePalti, MinasianKashani} is to assume the existence of a generic set of expansion forms on the internal manifold, defining the light 4d spectrum. These forms should satisfy just the minimal amount of constraints yielding a sensible 4d supergravity theory. In particular, they need not be closed; rather, the differential relations which are established among them define a set of `geometric charges' encoding the departure from the Calabi-Yau geometry.

The 4d theory resulting from this procedure corresponds to a gauged supergravity, where the charges associated with the gaugings are generated by the NS, RR and geometric fluxes \cite{GLW1, LouisMicuCYwithFlux, DallAgataFluxRev, GaugingHeisenberg, D'AuriaFerrTrig}. One of the fundamental features of gauged supergravities is that they contain a potential, and therefore their vacuum solutions are in general non-trivial. In particular, in contrast to the case in which the compactification is performed starting from a class of supersymmetric solutions of the 10d theory, in this off-shell approach the conditions for a supersymmetric vacuum are non-empty. 

Now, a basic question which arises is whether the supersymmetric solutions of the 4d effective theory lift to solutions of the 10d parent theory, and how one can find a correspondence between the supersymmetric vacuum conditions written in the 10d and in the 4d languages.

In this paper we will address these questions by considering off-shell compactifications of type II theories leading to $N=2$ supergravities in four dimensions. The $N=1$ vacuum conditions arising from the effective theory will then be compared with those obtained by a ten dimensional analysis. 

In \cite{GLW1} it has been argued that the most general class of 6d manifolds yielding a 4d \hbox{$N=2$} action admits a pair of $SU(3)$ structures. It turns out that these structures can be conveniently described in the framework of Hitchin's generalized geometry \cite{HitchinGenCY, GualtieriThesis} in terms of an \hbox{$SU(3)\times SU(3)$} structure living on $TM_6\oplus T^*M_6$, the sum of the tangent and the cotangent bundle of the 6d manifold. Such a generalized structure is characterized by a pair of pure $O(6,6)$ spinors (to be seen as formal sums of forms of even/odd degree) encoding the NS (metric and $B$-field) degrees of freedom on the internal manifold. Expanding the pure spinors and the RR supergravity fields on a basis of forms of the type outlined above identifies the light fields entering in the 4d effective action. A peculiar aspect of the generalized geometry approach is that the expansion forms can also be of mixed degree \cite{GLW2, GrimmBen}. A second remarkable point uncovered in \cite{GLW2} is that the general set of fluxes one allows for in this context can also be associated with backgrounds which are nongeometric \cite{HullGeoForNonGeo, SheltonTaylorWecht} (see also \cite{GrangeShafNameki, Ellwood, MicuPaltiTas} for the relation between generalized geometry and nongeometry). The resulting 4d effective theory is compatible with the general structure of $N=2$ gauged supergravity, and contains a set of (possibly massive) tensor multiplets.

Another fundamental point connecting $SU(3)\times SU(3)$ structures and 4d supergravity is that, as Hitchin shows \cite{HitchinGenCY, HitchinFunctionals}, the deformation space of both even and odd pure spinors at a point admits a special K\"ahler structure. This can be seen as a generalization of the Calabi-Yau moduli space, which consists of the product of two special K\"ahler manifolds parameterizing the K\"ahler- and complex-structure deformations.

In this paper we will adopt the off-shell approach and discuss the 4d effective action for $SU(3)\times SU(3)$ structure compactifications, generalizing certain results of \cite{GLW1}.

We begin in section \ref{GenGeoRev} by giving a brief introduction to relevant facts about \hbox{$SU(3)\times SU(3)$} structures and the associated pure spinors. After a discussion on the role played by a \hbox{generalized} diamond labeled by the relevant $SU(3)\times SU(3)$ representations, in section \ref{variations} we study the deformations of the pure spinors. We derive various results allowing us in particular to show that the natural supergravity metric on the moduli space of internal metric and $B$-field fluctuations coincides with the sum of Hitchin's special K\"ahler metrics on the spaces of even and odd pure spinors. This parallels what happens in the Calabi-Yau case and extends to a full $SU(3)\times SU(3)$ environment a previous analysis done in \cite{GLW1} for $SU(3)$ structures. 

In section \ref{KKreduct} we discuss the necessary properties and constraints to be imposed on the sets $\Sigma^\pm$ of expansion forms for the light degrees of freedom in order that this factorized structure of the moduli is then inherited by the 4d effective theory. These constraints have already been outlined in the literature \cite{MinasianKashani, GLW2}, and we just transpose them in a somehow more explicit form, emphasizing the relevance of the pure spinor deformations. 

We then study the role of the \hbox{$B$-twisted} Hodge star operator ($*_B$), and in particular we show how its action on the basis of forms generalizes to the \sst\ context the well-known expression for the usual Hodge $*$ acting on the harmonic three-forms of a Calabi-Yau manifold. The basic tool to get this result is again the decomposition of the supergravity fields in representations of \sst.

Following refs.$\:$\cite{GLW1, GLW2} we identify the data of the 4d $N=2$ gauged supergravity - in particular the $N=2$ Killing prepotentials - in terms of the pure spinors and internal fluxes expanded on the basis forms $\Sigma^\pm$. Starting from the expression of the Killing prepotentials, and using some general results about $N=2$ supergravity with tensor multiplets, we deduce then the fermionic shifts of the 4d theory.

In section \ref{VacuumCondit} we turn to the $N=1$ vacuum conditions. Again generalized geometry plays a key role: manifolds with $SU(3)\times SU(3)$ structure have been shown to represent the most general support for geometric $N=1$ backgrounds of type II theories. Indeed, the necessary and sufficient conditions for an $N=1$ vacuum can be formulated in the context of generalized geometry as differential (non)-integrability equations for the pure spinors characterizing the $SU(3)\times SU(3)$ structure \cite{GMPT1, GMPT2, GMPT3, JeschekWitt1}. This form of the 10d susy conditions is the most suitable one for a comparison with the $N=1$ constraints arising from the 4d effective action. In order to perform such a comparison, we rephrase the `pure spinor equations' in a 4d framework performing the integral over the internal manifold. A slight generalization of the differential operator acting on the pure spinors allows to take formally into account the general set of fluxes considered in the effective action approach, including the nongeometric ones.

By the way we remark that, still at the 10d level, the pure spinor equations for type IIA and type IIB acquire a perfectly symmetric form if one adopts a chirality assignement for the type IIA susy parameters being the opposite of the original one of \cite{GMPT1, GMPT2}.

Then we establish the $N=1$ vacuum conditions for the 4d $N=2$ gauged supergravity action by imposing the vanishing of the fermionic shifts under a single susy transformation. At this level, we don't need to specify the precise mechanism breaking $N=2\to N=1$, which could correspond to a spontaneous supersymmetry breaking but also to an explicit truncation of the action. By a direct inspection, we show that the $N=1$ vacua of the effective theory precisely satisfy the integrated version of the pure spinor equations.

In section \ref{N2toN1} we consider the 4d $N=1$ supergravity arising as a consistent truncation of the previously analysed $N=2$ theory. We revisit the way the superpotential can be obtained as a linear combination of the Killing prepotentials associated with the $N=2$ gaugings, in particular identifying the correct holomorphic variables, and on similar footing we derive an expression for the $N=1$ D-terms. 

We write again the supersymmetric vacuum equations, now in the $N=1$ language, as F- and D- flatness conditions. By considering the example of $N=2\to N=1$ truncation induced by an $O6$ orientifold, we recover the direct correspondence with the pure spinor equations.

Finally section \ref{conclusions} contains our conclusions, appendix$\;$\ref{conventions} resumes our conventions, appendix$\;$\ref{MukaiAndClifford} gives some technical details about the Mukai pairing and the Clifford map and appendix$\;$\ref{SpecialGeometry} collects some properties of special K\"ahler manifolds.\\

Note added: While we finished typing this manuscript, last week a paper by P.~Koerber and L.~Martucci \cite{KoerberMartucci} appeared on the arXiv, presenting some overlap with our work.

\section{Generalized structures in type II supergravity and their deformations }\label{GenStrAndDeform}

\setcounter{equation}{0}

\subsection{$SU(3)\times SU(3)$ structures and pure spinors}\label{GenGeoRev}

In dimensional reductions of type II theories, one considers a 10d spacetime given by the topological product $M_{9,1}=M_{3,1}\times M_6$, 
where $M_{3,1}$ is the 4d spacetime and $M_6$ is a 6d compact `internal' manifold. Each of the two $Spin(9,1)$ Majorana-Weyl supersymmetry parameters $\epsilon^{1,2}$ is then decomposed into the product of a spacetime $Spin(3,1)$ spinor and an internal $Spin(6)$ spinor. Focusing on type IIA, we will adopt the following decomposition ansatz, preserving the minimal $N=2$ supersymmetry in four dimensions:
\begin{eqnarray}
\nnb\epsilon^1 &=& \ep_{1}\otimes \eta^1_- + \ep^{1}\otimes \eta^1_+ \\
\label{eq:10dSpinorsIIA}\epsilon^2 &=& \ep_{2}\otimes \eta^2_+ + \ep^{2}\otimes \eta^2_-\;.
\end{eqnarray}
According to a standard notation in 4d $N=2$ supergravity, lower indices on $Spin(3,1)$ spinors $(\ep_{1}, \ep_{2})$ denote positive chirality, while upper indices $(\ep^{1}, \ep^{2})$ refer to negative chirality. Further, $\ep^{1,2}$ are defined as the charge conjugate of $\ep_{1,2}$, which in our conventions just amounts to complex conjugation (see App.$\,$\ref{conventions}). For the $Spin(6)$ spinors, instead, we indicate chirality by a $\pm$, so that $\eta_+^{1,2}$ has positive chirality and $\eta^{1,2}_-\equiv(\eta^{1,2}_+)^*$ has negative chirality. It follows that the $Spin(9,1)$ spinor $\epsilon^1$ has negative chirality, while $\epsilon^2$ has positive chirality.

The spinors $\ep_{1,2}$ parameterize the $N=2$ supersymmetry in 4d, while the \emph{Spin}(6) spinors $\eta^1$ and $\eta^2$ should be globally defined and nowhere vanishing on the compact manifold $M_6$. In \hbox{general}, any given globally defined spinor $\eta$ identifies a subgroup $SU(3)\subset SU(4)\cong Spin(6)$, and this determines an $SU(3)$ structure for $M_6$. This also implies the existence of a globally defined real 2-form $J$ and complex 3-form $\Om$ via the bilinears: $J_{ab}=i\eta_+^\dag\gamma_{ab}\eta_+\,$, $\Om_{abc}=-i\eta_-^\dag\gamma_{abc}\eta_+$. \hbox{$J$ represents} an almost symplectic structure, while $\Om$ determines an almost complex structure $I$. One therefore deduces that the decomposition (\ref{eq:10dSpinorsIIA}) implies the existence of a pair of $SU(3)$ structures, one for each of the two globally defined spinors $\eta^{1,2}$. Equivalently, we have two symplectic forms $J_1$ and $J_2$ and two almost complex structures $I_1$ and $I_2$. These are required to define the same (positive definite) metric $g_{mn}$ via the relation:
\be\label{eq:metricFromJandI}
J_{1,2\,mp} = g_{mn} I^{\;\;\;n}_{1,2\;\,p}\;.
\ee
Locally, the existence of the two globally defined spinors $\eta^{1,2}$ determines an $SU(2)$ structure. However, this is not necessarily true globally, as $\eta^1$ and $\eta^2$ may coincide at some points of $M_6$. The limiting case in which $\eta^1\equiv\eta^2$ everywhere on $M_6$ is also admitted, and then the structure group of $M_6$ is just $SU(3)$. These different cases can be seen as the different possible ways of intersecting the two $SU(3)$ structures defined above.

It turns out that a unifying description for the aforementioned cases can be obtained by considering \emph{generalized structures} living on $TM_6\oplus T^*M_6$, the sum of the tangent and the cotangent bundle of $M_6$. Such structures have been introduced in the mathematical literature by Hitchin \cite{HitchinGenCY} and further studied in \cite{GualtieriThesis, WittThesis}, in parallel with the development of the concept of generalized complex geometry. A physicists' review of generalized complex geometry can be found in \cite{GMPT3}, while applications to dimensional reductions of type II supergravity were first considered in \cite{GLW1}. In the following we recall some notions.

As a first thing we introduce the notion of \emph{generalized almost complex structure}. This is a map $\mathcal J: T\oplus T^*\to T\oplus T^*\,$ satisfying $\mathcal J^2 = -id\,$ (so it has $\pm i$ eigenvalues) together with the hermiticity condition $\mathcal J^T\mathcal I\mathcal J= \mathcal I$, where $\mathcal I = \left(\begin{array}{cc} 0 &  1 \\ [-1mm] 1 & 0   \end{array}\right)\;$ is the natural metric on $T\oplus T^*$ with $(6,6)$-signature. Now, the data contained in $J_{1,2}$, and $I_{1,2}$, as well as in the NS 2-form $B$, can all be encoded in a pair of generalized almost complex structures. Indeed, it can be checked that each of two matrices \cite{GualtieriThesis}:
\be\label{eq:CurlyJ+-}
\mathcal J^{\;\Lambda}_{\pm\;\Sigma} := \frac{1}{2}\left(\begin{array}{cc}1 & 0 \\ [1mm] -B & 1 \end{array}\right)         
\left(\begin{array}{cc} I_1 \mp I_2 & -(J_1^{-1} \pm J_2^{-1})  \\ [1mm] J_1 \pm J_2 & -(I^T_1 \mp I^T_2)   \end{array}\right)
\left(\begin{array}{cc} 1 & 0 \\ [1mm] B & 1 \end{array}\right)
\ee
satisfies the above requirements. The indices $\Lambda, \Sigma=1,\ldots,12$ run over the tangent and the cotangent spaces. Furthermore, $\mathcal J_\pm$ commute:
\be\label{eq:Jcommute}
[\mathcal J_+,\mathcal J_-]=0\;,
\ee
and also determine a positive definite metric on $T\oplus T^*$: it follows from (\ref{eq:metricFromJandI}) and (\ref{eq:CurlyJ+-}) that
\be\label{eq:TT*metric}
\mathcal G^{\Lambda}_{\;\;\Sigma}:= -\mathcal J_+ \mathcal J_- = -\mathcal J_- \mathcal J_+ =  \left(\begin{array}{cc}1 & 0 \\ [1mm] -B & 1 \end{array}\right) \left(\begin{array}{cc} 0 &  g^{-1} \\ [1mm] g & 0   \end{array}\right)\left(\begin{array}{cc} 1 & 0 \\ [1mm] B & 1 \end{array}\right) =  \left(\begin{array}{cc} g^{-1}B & g^{-1}  \\ [1mm] g- Bg^{-1}B & -Bg^{-1} \end{array}\right),
\ee
and it is readily checked that $\mathcal G_{\Lambda\Sigma}= \mathcal I_{\Lambda\Xi}\,\mathcal G^\Xi_{\;\;\Sigma}$ is symmetric and positive definite. Two commuting generalized almost complex structures yielding a positive definite metric on $T\oplus T^*$ are said to be compatible. Generically, the structure group of $T\oplus T^*$ with the natural metric $\mathcal I$ is $O(6,6)$. The existence of the compatible pair $\mathcal J_{\pm}$ determines a reduction to an $U(3)\times U(3)$ structure\footnote{The above discussion could also be reversed: the choice of an $U(3)\times U(3)$ structure for $T\oplus T^*$ - i.e. of a compatible pair of generalized almost complex structures - determines a positive definite metric $g_{mn}$ and a $B$-field on $M_6$.}.

The previous construction simplifies when the 6d manifold $M_6$ has $SU(3)$ structure. Indeed, in this case $\eta^1=\eta^2\;\Rightarrow\; J_1=J_2\equiv J\,,\,I_1=I_2\equiv I\,$. Then, assuming also $B=0$, the generalized almost complex structures $\mathcal J_\pm$ reduce to
\be
\mathcal J_{+} = \left(\begin{array}{cc} 0 & - J^{-1}  \\ [1mm] J & 0   \end{array}\right)
\qquad,\qquad
\mathcal J_{-} = 
\left(\begin{array}{cc} I &  0 \\ [1mm] 0 & -I^T   \end{array}\right)\;.
\ee

The generalized structures can also be conveniently encoded in pure spinors of $O(6,6)$ as we now summarize. As a first thing, we recall that the $O(6,6)$ spinors can be seen as elements of $\wedge^{\bullet}T^*$, the bundle of forms of every degree on $M_6$. Indeed, a Clifford action of $v + \zeta\in T\oplus T^*$ on $C\in \wedge^\bullet T^*$ is defined by
\be\label{eq:CliffAct}
(v + \zeta)\cdot C= (\iota_v + \zeta\wedge)C\;.
\ee
This is a Clifford action in that it squares to the norm with respect to the metric $\mathcal I$. As a consequence, the \emph{Cliff}(6,6) gamma matrices $\Gamma^\Lambda$ can be identified with the basis of $T\oplus T^*$:
\be\label{eq:Gamma66}
\Gamma^\Lambda\;=\; (\,d x^m\wedge \,,\, \iota_{\partial_m}\,)\qquad,   \qquad \{\Gamma^\Lambda,\Gamma^\Sigma\}=\mathcal I^{\Lambda\Sigma}\;, \qquad\Lambda,\Sigma=1,\ldots,12\;.\ee
The $Spin(6,6)$ spin representation decomposes in two irreducible Weyl representations, and this is reflected in the splitting $\wedge^\bullet T^*= \wedge^{\mathrm{even}}T^*\oplus \wedge^{\mathrm{odd}}T^*$. In this way, an even/odd form of mixed degree can be regarded as a Weyl spinor of $O(6,6)$ with positive/negative chirality\footnote{We just remark that the isomorphism between the $Spin(6,6)$-bundle and the bundle of forms is not canonical in that it requires the choice of a volume form on $M_6$ (see for instance \cite{GualtieriThesis, GLW1} for more details).}. 

A bilinear product between $O(6,6)$ spinors can be defined through the Mukai pairing:
\be\label{eq:DefMukai}
\langle A\,,C\rangle := \big[\,A\wedge \lambda(C)\,\big]_{\mathrm{top}}\;,
\ee
where $A,C\in\wedge^\bullet T^*\,,$ $[\;]_{\mathrm{top}}$ picks the 6-form component, while the involution $\lambda$ acts on a $k$-form $A_k$ as:
\be\label{eq:lambda}\lambda(A_k)=(-)^{[\frac{k}{2}]}\,A_k\;.\ee
In six dimensions $\langle\,,\,\rangle$ is antisymmetric; some other properties are collected in Appendix \ref{MukaiAndClifford}.

As already mentioned, a prominent role in relation with the generalized structures is played by \emph{pure} $O(6,6)$ spinors. If we introduce the annihilator space of a complex $O(6,6)$ spinor $\Phi$ as
\be L_\Phi: = \{v+\zeta \in (T\oplus T^*)\otimes \mathbb C \,:\, (v+\zeta)\cdot \Phi=0\}\;,\ee
then by definition we say that $\Phi$ is pure if $L_\Phi$ has maximal dimension $=6$. A one-to-one correspondence between pure spinors $\Phi$ and generalized almost complex structures $\mathcal J_\Phi$ can then be established by identifying the annihilator $L_\Phi$ of $\Phi$ with the $+i$ eigenbundle of $\mathcal J_\Phi$. More precisely, since a rescaling of $\Phi$ does not modify its annihilator space, the one-to one correspondence is between generalized almost complex structures and line bundles of pure spinors; furthermore, at each point of $M_6$ the pure spinor generating the complex line should satisfy the `nonvanishing norm' condition\footnote{A more precise definition for the norm of a pure spinor is $||\Phi||^2 := i\langle \Phi , \bar\Phi \rangle/vol_6\,$.} $\langle \Phi ,  \bar\Phi\rangle\neq 0$. An explicit formula for $\mathcal J_\Phi$ in terms of $\Phi$ which will be useful in the following is \cite{HitchinGenCY, GLW1, GMPT3}:
\be\label{eq:RelJPhi}
\mathcal J^{\;\Lambda}_{\Phi\;\Sigma} = \frac{4\langle \re\Phi , \Gamma^{\Lambda}_{\;\;\Sigma}\re \Phi \rangle}{i\langle \Phi , \bar\Phi \rangle}\;,
\ee
where the $T\oplus T^*$ indices are raised and lowered with the metric $\mathcal I$. The denominator ensures that $\mathcal J_\Phi$ doesn't depend on the choice of the volume form for $M_6$, nor on rescalings of $\Phi$ (about this last fact, see also subsect.$\:$\ref{variations}). 

In (\ref{eq:CurlyJ+-}) we introduced a pair of compatible generalized almost complex structures $\mathcal J_\pm$, and we stated they provide an $U(3)\times U(3)$ structure for $T\oplus T^*$. In general, thanks to the correspondence between generalized almost complex structures and line bundles of pure spinors, such a structure is equivalently characterized by the existence of a pair of pure spinors $\Phi_\pm$ satisfying the compatibility relation \cite{GualtieriThesis, GLW1}\footnote{See \cite{Tomasiello} for a proof of the equivalence between (\ref{eq:Jcommute}) and (\ref{eq:compatibility}).}:
\be
\label{eq:compatibility}\langle \Phi_+, \Gamma^\Lambda \Phi_-\rangle = \langle \bar \Phi_+, \Gamma^\Lambda \Phi_-\rangle \:= \:0
\ee
and defining a positive definite metric on $T\oplus T^*$. Furthermore, we also take
\be
\label{eq:equalNorm}\langle \Phi_+,  \bar\Phi_+\rangle = \langle \Phi_- ,  \bar\Phi_-\rangle\;.
\ee
If $\Phi_\pm$ are globally defined (i.e. if the line bundle of pure spinors has a global section), then the structure group is further reduced to $SU(3)\times SU(3)$. The pure spinors $\Phi_\pm$ are invariant under the action of the $SU(3)\times SU(3)$ structure they identify, much as a globally defined $Spin(6)$ spinor $\eta$ is invariant under the action of the $SU(3)\subset Spin(6)$ structure it determines.

The description in terms of pure spinors is particularly convenient for the applications to supergravity since they can be defined directly from the $Spin(6)$ spinors $\eta^{1,2}$ entering in the decomposition ansatz (\ref{eq:10dSpinorsIIA}) \cite{GMPT1}. This builds on the fact that one can send elements of $\wedge^\bullet T^*$ to $Spin(6)$ bispinors and vice versa by means of the Clifford map ``/" :
\be\label{eq:CliffMap}
C\,=\,\sum_k \frac{1}{k!}C^{(k)}_{m_1\ldots m_k}dx^{m_1}\wedge\ldots \wedge dx^{m_k}\qquad \longleftrightarrow \qquad /\!\!\!\!C \,=\, \sum_k \frac{1}{k!}C^{(k)}_{m_1\ldots m_k}\gamma^{m_1\ldots m_k}\;,
\ee
where the antisymmetrized products of \emph{Cliff}(6) gamma matrices $\gamma^{m_1\ldots m_k}$ represent the basis for bispinors. The correspondence with bispinors is better seen recalling the Fierz identity between two $Spin(6)$ spinors $\psi, \chi\,$:
\be
\label{eq:fierz} \psi\otimes \chi^\dag\;=\; \frac{1}{8}\sum_{k=0}^6\frac{1}{k!}\big( \chi^{\dag}\gamma_{m_k\ldots m_1}\psi \big)\gamma^{m_1\ldots m_k}\;,
\ee
Therefore out of two $Spin(6)$ spinors $\psi, \chi$ one can build a bispinor $\psi\otimes\chi^\dag$ and then map this to an element of $\wedge^\bullet T^*$  using the Fierz identity and then the Clifford map backwards. 

Applying this to the $Spin(6)$ spinors $\eta^{1,2}$ appearing in the decomposition (\ref{eq:10dSpinorsIIA}), one can introduce the globally defined $O(6,6)$ spinors \cite{GMPT2}:
\be
\label{eq:defPhi0}\Phi^0_\pm := 8 \eta^1_+\otimes\eta^{2\dag}_\pm\;,
\ee
where we assume the normalizations $\eta^{1\dag}_\pm\eta^1_\pm=\eta^{2\dag}_\pm\eta^2_\pm=1$, and the factor of 8 is introduced just for convenience. It is not difficult to see that $\Phi^0_+\in \wedge^{\mathrm{even}}T^*$ while $\Phi^0_-\in \wedge^{\mathrm{odd}}T^*$. Furthermore, it turns out \cite{GMPT3} that $\Phi^0_\pm$ define a compatible pair of pure spinors. Also, using the image (\ref{eq:MukaiUnderClifford}) of the Mukai pairing under the Clifford map and the fact that $\eta_\pm^{1,2}$ are normalized to 1 everywhere on $M_6$, one can see that their norms are equal and nowhere vanishing:
\be\label{eq:normPhi}
i\langle\,\Phi^0_\pm\,,\,\bar\Phi^0_\pm\,\rangle\;=\;8 (\eta^{1\dag}_\pm\eta^1_\pm)(\eta^{2\dag}_\pm\eta^2_\pm) vol_6 = 8 vol_6 \;,
\ee
where $vol_6$ is the volume form of $M_6$. Therefore we deduce that $\Phi^0_\pm$ identify an $SU(3)\times SU(3) \subset O(6,6)$ structure. It is also possible to include the NS 2-form $B$ degrees of freedom without losing any of the previous features by defining the $B$\emph{-transformed} spinors:
\be\label{eq:DefPhi_pm}
\Phi_+:=e^{-B}\Phi^0_+\qquad,\qquad \Phi_-:=e^{-B}\Phi^0_-\;,
\ee
where $e^{-B}=1-B+\frac{1}{2}B\wedge B - \frac{1}{6}B\wedge B\wedge B$ acts by the wedge product. In particular, thanks to the property (\ref{eq:Bdrops}) of the Mukai pairing, this does not change the norm of the pure spinors.

In the $SU(3)$ structure case, where $\eta^1\equiv \eta^2$, the $\Phi_\pm$ defined here above and the compatibility requirement (\ref{eq:compatibility}) take the form:
\be\label{eq:Phi+-SU3}
\Phi_+=e^{-B+iJ} \qquad,\qquad \Phi_-=-i\Om \qquad,\qquad (B-iJ)\wedge \Om=0\;,
\ee
where $J$ and $\Om$ are the invariant forms of the $SU(3)$ structure. In the general case in which $\eta^1\neq \eta^2$, the expression for $\Phi_\pm$ is more involved, and we refer for instance to \cite{GMPT3} for further details.

It is an instructive exercise to check that the generalized almost complex structures defined from the pure spinors (\ref{eq:DefPhi_pm}) via the formula (\ref{eq:RelJPhi}) correspond exactly to the matrices $\mathcal J_\pm$ provided in eq. (\ref{eq:CurlyJ+-}). We present the main steps of this computation at the end of appendix$\,$\ref{MukaiAndClifford}.

Before passing to consider deformations of the structures discussed above, two remarks are in order. 

1) When the $B$-field appearing in (\ref{eq:DefPhi_pm}) is non-trivial, instead of the tangent and cotangent bundles, one should consider an extended bundle in which on overlapping patches $B$ can be glued by gauge transformations \cite{HitchinGenCY, GualtieriThesis, GLW2}. We will implicitly assume this extension, but we'll keep on speaking of $T\oplus T^*$ for simplicity. 

2) The $T\oplus T^*$ bundle could also be generalized in another sense. Refs. \cite{GLW1, GLW2} adopted the strategy of reformulating type II supergravity on a background preserving eight supercharges only, but staying at a full ten-dimensional level and not even requiring a product structure $M_{9,1}=M_{3,1}\times M_6$ for the 10d spacetime. The actual dimensional reduction on a compact manifold $M_6$ was performed only as a second step. In this rewriting of 10d supergravity, the fields arrange however in 4d $N=2$-like multiplets, and the type II theory has the features of a 4d, $N=2$ supergravity. In order to achieve this reformulation, the authors of \cite{GLW1, GLW2} only had to require a splitting for the tangent bundle of the 10d spacetime of the type $T^{3,1}\oplus F\,$, where $T^{3,1}$ is a $SO(3,1)$ vector bundle and $F$ is a vector bundle admitting a pair of $SU(3)$ structures, not necessarily coinciding with the tangent bundle of a compact manifold $M_6$. We will follow only in part this approach: while in the next subsection we will avoid integrating over the internal manifold, postponing the Kaluza-Klein truncation to subsect.$\:$\ref{KKreduct}, we will however assume the 10d spacetime has the product structure $M_{9,1}=M_{3,1}\times M_6$; therefore, for us the bundle admitting an $SU(3)\times SU(3)$ structure will be just $TM_6\oplus T^*M_6$.

\subsection{Deformations}\label{variations}

In the previous subsection we discussed how any compactification of type II theories providing an $N=2$ effective supergravity in four dimensions requires the internal manifold $M_6$ to admit a pair of $SU(3)$ structures; we also recalled how this pair of structures can be encoded in an $SU(3)\times SU(3)$ structure on $TM_6\oplus T^*M_6$, characterized by the two pure spinors (\ref{eq:DefPhi_pm}).

In this subsection we want to study deformations of pure spinors, and so of $SU(3)\times SU(3)$ structures, putting them in relation with the kinetic terms for the internal metric and $B$-field fluctuations appearing in the 4d $N=2$ effective theory. With restriction to the $SU(3)$ structure case, a similar analysis has been performed in ref.$\:$\cite{GLW1}. Here we will extend the results of that paper, working with a general $SU(3)\times SU(3)$ structure for $T\oplus T^*$.

In order to do this, it will be useful to decompose the space of $O(6,6)$ spinors in representations of the $SU(3)\times SU(3)$ subgroup defined by the compatible pair $\Phi_+,\Phi_-$. Following \cite{GLW2}, we call $U_{\bf{r},\bf{s}}$ the set of forms transforming in the $(\bf{r},\bf{s})$ representation of $SU(3)\times SU(3)$, and we organize the different representations in a  ``generalized diamond''\footnote{$\;{\bf \bar 1}$ refers to the singlet coming from the decomposition under $SU(3)$ of the ${\bf \bar 4}$ of \emph{Spin}(6).} $\;$\cite{GualtieriThesis, GenHodgeDec}:
\be\label{eq:Udiamond}
\begin{array}{ccccccc}
&&& U_{\bf{1}, \bf{\bar 1}} &&&\\
&& U_{\bf{1}, \bf{3}} && U_{\bf{\bar 3}, \bf{\bar 1}}  &&\\
& U_{\bf{1}, \bf{\bar 3}} && U_{\bf{\bar 3}, \bf{3}}  && U_{\bf{3}, \bf{\bar 1}} & \\
U_{\bf{1}, \bf{1}} && U_{\bf{\bar3}, \bf{\bar 3}} && U_{\bf{3}, \bf{3}} && U_{\bf{\bar 1}, \bf{\bar 1}} \\
& U_{\bf{\bar 3}, \bf{1}} && U_{\bf{3}, \bf{\bar 3}}  && U_{\bf{\bar 1}, \bf{3}}  & \\
&& U_{\bf{3}, \bf{1}} && U_{\bf{\bar 1}, \bf{\bar 3}}  && \\
&&& U_{\bf{\bar 1}, \bf{1}} &&&
\end{array}
\ee
An important difference with respect to the usual $(p,q)$-decomposition of complex differential forms is that here the $U_{\bf{r},\bf{s}}$ contain forms of mixed degree. It turns out that $\wedge^{\mathrm{ev}}T^*$ and $ \wedge^{\mathrm{odd}}T^*$ transform differently under $SU(3)\times SU(3)$, i.e. the forms in $U_{\bf{r},\bf{s}}$ have definite parity:
\begin{eqnarray}
\nnb U_{\bf{1}, \bf{\bar 1}}\oplus U_{\bf{1}, \bf{\bar 3}} \oplus U_{\bf{\bar 3}, \bf{3}}  \oplus U_{\bf{3}, \bf{\bar 1}} \oplus  U_{\bf{\bar 3}, \bf{1}}  \oplus U_{\bf{3}, \bf{\bar 3}}  \oplus U_{\bf{\bar 1}, \bf{3}} \oplus   U_{\bf{\bar 1}, \bf{1}} &=& \wedge^{\mathrm{ev}}T^*\\   U_{\bf{1}, \bf{3}} \oplus U_{\bf{\bar 3}, \bf{\bar 1}} \oplus U_{\bf{1}, \bf{1}} \oplus  U_{\bf{\bar3}, \bf{\bar 3}} \oplus U_{\bf{3}, \bf{3}} \oplus U_{\bf{\bar 1}, \bf{\bar 1}} \oplus U_{\bf{3}, \bf{1}} \oplus U_{\bf{\bar 1}, \bf{\bar 3}} &=& \wedge^{\mathrm{odd}}T^*\;.
\end{eqnarray}
The $SU(3)\times SU(3)$ singlets $\Phi_\pm$, $\bar\Phi_\pm$ occupy the vertices of the diamond. More precisely, $\Phi_+$ spans $U_{\bf{1}, \bf{\bar 1}}$ while $\Phi_-$ spans $U_{\bf{1}, \bf{1}}$.

In the case of vanishing $B$, the $SU(3)\times SU(3)$ structure is defined by the $\Phi^0_\pm$ given in (\ref{eq:defPhi0}), and an explicit basis for the whole decomposition (\ref{eq:Udiamond}) can be built \cite{GMPT2, GMPT3} by exploiting the correspondence between differential forms and bispinors provided by the Clifford map. Starting from the lowest/highest weight states $\Phi^0_\pm$ and $\bar\Phi^0_\pm$, and acting with holomorphic/antiholomorphic \emph{Cliff}$(6)$ gamma matrices (to be seen as lowering/raising operators), one can reconstruct the whole decomposition of the $O(6,6)$ spinors under $SU(3)\times SU(3)$, with the result (for further details see App.$\:$A in ref. \cite{GMPT3}):
\begin{equation}\label{eq:BasisDiamond}
  \begin{array}{c}\vspace{.1cm}
\Phi^0_+ \\ \vspace{.1cm}
\Phi^0_+\gamma^{i_2}  \hspace{1cm} \gamma^{\bar \imath_1}\Phi^0_+ \\ 
\Phi^0_-\gamma^{\bar \imath_2} \hspace{1cm} \gamma^{\bar \imath_1} \Phi^0_+\gamma^{i_2} 
\hspace{1cm} \gamma^{i_1} \bar \Phi^0_-\\
\Phi^0_- \hspace{1.2cm}\gamma^{\bar \imath_1}\Phi^0_-\gamma^{\bar \imath_2} 
\hspace{1cm} \gamma^{i_1}\bar\Phi^0_-\gamma^{i_2} 
\hspace{1.2cm}\bar\Phi^0_-\\
 \gamma^{\bar \imath_1} \Phi^0_-\hspace{1cm} \gamma^{i_1} \bar \Phi^0_+ \gamma^{\bar \imath_2} 
\hspace{1cm}\bar\Phi^0_-\gamma^{i_2}\\
\gamma^{i_1} \bar\Phi^0_+ \hspace{1cm}  \bar\Phi^0_+\gamma^{\bar \imath_2}\\
 \bar\Phi^0_+\\
  \end{array}\ 
\end{equation}
The basis elements can be seen either as bispinors, or, using the Clifford map backwards, as differential forms. In this last case, the \emph{Cliff}(6) gamma matrices are mapped to elements of $T\oplus T^*$, acting as in (\ref{eq:CliffAct}).

Actually, here we are interested in $O(6,6)$ spinors containing also the NS 2-form $B$. This means that we consider the $SU(3)\times SU(3)$ structure defined by $\Phi_\pm = e^{-B}\Phi^0_\pm$ (which is different from the one considered above). A basis for the decomposition (\ref{eq:Udiamond}) under this $SU(3)\times SU(3)$ is simply obtained by multiplying by $e^{-B}$ the basis (\ref{eq:BasisDiamond}). Indeed, this is just the result of doing the following $B$-transformation: for the pure spinors one has
\be\nnb \Phi^0_\pm \;\;\overset{B\mathrm{-transf}}{\longrightarrow}\;\; \Phi_\pm = e^{-B}\Phi^0_\pm\;,\ee 
while the raising/lowering operators $\buildrel\to\over\gamma\!{}^{i_1},\,\buildrel\to\over\gamma\!{}^{\bar \imath_1},\,\buildrel\leftarrow\over\gamma\!{}^{i_2},\,\buildrel\leftarrow\over\gamma\!{}^{\bar \imath_2}\,$, viewed as elements of $T\oplus T^*$ (see \cite{GMPT3} for their expression), are shifted as$\,$\footnote{We recall that a generic $v+\zeta \in T\oplus T^*$ gets $B$-transformed into $v+\zeta + \iota_v B\;$ (with a positive sign in front of $\iota_v B$ if the pure spinors transform with $e^{-B}$ \cite{GualtieriThesis} ).} 
\be\nnb
\buildrel\to\over\gamma{}^i= P_1^{\,i}{}_{\,n}(dx^n+ iJ_1^{np}\partial_p)    \quad\overset{B\mathrm{-transf}}{\longrightarrow}\quad \buildrel\to\over\gamma_{\!B}\!\!\!{}^{i}= P_1^{\,i}{}_{\,n}\big(dx^n+ iJ_1^{np}(\partial_p + B_{pq}dx^q)\,\big)\;,\quad\begin{array}{c}\mathrm{(analogous\; for}\\ \mathrm{the\; others).}\end{array}
\ee
$P_{1}$ is the holomorphic projector with respect to the almost complex structure $I_1$. We deduce that, for instance, $\buildrel\to\over\gamma_{\!B}\!\!\!\!{}^{i}\,\bar \Phi_+ = e^{-B}\buildrel\to\over\gamma\!\!{}^i \bar \Phi^0_+$, and similarly for all the other basis elements.

Disposing of an explicit basis, it is easy to check that the generalized diamond is orthogonal with respect to the Mukai pairing, i.e. one can obtain nonvanishing pairings only between forms transforming in conjugate representations $(\bf{r},\bf{s})$ and $(\bf{\bar r},\bf{\bar s})$. This is best seen in the bispinor picture, using the image (\ref{eq:MukaiUnderClifford}) of the Mukai pairing under the Clifford map.
\vskip .3cm
Having introduced the previous technical tools, we can now discuss the moduli space of pure spinors and their relevance for compactifications. Building on a previous work \cite{HitchinFunctionals}, in ref.$\:$\cite{HitchinGenCY} Hitchin showed that the space of even/odd pure spinors at a point admits a rigid special K\"ahler structure. This result was first transposed in the context of supergravity in \cite{GLW1}, to which we also refer for a review of Hitchin's work. Here we just recall that starting from the rigid special K\"ahler structure defined by Hitchin, one can obtain a local special K\"ahler manifold taking the quotient by the $\mathbb C^*$ action corresponding to a rescaling of the pure spinors. Clearly, it is this local special K\"ahler structure which is relevant for the supergravity applications. The K\"ahler potentials $K_\pm$ yielding the local special K\"ahler metrics on the spaces of even/odd pure spinors $\Phi_\pm$ turn out to be \cite{GLW1, GLW2}:
\be\label{eq:Kpm}
e^{-K_\pm}=i\langle\Phi_\pm,\bar\Phi_\pm\rangle\;.
\ee
We stress that this result is valid \emph{at a point} of the 6d manifold $M_6$. Indeed, in (\ref{eq:Kpm}) no integral is performed over the compact space. Put in the context of type II compactifications, this means that we are keeping a full dependence of the higher dimensional fields on both the external spacetime as well as the internal coordinates. We will come back on this issue at the beginning of the next section.

In the $SU(3)$ structure case, substituting the pure spinors (\ref{eq:Phi+-SU3}) into (\ref{eq:Kpm}), one gets for $K_\pm\,$:
\be\label{eq:KahlerPotSU3}
e^{-K_+} = \frac{4}{3}J\wedge J\wedge J\qquad,\qquad e^{-K_-} = -i\Om\wedge \bar \Om\;,
\ee
expressions which are well-known for instance from the analysis of the moduli space of Calabi-Yau manifolds\footnote{However, here we are not integrating over the internal manifold.} \cite{CandelasOssa}.

When dimensional reducing 10d supergravities, the kinetic terms of the 4d scalars associated with the fluctuations of the internal metric and $B$-field are defined by a $\sigma$-model whose target space metric can be written as 
\be
\label{eq:sugraSigma} ds^2\;\sim\;\int_{M_6}  g^{mn}g^{pq}(\delta g_{mp}\delta g_{nq} + \delta B_{mp}\delta B_{nq})vol_6\;.
\ee 
With restriction to the $SU(3)$ structure case, in \cite{GLW1} it was shown that such kinetic terms are reproduced by the sum of the special K\"ahler metrics obtained by variation of the K\"ahler potentials (\ref{eq:KahlerPotSU3}). We now extend this result to the more general $SU(3)\times SU(3)$ structure context.

Let's start discussing deformations of pure spinors. Following Hitchin \cite{HitchinGenCY}\footnote{Related discussions can be found in \cite{GualtieriThesis} (where deformations of generalized complex structures are studied) and in \cite{Tomasiello} (in connection with the landscape of supersymmetric string backgrounds).}, we write the generic infinitesimal variation $\delta\Phi$ of a pure spinor $\Phi$ at a point of $M_6$ as:
\be
\delta\Phi = c\Phi +  \sigma\cdot\Phi\qquad,\qquad\qquad \sigma\cdot \equiv \sigma_{\Lambda\Sigma}\Gamma^{\Lambda\Sigma}\;,
\ee
where $c\in \mathbb C$ is small and $\sigma\cdot$ is an element of the complexified $O(6,6)$ algebra with (infinitesimal) complex parameters $\sigma_{\Lambda\Sigma}\,$. Recalling (\ref{eq:Gamma66}) we can write the $\Gamma^{\Lambda\Sigma}$ as:
\be
\Gamma^{\Lambda\Sigma} = \big(\, dx^m\wedge dx^n\wedge\;,\; \frac{1}{2} [dx^m\wedge, \iota_{\partial_n}] \;,\; \frac{1}{2} [\iota_{\partial_m} , dx^n\wedge]\;,\;\iota_{\partial_m}\iota_{\partial_n}\, \big)\;.\ee
We can also express $\sigma\cdot$ in terms of a basis of creators and annihilators for $\Phi$.  The nonzero variations are obtained acting with the antisymmetrized product of two creators, or of a creator and the associated annihilator (in this case the result is proportional to $\Phi$, and we could absorb it in the parameter $c$). 

Consider the two pure spinors $\Phi_\pm$ together with the $SU(3)\times SU(3)$ structure they identify. Decomposing their variations $\delta\Phi_\pm$ in representations of $SU(3)\times SU(3)$, and referring to the diamond (\ref{eq:Udiamond}), we deduce that:
\be
\delta\Phi_- \;\in\; U_{\bf{1},\bf{1}} \oplus U_{\bf{1},\bf{3}}\oplus U_{\bf{\bar 3},\bf{\bar 3}}\oplus U_{\bf{3},\bf{1}}\qquad, \qquad
\delta\Phi_+ \;\in\; U_{\bf{1},\bf{\bar 1}} \oplus U_{\bf{1},\bf{\bar 3}}\oplus U_{\bf{\bar 3},\bf{3}}\oplus U_{\bf{3},\bf{\bar 1}} \;.
\ee
However, we require the deformed pure spinors $\Phi_\pm + \delta\Phi_\pm$ again be compatible, and this imposes constraints on the allowed variations. Indeed, varying the compatibility condition (\ref{eq:compatibility}) we get:
\be\label{eq:varCompat}
\langle \delta\Phi_+, \Gamma^\Lambda \Phi_-\rangle + \langle \Phi_+, \Gamma^\Lambda \delta\Phi_-\rangle = 0\qquad,\qquad  \langle \delta\bar \Phi_+, \Gamma^\Lambda \Phi_-\rangle +\langle \bar \Phi_+, \Gamma^\Lambda \delta\Phi_-\rangle = 0\;.
\ee
Here we want to avoid imposing any relation between the $\Phi_+$- and $\Phi_-$-deformations, so we demand that each Mukai pairing in (\ref{eq:varCompat}) vanishes separately; as a consequence, all the variations of $\Phi_\pm$ transforming in the vector of $O(6,6)$ (corresponding to the $({\bf 3},{\bf 1})\oplus ({\bf {\bar 3}},{\bf 1})\oplus ({\bf 1},{\bf 3})\oplus ({\bf 1},{\bf{\bar 3}})$ of $SU(3)\times SU(3)\,$) are removed. We argue in this way that $\delta\Phi_- \in U_{\bf{1},\bf{1}} \oplus U_{\bf{\bar 3},\bf{\bar 3}}$ and $\delta\Phi_+ \in U_{\bf{1},\bf{\bar 1}} \oplus U_{\bf{\bar 3},\bf{3}}\;$. This generalizes an analogous argument proposed in ref.$\:$\cite{GLW1} in the context of $SU(3)$ structures. Furthermore, it supports the prescription given in ref.$\:$\cite{GLW2} of projecting out all the fields transforming in the vector representation of $O(6,6)$ when decomposing the 10d supergravity fields on the basis (\ref{eq:Udiamond}), due to the fact that they would assemble to define spin $3/2$ multiplets in 4d, which correspond to non-standard couplings of $N=2$ supergravity.

Preserving the `equal norm' condition (\ref{eq:equalNorm}) is not strictly necessary; however, in order to achieve this, it is sufficient to equate the real parts of the coefficients parameterizing the rescaling piece of the $\Phi_+$ and $\Phi_-$ deformations. As it will be clear in the following, this does not introduce any relation between the moduli spaces of $\Phi_+$ and $\Phi_-\,$.

We eventually rewrite the infinitesimal deformations of $\Phi_\pm$ at a point in a notation reminding the Kodaira formula for the holomorphic 3-form $\Om$ of a Calabi-Yau manifold (see eq. (\ref{eq:KodairaOm}) below):
\be\label{eq:variationPhi}
\delta \Phi_\pm = \delta \kappa_\pm \Phi_\pm + \delta \chi_\pm\;. 
\ee 
where $\delta\kappa_\pm$ are complex parameters, while $\delta\chi_- \in U_{{\bf \bar 3},{\bf \bar 3}}$ and $\delta\chi_+ \in U_{{\bf \bar 3},{\bf 3}}$ can be expanded on the basis (\ref{eq:BasisDiamond}) as $\delta\chi_\pm = e^{-B}\delta\chi^0_\pm\,$, with $\delta\chi^0_\pm = (\delta \chi_\pm)_{mn} \gamma^m \Phi^0_\pm \gamma ^n$.

As a first application of the above discussion, we can give an expression for the special K\"ahler metrics on the space of pure spinors evaluating the holomorphic and antiholomorphic variations of the K\"ahler potentials (\ref{eq:Kpm}). Using (\ref{eq:variationPhi}), we obtain:
\be\label{eq:MetricOnPhi+-}
ds^2_\pm \,=\, \delta^{\mathrm{holo}}\delta^{\mathrm{anti}}K_\pm \;=\; \frac{\langle \Phi_\pm, \delta\bar\Phi_\pm\rangle}{\langle \Phi_\pm, \bar\Phi_\pm\rangle}\frac{\langle \delta\Phi_\pm, \bar\Phi_\pm\rangle}{\langle \Phi_\pm, \bar\Phi_\pm\rangle} - \frac{\langle \delta\Phi_\pm, \delta\bar\Phi_\pm\rangle}{\langle \Phi_\pm, \bar\Phi_\pm\rangle}\;\,=\;\, - \frac{\langle \delta\chi_\pm, \delta \bar\chi_\pm\rangle}{\langle \Phi_\pm, \bar\Phi_\pm\rangle}\;.
\ee
Notice that the rescalings of the pure spinors don't contribute to the metric.

We now analyse the relation of the pure spinor deformations with the supergravity \hbox{$\sigma$-model} (\ref{eq:sugraSigma}). We will show that this last can be expressed as the sum of two independent contributions, associated with the variations of the two generalized almost complex structures $\mathcal J_\pm$ given in (\ref{eq:CurlyJ+-}). These two terms will turn out to be the special K\"ahler metrics (\ref{eq:MetricOnPhi+-}). 

Starting from (\ref{eq:TT*metric}), we observe that the integrand of (\ref{eq:sugraSigma}) can also be written in terms of fluctuations of the $T\oplus T^*$ metric $\mathcal G$:
\be\label{eq:metricModuliSpace}
g^{mn}g^{pq}(\delta g_{mp}\delta g_{nq} + \delta B_{mp}\delta B_{nq}) = -\frac{1}{2}\Tr \big(\delta \mathcal G\delta \mathcal G\big)\;,
\ee
where the trace is taken over the $T\oplus T^*$ indices $\Lambda, \Sigma$. This can be expressed in terms of deformations of the generalized almost complex structures $\mathcal J_\pm$. Indeed, recalling (\ref{eq:TT*metric}) we have $\,\delta \mathcal G = -(\delta\mathcal J_+) \mathcal J_- - \mathcal J_+(\delta \mathcal J_-)\,$, and hence
\be\label{eq:deltaGAsDeltaJ}
\Tr(\delta \mathcal G\delta \mathcal G) = \Tr\big[(\delta\mathcal J_+) \mathcal J_- + \mathcal J_+(\delta \mathcal J_-)\big]\big[(\delta\mathcal J_+) \mathcal J_- + \mathcal J_+(\delta \mathcal J_-)\big]\;.
\ee
To evaluate the variations of $\mathcal J_\pm$ we put them in relation with the pure spinor deformations. From (\ref{eq:RelJPhi}) we have (omitting the $\pm$ label for simplicity):
\be\label{eq:VarJstep1}
\delta\mathcal J_{\Lambda\Sigma}=\frac{8\langle \re(\delta\Phi) , \Gamma_{\Lambda\Sigma}\re \Phi \rangle}{i\langle \Phi , \bar\Phi \rangle} -\mathcal J_{\Lambda\Sigma} \frac{\delta\langle \Phi , \bar\Phi \rangle}{\langle \Phi , \bar\Phi \rangle}\;,
\ee
where we collected the two terms containing $\re(\delta\Phi)$ using (\ref{eq:GammaPass}). Using (\ref{eq:variationPhi}), we write $\re(\delta\Phi)$ as $\re(\delta\kappa)\re \Phi - \im(\delta\kappa)\im\Phi + \re(\delta\chi)$. Now, the contribution of $\re(\delta\kappa)\re \Phi$ in (\ref{eq:VarJstep1}) compensates exactly the variation of $\delta\langle \Phi , \bar\Phi \rangle$, while it is not difficult to see that $\im(\delta\kappa) \langle\im\Phi, \Gamma_{\Lambda\Sigma}\re\Phi\rangle$ vanishes. Therefore the piece of the variation of $\Phi$ consisting in a rescaling drops out. This was expected, since, as we already discussed, a generalized almost complex structure is in one-to-one correspondence with a complex \emph{line} of pure spinors. Therefore we obtain
\be\label{eq:VarJ}
\delta\mathcal J_{\pm\Lambda\Sigma} = \frac{8\langle \re(\delta\chi_\pm) , \Gamma_{\Lambda\Sigma}\re\Phi_\pm\rangle}{i\langle \Phi_\pm , \bar \Phi_\pm \rangle}\;.
\ee
Since $\re(\delta\chi_-) \in U_{{\bf \bar 3},{\bf \bar 3}}\oplus U_{{\bf  3},{\bf  3}}$ and $\re(\delta\chi_+) \in U_{{\bf \bar 3},{\bf 3}}\oplus U_{{\bf 3},{\bf \bar 3}}$, the only nonzero contributions to $\delta \mathcal J_\pm$ come from the components of $\Gamma_{\Lambda\Sigma}\re\Phi_\pm$ being in the same representations. For $B=0$ these are of the form $\gamma^m\re \Phi^0_\pm\gamma^n$ (see eq.$\:$\ref{eq:2GammaUnderCliff}), while for nonvanishing $B$ there are extra contributions yielding the matrices ${1\; 0 \choose -B\; 1}$ and ${1\;\; 0 \choose B\; 1}$ as discussed at the end of App.$\:$\ref{MukaiAndClifford}.

We are now ready to evaluate (\ref{eq:deltaGAsDeltaJ}). By a quite long but straightforward computation one can see that the terms mixing the variations of $\mathcal J_+$ and $\mathcal J_-$ vanish:
\be 
\Tr\big[\mathcal J_-\mathcal J_+(\delta \mathcal J_-)(\delta\mathcal J_+)  \big]\;\equiv\; -\Tr\big[\mathcal G(\delta \mathcal J_-)(\delta\mathcal J_+)  \big]\; =\;0\;.
\ee
In order to evaluate this we used the bispinor picture, in particular the image under the Clifford map of the Mukai pairing and of $\Gamma_{\Lambda\Sigma}$, given by eqs. (\ref{eq:MukaiUnderClifford}) and (\ref{eq:2GammaUnderCliff}) respectively; we found cancellation between all the nonzero terms involved in the trace. Therefore the metric (\ref{eq:metricModuliSpace}) factorizes into the sum of two contributions, parameterizing the independent deformations of $\mathcal J_-$ and $\mathcal J_+$ (or, equivalently, of the associated pure spinors $\Phi_\pm$):
\be
-\frac{1}{2}\Tr(\delta \mathcal G\delta \mathcal G) \;=\;  -\frac{1}{2}\Tr\big[\mathcal J_+(\delta \mathcal J_-)\mathcal J_+(\delta \mathcal J_-)\big] -\frac{1}{2} \Tr\big[(\delta\mathcal J_+) \mathcal J_-(\delta\mathcal J_+) \mathcal J_-\big]\;.
\ee
Again we can rewrite these terms using the bispinor picture. For the first one we find (omitting the slashes in order not to clutter the formulas)
\begin{eqnarray}
\nnb-\frac{1}{2}\Tr\big[\mathcal J_+(\delta \mathcal J_-)\mathcal J_+(\delta \mathcal J_-)\big] & = & \frac{1}{ 8^6} \tr[\gamma(\re\Phi^0_+)^{T} \gamma_{mp}\re\Phi^0_+] \tr[\gamma(\re\Phi^0_+)^{T} \re\Phi^0_+\gamma_{nq} ]\cdot\\ 
\nnb &&\cdot\tr[\gamma\re(\delta\chi^0_-)^{T} \gamma^{p}\re\Phi^0_-\gamma^n] \tr[\gamma \re(\delta\chi^0_-)^{T} \gamma^{m}\re\Phi^0_-\gamma^q] \\ [2mm]
\nnb &=& 8(\delta\chi_-)_{mn}  (\delta\bar\chi_-)_{pq}(g^{mp}+iJ_1^{mp})(g^{nq}+iJ_2^{nq})\\ [1mm]
&=& - 8\frac{\langle \delta\chi_-, \delta \bar\chi_-\rangle}{\langle \Phi_-, \bar\Phi_-\rangle}\;.
\end{eqnarray}
The computation for the term involving the variation of $\delta \mathcal J_+$ is completely analogous. We conclude that $\Tr(\delta \mathcal G\delta \mathcal G)$, capturing the internal metric and $B$-field fluctuations (recall (\ref{eq:metricModuliSpace})), is expressed as
\be
-\frac{1}{16}\Tr(\delta \mathcal G\delta \mathcal G)  \;=\; - \frac{\langle \delta\chi_-, \delta \bar\chi_-\rangle}{\langle \Phi_-, \bar\Phi_-\rangle} - \frac{\langle \delta\chi_+, \delta \bar\chi_+\rangle}{\langle \Phi_+, \bar\Phi_+\rangle} \;\equiv\; ds^2_- \,+\, ds^2_+  \;.
\ee
This indeed coincides with the sum of the special K\"ahler metrics (\ref{eq:MetricOnPhi+-}) for the $\Phi_-$- and $\Phi_+$- deformation spaces.

As a last remark, we recall that in the evaluation of $\delta\mathcal J_\pm$ we have discarded the deformations of $\Phi_\pm$ in the vector representation of $O(6,6)$. Starting from a similar argument given in ref.$\:$\cite{GLW1} in the context of $SU(3)$ structures, it would be interesting to explicitly verify whether such deformations correspond to variations of the generalized almost complex structures which don't modify the metric on $T\oplus T^*$.

\section{$N=2$ effective theory in four dimensions}\label{KKreduct}

\setcounter{equation}{0}

In the previous subsection we worked at a point of the internal manifold $M_6$, and nowhere we took the integral over it. This implies that the above expressions (in particular the K\"ahler potentials (\ref{eq:Kpm}) and the metrics (\ref{eq:MetricOnPhi+-}) on the space of even/odd pure spinors) are not yet associated with an actual 4d effective theory. For this to be defined, one should expand the higher dimensional fields in eigenforms of appropriate mass operators and then perform a truncation of the KK spectrum, keeping in this way just a finite set of light modes.

However, as discussed in the introduction, in the presence of fluxes a precise characterization of the expansion forms defining the light 4d degrees of freedom is not known. Adopting the off-shell approach to the compactification one definitely does not try to obtain such characterization, which would require to fix at least some properties of the background around which to study the fluctuations. Rather, one works with a finite basis of forms for the light modes whose properties stay unspecified, but which is assumed to satisfy a set of constraints allowing to define a consistent 4d $N=2$ effective supergravity action. Crucially, these forms are not necessarily closed, and one even is led to consider cases in which they are of mixed degree \cite{GLW2}.

The safest way to proceed in order that the result of the compactification displays the features of an $N=2$ supergravity in 4d is to stay as close as possible to the well-known path of the Calabi-Yau case. For example, one of the features of Calabi-Yau compactifications one wants to reproduce is the fact that the $\sigma$-model governing the kinetic terms of the scalars associated with the metric and $B$-field fluctuations has a target space consisting of the product of two special K\"ahler manifolds (this is an essential aspect of mirror symmetry, but it's not strictly required by 4d $N=2$ supergravity: the quaternionic manifold needs not contain a special K\"ahler subspace). As we have seen in the previous subsection, $SU(3)\times SU(3)$ structure backgrounds have this property: the spaces of both even and odd pure spinors are special K\"ahler, and their metric precisely describes the internal metric and $B$-field fluctuations. One of the aims of the mentioned constraints on the basis of expansion forms is to guarantee that the same structure is inherited by the 4d theory for a finite set of modes.

In the generalized geometry context, conditions for the reduction to go through similarly to the Calabi-Yau case have been discussed in \cite{GLW1, GLW2}. For the $SU(3)$ structure case, a thorough analysis with a complete list of the constraints on the basis forms has been presented in  \cite{MinasianKashani}. In the next subsection we give a somehow more explicit version of the analysis of \cite{GLW1, GLW2}, partially to fix our notations. In addition, we stress the relevance of decomposing the pure spinor deformations in representations of $SU(3)\times SU(3)$.

\subsection{Ansatz for the basis forms and special K\"ahler geometry on the truncated space of pure spinors}\label{AnsatzForms}

Henceforth we will integrate over the internal manifold $M_6$. The integrated Mukai pairing $\int_{M_6}\langle\,\cdot\,,\,\cdot\,\rangle\;$ will then provide 4d scalars\footnote{Since there is no risk of confusion, in the following we will omit the label $M_6$ for the integral.}. We denote with $\mathscr M_-$ and $\mathscr M_+$ the K\"ahler manifolds describing the truncated spaces of odd and even pure spinors, and we take their (finite) dimensions to be $b^-$ and $b^+$ respectively.

In the notation of \cite{GLW2}, we split the finite basis of forms in two subsets $\Sigma^-$ and $\Sigma^+$, composed of odd and even \emph{real} forms respectively, not necessarily of pure degree. We require that a symplectic structure is defined by means of the integrated Mukai pairing, i.e. that
\begin{eqnarray}
\nnb\Sigma^- &=& \{\alpha_I\,,\,\beta^J\}\;\;,\quad\qquad I,J\,=\,0,1,\ldots, b^- \\ [1mm]
\nnb\Sigma^+ &=&\{\om_A, \tilde \om^B\}\;\;,\quad\qquad A,B=0,1,\ldots,b^+
\end{eqnarray}
satisfying
\be \label{eq:SymplStrM-}\; \left(\begin{array}{cc} \int\langle \alpha_I,\alpha_J\rangle \;&\; \int\langle \alpha_I,\beta^J\rangle\\ [2mm]
\int\langle \beta^I,\alpha_J\rangle \;&\; \int\langle \beta^I , \beta^J \rangle  \end{array}\right)\;\; =\;\;
\left(\begin{array}{cc} 0 & \delta_I^{\;\;J} \\ [2mm] -\delta^I_{\;\;J} & 0
\end{array}\right)\;:=\;\mathbb{S}_-\;,
\ee
\vskip 2mm
\be\label{eq:evenSymplecticStr} \left(\begin{array}{cc} \int\langle \om_A,\om_B\rangle \;&\; \int\langle \om_A,\tilde\om^B\rangle\\ [2mm]
\int\langle \tilde\om^A,\om_B\rangle \;&\; \int\langle \tilde\om^A,\tilde\om^B\rangle  \end{array}\right)\; =\;
\left(\begin{array}{cc} 0 & \delta_A^{\;\;B} \\ [2mm] -\delta^A_{\;\;B} & 0
\end{array}\right)\;:=\; \mathbb{S}_+ \;.
\ee
Hence $\mathbb{S}_\pm$ are the symplectic metrics of $Sp(2b^\pm + 2, \mathbb R)$.

As illustrated in \cite{GLW2}, a first condition to be imposed on $\Sigma^{\pm}$ is
\be
\label{eq:StrongCompatibility}\langle \om_A , \Gamma^\Lambda \alpha_I\rangle = \langle \om_A , \Gamma^\Lambda \beta^I\rangle = \langle \tilde\om^A , \Gamma^\Lambda \alpha_I\rangle =\langle \tilde\om^A , \Gamma^\Lambda \beta^I\rangle = 0\;.
\ee
Its relevance is twofold: on the one hand, since in the following the pure spinors $\Phi_\pm$ will be expanded on the basis $\Sigma^\pm$, eq. (\ref{eq:StrongCompatibility}) corresponds to requiring that the compatibility condition (\ref{eq:compatibility}) is respected already at the level of the basis forms, preventing in this way a relation between the moduli of $\Phi_+$ and $\Phi_-$. On the other hand, (\ref{eq:StrongCompatibility}) ensures that none of the modes we keep transforms in the $({\bf 3}, {\bf 1}) \oplus ({\bf \bar 3}, {\bf 1}) \oplus ({\bf 1} , {\bf 3}) \oplus ({\bf 1} , {\bf \bar 3})$ of $SU(3)\times SU(3)$; therefore no massive spin-$3/2$ multiplets appear in the effective action for the light degrees of freedom. It follows that the $SU(3)\times SU(3)$ representations relevant to the definition of the $N=2$ effective action reside in the horizontal and vertical axis of the diamond (\ref{eq:Udiamond}). This is somehow analogous to the Calabi-Yau case, where however the diamond is a true Hodge diamond, in that it consists of $(p,q)$-\emph{harmonic} forms.

\vskip .5cm

\underline{\emph{Special K\"ahler geometry for $\mathscr M_-$}}\\ [-2mm]

\noindent Using the basis forms and the Mukai pairing we can define the periods of $\Phi_-$:
\be
\label{eq:symplSectPhi-} Z^I:=\int\langle\Phi_-,\beta^I\rangle\qquad,\qquad \mathcal G_I:=\int\langle \Phi_-,\alpha_I\rangle\;.
\ee
Then $\Phi_-$ can be expanded on the basis forms as: 
\be
\label{eq:expansPhi-}\Phi_-=Z^I\alpha_I-\mathcal G_I\beta^I\;.
\ee
{From} (\ref{eq:symplSectPhi-}) we see that performing a constant rescaling $\Phi_-\to\lambda\Phi_-$ implies $Z^I\to\lambda Z^I$ and $\mathcal G_I\to\lambda \mathcal G_I\,$. We would like to conclude that $\Phi_-$ is a homogeneous function of degree 1 in the $Z^I$ variables, and then see these as projective coordinates for $\mathscr M_-\,$. For this to be true, we need that the $Z^I$ define $b^-$ independent functions on $\mathscr M_-$ (then the $\mathcal G_I$ are holomorphically determined by the $Z^J$), and that the basis forms are homogeneous of degree 0 in the $Z^I$. Once this is satisfied, away from the $Z^0=0$ locus we can also introduce special coordinates $z^i=Z^i/Z^0,\, i = 1,2,\ldots,b^-$ for $\mathscr M_-\,$.

Given (\ref{eq:expansPhi-}), the K\"ahler potential $K_-$ written in (\ref{eq:Kpm}) takes the standard form of special geometry:
\be\label{eq:K-standardForm} K_- \equiv -\ln i\int\langle\Phi_-,\bar\Phi_-\rangle=-\ln i(\bar Z^I \mathcal G_I-Z^I\bar{\mathcal G}_I)\;.\ee
In the Calabi-Yau case, an essential tool to show that the space of complex structure deformations is special K\"ahler consists in the Kodaira formula (see, for instance, ref.$\:$\cite{CandelasOssa}). If $\tilde\Om(z^i)$ is the holomorphic $(3,0)$-form of the Calabi-Yau 3-fold, then this formula states that its variation with respect to the moduli $z^i$ can be written as 
\be\label{eq:KodairaOm}
\pd{\tilde\Om}{z^{i}}= \tilde \kappa_i \tilde\Om +\tilde \chi_i\qquad\qquad i=1,\ldots, h^{2,1}\;,
\ee
where $\tilde \kappa_i$ are coefficients which can depend on $z$ but not on the coordinates of $M_6$, and $\{\tilde \chi_i\}$ is a basis for the $(2,1)$-harmonic forms. Introducing $\Om(Z^I)=Z^0\tilde\Om(z^i)\,$, we can also rewrite (\ref{eq:KodairaOm}) in terms of projective coordinates $Z^I=(Z^0, \,Z^i=Z^0z^i)$  as
\be\label{eq:KodairaBis}
\pd{\Om}{Z^{I}}= \kappa_I \Om +\chi_I\qquad\qquad I=0,1,\ldots, h^{2,1}\;,
\ee
where $\,\kappa_I = (\kappa_0,\kappa_i) =  \frac{1}{Z^0}(1-z^i\tilde \kappa_i \,,\, \tilde \kappa_i) $ and $ \chi_I =(\chi_0, \chi_i)=(-z^i\tilde\chi_i \,,\, \tilde \chi_i) \;$. Notice that $\chi_i=\tilde\chi_i$ is homogeneous of degree 0.

We now reconsider deformations of pure spinors, which in subsect.$\:$\ref{variations} we wrote in the form (\ref{eq:variationPhi}), and we rephrase them in a form analogous to (\ref{eq:KodairaOm}) and (\ref{eq:KodairaBis}). Parameterizing the truncated space of pure spinors $\mathscr M_-$ by the moduli $z^i$, or alternatively by the projective coordinates $Z^I$, we can write:
\be\label{eq:KodairaPhi-}
\pd{\tilde \Phi_-}{z^i}\,\sim\, \tilde\kappa^-_i \tilde\Phi_- + \tilde\chi^-_i\;\;, \;\qquad\textrm{or}\;\qquad \pd{\Phi_-}{Z^I}\,\sim\, \kappa^-_I \Phi_- + \chi^-_I\;,
\ee
where the tildes have the same meaning as above, and the relations between the $\kappa_I,\;\chi_I$ and the $\tilde \kappa_i,\;\tilde\chi_i$ are also the same. Referring to (\ref{eq:variationPhi}), we identify $\delta \kappa_-= \tilde\kappa^-_i\delta z^i\;$, $ \delta \chi_- = \tilde\chi^-_i \delta z^i\,$, and therefore we have $\chi^-_I\in U_{\bf{\bar 3},\bf{\bar 3}}\,$.

Adopting the notation of \cite{GLW2}, here and in the following by the symbol $\sim$ we mean `equality up to terms that vanish in the symplectic pairing'. In the above expression it is required because in principle the pure spinor variations contain a term transforming in the triplets of $SU(3)\times SU(3)$, and we are preventing its presence in the light spectrum by assumption (\ref{eq:StrongCompatibility}).

Since (\ref{eq:KodairaPhi-}) does not contain a term proportional to $\bar \Phi_-$, we have
\be
\qquad\qquad\label{eq:PhiDePhi}\int\langle\Phi_-, \partial_I \Phi_-\rangle=0\qquad\qquad\qquad\qquad {\textstyle(\partial_I\equiv \pd{\,}{Z^I})}\;,
\ee
which indeed is a necessary condition for special K\"ahler geometry. From the expansion (\ref{eq:expansPhi-}) we have
\be\label{eq:DelPhi-} 
\partial_I\Phi_- = \alpha_I - \partial_I\mathcal G_{K}\beta^K + Z^K\partial_I\alpha_K - \mathcal G_K\partial_I \beta^K\;,
\ee
where the last two terms have been taken into account because in general the expansion forms are moduli dependent. This is true also when considering a Calabi-Yau 3-fold, but in this case $\partial_I\alpha_J$ and $\partial_I\beta^J$ are exact and don't contribute to the integral. In the more general case this is not automatic, and we are led to require
\be\label{eq:AlphaDeAlpha}
\int\langle \alpha_J,\partial_I\alpha_K \rangle = \int\langle \alpha_J,\partial_I\beta^K \rangle = \int\langle \beta^J,\partial_I\beta^K \rangle =0\;.
\ee
This also guarantees constancy of the symplectic structure (\ref{eq:SymplStrM-}). Analogously to the Calabi-Yau case, (\ref{eq:PhiDePhi}) then gives 
\be 2\mathcal G_I=\partial_I(Z^K\mathcal G_K)\;,\ee
which implies that $\mathcal G:=\frac{1}{2}Z^K\mathcal G_K$ is a homogeneous function of degree 2 in the $Z$ variables (the prepotential), and $\mathcal G_I=\partial_I\mathcal G$. Then $\mathcal G_I$ is homogeneous of degree 1: $\mathcal G_I = Z^K \partial_K \mathcal G_I$. We will denote $\mathcal G_{IJ}:=\partial_I\mathcal G_J=\partial_I\partial_J \mathcal G$.

We can now derive an expression for the coefficient $\kappa^-_I$ appearing in (\ref{eq:KodairaPhi-}) in terms of the special geometry data. Assuming that $\kappa^-_I$ does not depend on the coordinates of $M_6$ (this condition is automatically verified in the Calabi-Yau case), we obtain
\be\label{eq:kappaI}
\kappa^-_I= \frac{\int \langle \partial_I\Phi_- , \bar\Phi_- \rangle}{\int\langle \Phi_- , \bar\Phi_- \rangle} = \frac{\im\mathcal G_{IJ}\bar Z^J}{Z^K\im\mathcal G_{KL}\bar Z^L}\;.
\ee
where for the first equality we used the orthogonality of the different representations in (\ref{eq:Udiamond}). Notice that $\kappa^-_I = - \partial_I K_-$ and therefore from (\ref{eq:KodairaPhi-}) $\chi^-_I \sim D_I\Phi_- \sim D_I Z^J\alpha_J - D_I\mathcal G_J \beta^J$, where $D_I= \partial_I + \partial_IK_-$. Again, these are direct generalizations of expressions valid in the Calabi-Yau case (see e.g. \cite{CandelasOssa}).\\ [-2mm]

Provided the whole set of conditions summarized in this subsection is satisfied (with possible additional conditions along the lines of \cite{MinasianKashani}), we can conclude that $\mathscr M_-$ has a local special K\"ahler structure. From (\ref{eq:MetricOnPhi+-}) it follows that the metric $g^-_{i\bar \jmath}$ on $\mathscr M_-$ is given by
\be\label{eq:metricOnM-}
g^-_{i\bar \jmath}\,=\,\pd{\,}{z^i}\pd{\,}{\bar z^{\bar \jmath}} K_- \,=\, - \frac{\int\langle \chi^-_i\,,\, \bar\chi^-_{\bar \jmath}\rangle}{\int\langle \Phi_-, \bar\Phi_-\rangle}\;.
\ee
In the Calabi-Yau case, (\ref{eq:metricOnM-}) reduces to the well-known expression $g_{i\bar\jmath} = -\frac{\int \chi_i\wedge \bar \chi_{\bar \jmath}}{\int \Om\wedge\overline{\Om}}$, where $\chi_i$ are the harmonic (2,1)-forms \cite{CandelasOssa}.

As briefly reviewed in App.$\,$\ref{SpecialGeometry}, certain important properties of special K\"ahler geometry are expressed in terms of the period matrix - that in the present case we call $\mathcal M$ - defined via the relations
\be \mathcal G_I=\mathcal M_{IJ}Z^J\qquad,\qquad D_k\mathcal G_I=\overline {\mathcal M}_{IJ}D_kZ^J\;,\ee
where the K\"ahler covariant derivative $D_k$ acts on the periods as $D_{k}=\partial_{z^k}+\partial_{z^k}K_-\,$. $\mathcal M$ is also an important ingredient of the compactification, since it appears explicitly in the 4d $N=2$ effective action.

\vskip .4cm

\underline{\emph{Special K\"ahler geometry for $\mathscr M_+$}}\\ [-2mm]

\noindent Parallel arguments and similar requirements can be adopted to ensure the special K\"ahler structure of $\mathscr M_+$. Here we summarize the important relations, mainly to fix our notation.\\
The periods of $\Phi_+$ are defined as
\be
\label{eq:symplSectPhi+} X^A:=\int\langle\Phi_+,\tilde\om^A\rangle\qquad,\qquad \mathcal F_A:=\int\langle\Phi_+,\om_A\rangle\;.
\ee
$\Phi_+$ is then expanded on the truncated basis of forms as: 
\be
\label{eq:expansPhi+}\Phi_+=X^A\om_A-\mathcal F_A\tilde\om^A\;.
\ee
The $\mathcal F_A$ are holomorphic functions of the $X^A$, and can be obtained by $\partial_{X^A}\mathcal F$, where $\mathcal F$ is the prepotential (holomorphic and homogeneous of degree two in the $X^A$). We denote the special coordinates for $\mathscr M_+$ as $t^a=X^a/X^0$. The K\"ahler potential $K_+$ is expressed as
\be\label{eq:K+standardForm}
K_+ = -\ln i\int\langle\Phi_+,\bar\Phi_+\rangle=-\ln i(\bar X^A\mathcal F_A-X^A\bar{\mathcal F}_A)\;.
\ee
The metric $g^+_{a \bar b}$ on $\mathscr M_+$ can be obtained from $K_+$ by:
\be
g^+_{a \bar b}=\pd{\,}{t^a}\pd{\,}{\bar t^{\bar b}} K_+ \,=\,- \frac{\int\langle \chi^+_a\,,\, \bar\chi^+_{\bar b}\rangle}{\int\langle \Phi_+, \bar\Phi_+\rangle}\;.
\ee
The period matrix $\mathcal N$ for the special geometry on $\mathscr M_+$ is given by
\be\label{eq:DefMathcalN}
\mathcal F_A=\mathcal N_{AB}X^B\qquad, \qquad D_a \mathcal F_B=\overline{\mathcal N}_{BC}D_aX^C\;,
\ee
where here the K\"ahler covariant derivative is $D_a=\partial_{t^a}+\partial_{t^a}K_+\;$.\\ [-2mm]

The $z^i$ and $t^a$ coordinates for $\mathscr M_-$ and $\mathscr M_+$ are the light moduli associated with the internal metric and $B$-field fluctuations. From the above discussion, together with the results of subsect.$\:$\ref{variations}, we conclude that their kinetic terms entering the 4d effective lagrangian are\footnote{The notation should not lead to confusion: $g_{mn}$ is the metric on $M_6$, while $g^-_{i\bar \jmath}$ and $g^+_{a\bar b}$ are the special K\"ahler metrics on the moduli spaces $\mathscr M_-$ and $\mathscr M_+$.}:
\be
\frac{1}{8\int vol_6}\int g^{mn}g^{pq}(\partial_\mu g_{mp}\partial^\mu g_{nq} + \partial_\mu B_{mp}\partial^\mu B_{nq})vol_6 \;=\; g^-_{i\bar \jmath}\partial_\mu z^i\partial^\mu \bar z^{\bar \jmath} \,+\, g^+_{a\bar b}\partial_\mu t^a\partial^\mu \bar t^{\bar b}\;.
\ee

\subsection{The twisted Hodge star $*_B$}\label{*B}

Another piece of information about the 4d $N=2$ effective theory for $SU(3)\times SU(3)$ compactifications can be extracted from the study of the $B$-twisted Hodge star operator \cite{JeschekWitt1, WittThesis, GrimmBen}:
\be\label{eq:def*B} *_B=: e^{-B}*\lambda\, e^B\;,\ee
which is the covariant generalization of the usual Hodge $*$ when considering $O(6,6)$ spinors containing the $B$-field, as $\Phi_\pm=e^{-B}\Phi^0_\pm$.

In particular, we are interested in identifying the action of $*_B$ on the basis of forms $\Sigma^{\pm}$ in terms of the special geometry data. Besides its importance for obtaining the $N=2$ effective action, this will be needed in subsection \ref{ExpPureSpEq} when expanding the pure spinor equations.

We start with a couple of remarks. It is easy to check that $(*_B)^2=-id$; therefore its eigenvalues are $\pm i$ and an almost complex structure is defined on $\wedge^\bullet T^*$. At least when the bispinor picture can be used, one can readily verify that the $U_{\bf{r},\bf{s}}$ defined in (\ref{eq:Udiamond}) are $\pm i$ eigenspaces for $*_B$. This can be seen as follows: in the differential form picture, consider the $B$-transformed of (\ref{eq:BasisDiamond}), and act on it with $*_B$; then pass to the bispinor picture, using (\ref{eq:action*lambda}) to evaluate $*\lambda$ under the Clifford map. One obtains the eigenvalues\footnote{For $B=0$, this can be found in \cite{Tomasiello}.}:

\be\label{eq:iDiamond}
\begin{array}{ccccccc}
&&& i &&&\\
&& i && -i &&\\
& i && -i  && i & \\
i && -i  && i && -i\\
& -i && i  && -i & \\
&& i && -i  && \\
&&& -i &&&
\end{array}\ee
In particular, we have $*_B \Phi_\pm= i\Phi_\pm$, and therefore
\be*_B\re(\Phi_\pm)= -\im(\Phi_\pm).\ee
So we can conclude that once the metric has been fixed, $*_B$ behaves as the derivative of the Hitchin function \cite{HitchinGenCY}, since acting on the real part of the pure spinor it gives minus its imaginary part\footnote{The same holds for the derivative of the Hitchin function $H$: $\pd{H(\re \Phi)}{(\re \Phi)}=-\im \Phi$. Anyway, the operator defined by Hitchin is more general in that it does not need the metric.}.

Let us now determine the action of the $*_B$ operator on the elements of the basis $\Sigma^\pm$. In doing so, we will generalize the well known result of refs. \cite{Suzuki, CeresoleEtAl1} for the action of the usual Hodge $* $ on the harmonic 3-forms of a Calabi-Yau 3-fold (see also \cite{MinasianKashani} for the $SU(3)$ structure case). In the Calabi-Yau case, such a result is obtained starting from the simple observation that the Hodge $*$ acts\footnote{See (\ref{eq:defHodge*}) for our convention on the Hodge $*$.} as $-i$ on $(3,0)$-forms and as $+i$ on the $(2,1)$-harmonic forms which parameterize the complex structure deformations.

Our generalization employs the decomposition of $\wedge^\bullet T^*$ in terms of $SU(3)\times SU(3)$ representations instead of the $(p,q)$-decomposition of complex forms with fixed degree.

As a starting point we need the assumption that the action of $*_B$ on the elements of $\Sigma^{\pm}$ can still be expanded on $\Sigma^\pm$. Focusing on $\Sigma^-$:
\be
\label{eq:Expansion*B} *_B\alpha_I\sim \mathcal A_{I}^{\;J}\alpha_J + \mathcal B_{IJ}\beta^J \qquad,\qquad *_B\beta^I \sim \mathcal C^{IJ}\alpha_J\ +\mathcal D^I_{\;J}\beta^J \;.
\ee
In particular we require that the matrices $\mathcal A,\mathcal B,\mathcal C,\mathcal D$ do not depend on the coordinates of $M_6$.\\
For a Calabi-Yau (\ref{eq:Expansion*B}) is not an assumption but a matter of fact since $\Sigma^-$ consists of harmonic 3-forms.

Using (\ref{eq:SymplStrM-}) and the fact that for every $A, C\in \wedge^\bullet T^*$  $\langle A,*_BC\rangle=-\langle *_B A,C\rangle\,$ (this descends from eqs.$\:$(\ref{eq:Bdrops})-(\ref{eq:lambdaPasses})), we see immediately that ($I,J$ indices are understood):
\be
\mathcal B^T= \mathcal B=\int\langle\alpha,*_B\alpha\rangle\quad,\quad \mathcal C^T=\mathcal C=-\int\langle\beta,*_B\beta\rangle\quad,\quad -\mathcal A^T=\mathcal D=\int\langle\alpha,*_B\beta\rangle\;.
\ee
Applying $*_B$ to (\ref{eq:Expansion*B}), using $(*_B)^2=-id\;$ and (\ref{eq:SymplStrM-}), one can see that the matrix 
\be
\mathbb{M}:=\;\left(\begin{array}{cc}  \int\langle \alpha,*_B\beta\rangle  \; &\; -\int\langle \beta,*_B\beta\rangle \\
\int\langle \alpha,*_B\alpha\rangle \;& \;  -\int\langle \beta,*_B\alpha\rangle \end{array}\right)   \; =\;
\left(\begin{array}{cc}
\mathcal D & \mathcal  C \\ \mathcal B & \mathcal A
\end{array}\right)
\ee
is symplectic (i.e. $\mathbb M^T\mathbb S_-\mathbb M=\mathbb S_-$, with $\mathbb S_-$ given in (\ref{eq:SymplStrM-})) and satisfies $\mathbb M^2=-1$.

Now, the key observation is that, as one sees from (\ref{eq:iDiamond}), $*_B$ acts as $+i$ on $\Phi_-\in U_{\bf{1},\bf{1}}$ and as $-i$ on $\chi^-_I\in U_{\bf{\bar 3},\bf{\bar 3}}$, so that, referring to eq. (\ref{eq:KodairaPhi-}), we have
\be\label{eq:*BonDelPhi}
*_B(\partial_I\Phi_-) \sim -i (\partial_I\Phi_- - 2\kappa^-_I\Phi_-)
\ee
On the other hand, recalling (\ref{eq:DelPhi-}) and (\ref{eq:AlphaDeAlpha}),
\be\label{eq:DelPhi}
\partial_I\Phi_- \sim \alpha_I - \mathcal G_{IJ}\beta^J\;.
\ee
Substituting (\ref{eq:DelPhi}) into (\ref{eq:*BonDelPhi}) and using (\ref{eq:Expansion*B}) we get
\be
(\mathcal A_I^{\;\;J}-\mathcal G_{IK}\mathcal C^{KJ})\alpha_J + (\mathcal B_{IJ} + \mathcal A_J^{\;\;K}\mathcal G_{KI})\beta^J  \sim -i(\delta_I^{\;\;J} - 2\kappa^-_IZ^J)\alpha_J + i(\mathcal G_{IJ} - 2\kappa^-_I\mathcal G_J)\beta^J\;.
\ee
Taking the Mukai pairing of this expression with the basis elements $\alpha, \beta$, separating into real and imaginary parts, and using the expression (\ref{eq:kappaI}) for $\kappa^-_I$ we arrive at
\begin{eqnarray}
\nnb \mathcal C^{IJ} &=& (\im \mathcal G)^{-1\,IJ} - \frac{Z^I\bar Z^J + \bar Z^IZ^J}{Z^K\im\mathcal G_{KL}\bar Z^L} \,\;=\;\, -(\im \mathcal M)^{-1\,IJ}\\ [1mm]
\mathcal A_I^{\;\;J} &=& [\re\mathcal G(\im \mathcal G)^{-1}]_I^{\;\;J} - \frac{\mathcal G_I\bar Z^J + \bar{\mathcal G}_I Z^J}{Z^K\im\mathcal G_{KL}\bar Z^L} \,\;=\;\, -[\re\mathcal M (\im\mathcal M)^{-1}]_I^{\;\;J}\\ [1mm]
\nnb \mathcal B_{IJ} &=& -[\im\mathcal G +\re\mathcal G(\im\mathcal G)^{-1}\re\mathcal G]_{IJ} + \frac{\mathcal G_I\bar{\mathcal G}_J + \bar{\mathcal G}_I\mathcal G_J}{Z^K\im\mathcal G_{KL}\bar Z^L} \,\;=\;\, [\im\mathcal M+\re\mathcal M(\im\mathcal M)^{-1}\re\mathcal M]_{IJ}\;,
\end{eqnarray}
where to write the second equalities we use (\ref{eq:RelGM}).
So the matrices $\mathcal A,\mathcal B,\mathcal C,\mathcal D$ are expressed in terms of the period matrix $\mathcal M$, and the result can be summarized in
\vskip 1mm
\be
\label{eq:mathbbM}\mathbb M\equiv \left(\begin{array}{cc}\int \langle\alpha,*_B\beta\rangle  & -\int \langle\beta,*_B\beta\rangle \\ [2mm]
\int\langle\alpha, *_B\alpha\rangle  \;& -\int \langle\beta,*_B\alpha\rangle  \end{array}\right) =\left(\begin{array}{cc}
(\im\mathcal M)^{-1}\re\mathcal M    \;\;&\;\;  -(\im\mathcal M)^{-1}\\ [2mm]
\im\mathcal M+\re\mathcal M(\im\mathcal M)^{-1}\re\mathcal M \;&\; -\re\mathcal M(\im\mathcal M)^{-1}  \end{array}\right)\,.
\ee
The symmetric matrix
\be\label{eq:tildeM}
\tilde{\mathbb M}:=\mathbb S_-\mathbb{M}\;=\;\left(\begin{array}{cc} 1 & -\re\mathcal M \\ [1mm] 0 & 1 \end{array}\right)
\left(\begin{array}{cc} \im \mathcal M & 0 \\ [1mm] 0 & (\im \mathcal M)^{-1} \end{array}\right)
\left(\begin{array}{cc} 1 & 0 \\ [1mm] -\re\mathcal M & 1 \end{array}\right)\;
\ee
is an important piece of information in the definition of $N=2$ effective actions by compactification of type II theories to four dimensions. In particular, for type IIA compactifications it appears in the kinetic terms for the scalars $\xi^I, \tilde\xi_I$ coming from the expansion of the RR potentials (see eqs.$\:$(\ref{eq:expRRflAndPot}) and (\ref{eq:quaternionicMetric}) below). Namely, it is one of the special geometry data that determine, via the c-map, the quaternionic metric for the $N=2$ hypermultiplets $\sigma$-model. While this is familiar for dimensional reductions on a Calabi-Yau, we have shown that the same structure can be extended to more general settings, for instance to cases in which the basis forms are not of pure degree.

It is readily checked that when considering Calabi-Yau (or more generally $SU(3)$ structure \cite{MinasianKashani}) compactifications, (\ref{eq:mathbbM}) reduces to the well known expression for the action of the Hodge $*$ on the harmonic 3-forms. Indeed in this case, because of the constraint $B\wedge \alpha_I = B\wedge \beta^I =0$, the action of $*_B$ on $\Sigma^-$ simplifies to the action of the usual Hodge $*\;$, so that $\int\langle \alpha ,*_B \beta \rangle = \int \alpha\wedge *\beta\;$ (similarly for the other pairings). Therefore (\ref{eq:mathbbM}) coincides with the result of \cite{Suzuki,CeresoleEtAl1}.\\ [-1mm]

One can now proceed in a completely parallel fashion to get the action of $*_B$ on the even basis $\Sigma^+$. In this case, $\Phi_+\in U_{\bf{1},\bf{\bar 1}}$ and its deformations are in $U_{\bf{\bar 3},\bf{3}}$ (deformations in $U_{\bf{3},\bf{\bar 1}}\oplus U_{\bf{1},\bf{\bar{3}}}$ are assumed to vanish in the Mukai pairing due to condition (\ref{eq:StrongCompatibility})). Again, these two sets are eigenspaces of $*_B$ corresponding to opposite eigenvalues.\\
Repeating the steps done for the odd forms, and adopting analogous assumptions, we find
\be
\label{eq:SymplecticN} \mathbb{N}:=\:\left(\begin{array}{cc}  \int\langle \om,*_B\tilde\om\rangle  \;&\; -\int\langle \tilde\om,*_B\tilde\om\rangle \\ [2mm]
\int\langle \om,*_B\om\rangle \;&\;  -\int\langle \tilde\om,*_B\om\rangle \end{array}\right) = \left(\begin{array}{cc}  
(\im\mathcal N)^{-1}\re\mathcal N \;\;&\;\; -(\im \mathcal N)^{-1} \\ [2mm]
\im \mathcal N +\re \mathcal N(\im \mathcal N)^{-1}\re \mathcal N   \;\;&\;\; -\re\mathcal N(\im\mathcal N)^{-1}   
\end{array}\right)\,,
\ee
where $\mathbb N\in Sp(2b^++2,\mathbb{R})\;$ and satisfies $\mathbb N^2=-1$. The analog of (\ref{eq:tildeM}) is:
\be
\tilde{\mathbb{N}}\;:=\;\mathbb S_+\mathbb N=\left(\begin{array}{cc} 1 & -\re\mathcal N \\ 0 & 1 \end{array}\right)
\left(\begin{array}{cc} \im \mathcal N & 0 \\ 0 & (\im \mathcal N)^{-1} \end{array}\right)
\left(\begin{array}{cc} 1 & 0 \\ -\re\mathcal N & 1 \end{array}\right)\;,
\ee
which is symmetric and negative definite whenever $\im\mathcal N$ is (in IIA compactifications, indeed, $\im\mathcal N$ defines the vector kinetic matrix of the $N=2$ effective action, and as such should be negative definite).

Note that in the particular case of Calabi-Yau 3-folds one can check (\ref{eq:SymplecticN}) explicitly by evaluating its LHS and RHS by two separate computations. In order to evaluate the LHS of (\ref{eq:SymplecticN}) one can choose a basis for the harmonic forms of even degree $\Sigma^+: \{\om_0=1,\om_a:\textrm{2-forms},\,\tilde\om^a,\tilde\om^0\}$ in such a way that (\ref{eq:evenSymplecticStr}) is satisfied, then expand $B+iJ=\frac{X^a}{X^0}\om_a$ and use $\frac{1}{4\mathcal V}\int\langle \om_a,*\om_b\rangle=g^+_{ab}$ ($\mathcal V$ is the internal volume). On the other hand, the period matrix $\mathcal N_{AB}$ appearing in the RHS can be obtained starting from the usual cubic prepotential
\be\label{eq:cubicPrepot}
\mathcal{F}= -\frac{1}{6}\mathcal{K}_{abc}\frac{X^aX^bX^c}{X^0}\;,\qquad \qquad \big(\,\mathcal K_{abc}=\int_{M_6} \om_a\wedge\om_b\wedge\om_c\,\big)\;,
\ee
and using the special geometry formula (\ref{eq:relMandG}) (translated in the notation for $\mathscr M_+$).

\subsection{Differential conditions, RR fields and general fluxes}\label{DiffConAndFluxes}

In this subsection we introduce a general set of charges coming from the NS, RR, geometric as well as nongeometric fluxes associated to the type II theory. From a 4d viewpoint, these correspond to electric and magnetic Fayet-Iliopoulos charges for the $N=2$ gauged supergravity. This will be apparent in the next subsection, where we will express the $N=2$ Killing prepotentials in terms of the flux charges. Our discussion mainly follows refs. \cite{GLW1, GLW2}.
\vskip .2cm
Let's start considering a 6d manifold with $SU(3)$ structure. As already remarked, the expansion forms need not be closed. Seeing this as a deformation of the Calabi-Yau case, one can choose $\Sigma^-$ as composed of 3-forms only, and take $\Sigma^+: \{\om_0=1,\om_a:\textrm{2-forms},\,\tilde\om^a,\tilde\om^0\}$ in such a way that (\ref{eq:evenSymplecticStr}) is satisfied. The analog of (\ref{eq:StrongCompatibility}) is now $\om_a \wedge\alpha_I=0= \om_a\wedge\beta^I\,$, which implies the usual $SU(3)$ structure constraint $J\wedge \Om=0$. We separate the internal NS 3-form $H$ (satisfying the Bianchi identity $d H=0$) into an exact and a flux piece:
\be
H=H^{\mathrm{fl}} + d B\;,
\ee
and we introduce the `twisted' differential
\be\label{eq:dHfl}
d_{H^{\mathrm{fl}}}=d-H^{\mathrm{fl}}\!\wedge\;.
\ee
Expanding the NS flux as $H^{\mathrm{fl}} = m_0^I\alpha_I - e_{I0}\beta^I$, and demanding closure of $\Sigma^\pm$ under the action of $d_{H^{\mathrm{fl}}}$, one is led to assume \cite{GurLouisMicuWaldr, GLW1, TomasMirrorSymFl} (see \cite{GaugingHeisenberg} for a 4d sugra interpretation):
\begin{eqnarray}
\nnb d_{H^{\mathrm{fl}}} \alpha_I = e_{IA}\tilde\om^A \qquad &,& \qquad d_{H^{\mathrm{fl}}} \beta^I = m_{A}^I\tilde\om^A \\
\label{eq:DiffCondSU3} d_{H^{\mathrm{fl}}} \om_A = m_A^I\alpha_I -e_{IA}\beta^I \qquad &,&\qquad d_{H^{\mathrm{fl}}}\tilde\om^A = 0\;.
\end{eqnarray}
The $m_a^I$ and $e_{aI}$ charges can be put in relation with the torsion classes of the $SU(3)$ structure under consideration \cite{ChiossiSalamon}. Note that $(d_{H^{\mathrm{fl}}})^2=0$ implies $m^I_A e_{IB} - e_{IA}m^I_B=0\,$.

On more general backgrounds, the basis forms are not necessarily of pure degree. Furthermore, one may allow for a formal extension of the $d_{H^{\mathrm{fl}}}$ operator to include non-geometric fluxes \cite{SheltonTaylorWecht}: 
\be\label{eq:CurlyD} d_{H^{\mathrm{fl}}}\,\rightarrow\,  \mathcal D:=d_{H^{\mathrm{fl}}} -Q\cdot -R\llcorner\;,\ee 
where in the notation of \cite{SheltonTaylorWecht} the $Q$ and $R$ operators act on a differential $k$-form $C$ as 
\be
(Q\cdot C)_{m_1\ldots m_{k-1}}= Q^{ab}_{\phantom{ab}[m_1}C_{|ab|m_2\ldots m_{k-1}]}\qquad,\qquad (R\llcorner C)_{m_1\ldots m_{k-3}} = R^{abc}C_{abcm_1\ldots m_{k-3}}\;,
\ee
and so they lower its degree by 1 and 3 respectively. Therefore, $\mathcal D$ still sends odd/even forms into even/odd forms. Without specifying the details of the model, we are led to adopt the following general differential conditions\footnote{We recall that $\sim$ means equality up to terms vanishing inside the symplectic pairing.} for the basis $\Sigma^\pm$ \cite{GLW2}:
\begin{eqnarray}
\nnb \mathcal D \alpha_I \,\sim\, p^A_I\om_A + e_{IA}\tilde\om^A \qquad &,& \qquad \mathcal D \beta^I \,\sim\, q^{IA}\om_A + m_{A}^I\tilde\om^A\ \\
\label{eq:GenDiffCond}\mathcal D \om_A \,\sim\, m_A^I\alpha_I -e_{IA}\beta^I \qquad &,&\qquad \mathcal D\tilde\om^A \,\sim\, -q^{IA}\alpha_I+p^A_I\beta^I\;.
\end{eqnarray}
In \cite{GLW2} it has been argued that in order to switch on the whole set of charges in (\ref{eq:GenDiffCond}), the background should necessarily be non-geometric. When considering the specific case of the $SU(3)$ structure basis, one can identify $q^{Ia}$ and $p^a_I$ as arising from the action of $Q\cdot\,$, while $q^{I0}$ and $p^0_I$ as being generated by $R\llcorner\,$.

Again following \cite{GLW2}, by introducing the $(2b^-+2)\times (2b^+ + 2)$ rectangular charge matrix :
\be
\label{eq:ChargeMatrixQ}\mathbb Q:=\left(\begin{array}{cc} m^I_{\;\;A} &  q^{IA} \\ e_{IA} & p_I^{\;\;A}   \end{array}\right)\;,
\ee
one can summarize (\ref{eq:GenDiffCond}) in 
\be \mathcal D \Sigma^- \sim \mathbb Q \Sigma^+\qquad,\qquad \mathcal D \Sigma^+ \sim (\mathbb S_+)^{-1}\mathbb Q^T \mathbb S_- \Sigma^-\;,\ee 
where here $\Sigma^\pm$ should be seen as the vector of forms
\be\label{eq:Sigma+-AsVectors}
\Sigma^+ := \left(\begin{array}{c} \tilde\om^A \\ \om_A \end{array}\right)\qquad , \qquad \Sigma^- := \left(\begin{array}{c} \beta^I \\ \alpha_I \end{array}\right)
\ee
The differential $d_{H^{\mathrm{fl}}}$ satisfies the nilpotency condition $(d_{H^{\mathrm{fl}}})^2 = 0$, and the natural extension $\mathcal D^2=0$ should be required for the $\mathcal D$ operator \cite{SheltonTaylorWecht}. It can be seen from (\ref{eq:GenDiffCond}) that this imposes the quadratic relations among the charges: $\mathbb Q  (\mathbb S_+)^{-1}\mathbb Q^T=0\;,\;\mathbb Q^T  \mathbb S_-\mathbb Q=0\;$. It turns out that these are important to guarantee the consistency of the 4d $N=2$ effective action \cite{N=2withTensor1, N=2withTensor2, D'AuriaFerrTrig}.
\vskip .3cm
Using the expressions introduced above, we can now define the expansion of the internal RR field strengths on $\Sigma^\pm$. Henceforth we will focus on a type IIA context.

We consider the sum of internal RR field strengths $G=G_0+G_2+G_4+G_6$ belonging to $\wedge^{\mathrm{ev}} T^*$ and we express it in terms of the fluxes and the sum $A$ of the RR potentials as
\be\label{eq:Gstrengths}
G= G^{\mathrm{fl}} + d_{H^{\mathrm{fl}}}A \;.
\ee
In the absence of localized sources, $G$ satisfies the Bianchi identity $d_{H^{\mathrm{fl}}}G=0$. The sum $F$ of the usual modified field strengths appearing in the 10d supergravity action can be written as 
\be\label{eq:F=eBG} F= e^{B}G\qquad,\qquad\textrm{with}\quad (d - H\wedge)F =0\;.
\ee 
When considering the $d_{H^{\mathrm{fl}}}\,\rightarrow\,  \mathcal D$ extension, the RR field strengths (\ref{eq:Gstrengths}) are formally modified to
\be\label{eq:Gfl+DA}
G=G^{\mathrm{fl}}+\mathcal DA\;
\ee
in such a way that the associated Bianchi identity is $\mathcal D G=0\,$ \cite{{SheltonTaylorWecht}}.

We expand the internal RR fluxes and potentials on the basis of forms on $M_6$ as:
\be\label{eq:expRRflAndPot}
G^{\mathrm{fl}} = \sqrt 2(m^A_{\mathrm{RR}}\om_A+e_{\mathrm{RR}A}\tilde\om^A)\qquad,\qquad A=\sqrt 2(\xi^I\alpha_I-\tilde\xi_I\beta^I)\;,
\ee
where $\xi^I, \tilde\xi_I$ are 4d scalar fields, and the $\sqrt 2$ is introduced in order to avoid some cumbersome factors in the expressions of subsect.$\:$\ref{4dsugraPicture} below. Using (\ref{eq:GenDiffCond}), we can write
\be\label{eq:ExpRRstrengths}
G\sim G^A\om_A+\tilde G_A\tilde\om^A\;,
\ee
with $\;\; G^A=\sqrt 2(m^A_{\mathrm{RR}} + \xi^I p^A_{I} - \tilde\xi_I q^{IA})\quad$ and $\quad\tilde G_A=\sqrt 2(e_{\mathrm{RR}A}+ \xi^I e_{IA} -\tilde\xi_I  m_{A}^{I}) \;$.\\

\subsection{$N=2$ Killing prepotentials from the dimensional reduction}\label{KillingPrep}

In this subsection we briefly summarize how refs.$\:$\cite{GLW1, GLW2} determined the $N=2$ Killing prepotentials $\mathcal P^x\,,\;x=1,2,3$ of the 4d $N=2$ effective theory. The Killing prepotentials are a basic element of gauged supergravities (see e.g. \cite{N=2review} for a review); in particular, they are related to the fermionic susy variations which determine the potential of the $N=2$ theory. The consistency with the $N=2$ formalism of the expressions we will obtain is discussed in the forthcoming subsect.$\:$\ref{4dsugraPicture}.

Generalizing a previous analysis \cite{GLW1} done for $SU(3)$ structures, the authors of \cite{GLW2} reduced the type II gravitino susy variations on $SU(3)\times SU(3)$ structure backgrounds, and determined in this way an expression for the susy transformation of the 4d $N=2$ gravitini in terms of the higher dimensional fields. Then this expression was confronted with the generic form of the 4d $N=2$ gravitino susy transformation law, whose relevant part reads:
\be
\label{eq:N=2GravVar}\delta\psi_{\mathcal A\mu}=\ldots + \nabla_\mu\ep_{\mathcal A}- S_{\mathcal A\mathcal B}\gamma^{(4)}_\mu\ep^{\mathcal B}\;,
\ee
where $\psi_{\mathcal A\,\mu}\;\,(\mathcal A,\mathcal B=1,2)$ are the $N=2$ gravitini\footnote{Our $R$-symmetry $SU(2)$ indices are $\mathcal A,\mathcal B=1,2$, while we reserved the letters $A,B, \ldots$ (running over $0,1, \ldots,b^+$) for the symplectic sections of $\mathscr M_+\,$.}, $\ep_{\mathcal A}$ and $\ep^{\mathcal B}$ are the 4d $N=2$ susy parameters as in (\ref{eq:10dSpinorsIIA}), $\nabla_\mu$ is the usual 4d spacetime covariant derivative for $Spin(3,1)$ spinors, $\gamma_\mu^{(4)}$ is the \emph{Cliff}$(3,1)$ gamma matrix associated with the 4d metric $g^{(4)}_{\mu\nu}$ defined in (\ref{eq:g4}) here below, and $S_{\mathcal A\mathcal B}$ is the gravitino mass matrix containing the Killing prepotentials $\mathcal P^x\,,\;x=1,2,3\,$. The comparison allowed to extract an expression for $S_{\mathcal A\mathcal B}\,$. For type IIA this reads\footnote{A few remarks are in order for the comparison with ref.$\:$\cite{GLW2}. Our matrix $S_{\mathcal A\mathcal B}$ corresponds to the matrix called $S^{(4)}_{AB}(\mathrm{IIA})$  there. The differences in the numerical factors are due to different choices of normalization for the pure spinors. Finally, here we have already taken the integral over $M_6$.}:
\be
\label{eq:massmatrix} S_{\mathcal A\mathcal B}=ie^{\frac{K_+}{2}}\left( \begin{array}{cc} e^{\frac{K_-}{2}+\varphi}\int\langle \Phi_+ \,,\, d_{H^{\mathrm{fl}}}\Phi_-  \rangle  &  \frac{e^{2\varphi}}{\sqrt 8}\int\langle \Phi_+ \,,\, G \rangle \\ [1mm]
\frac{e^{2\varphi}}{\sqrt 8}\int\langle \Phi_+ \,,\, G  \rangle   &   -e^{\frac{K_-}{2}+\varphi}\int\langle \Phi_+ \,,\, d_{H^{\mathrm{fl}}}\bar\Phi_-  \rangle     \end{array} \right)\;.
\ee
Notice that the flux part $H^{\mathrm{fl}}$ of the NS field strength is contained in the $d_{H^{\mathrm{fl}}}$ operator defined in (\ref{eq:dHfl}), while the $B$-field is included in the pure spinors $\Phi_\pm$ as in (\ref{eq:DefPhi_pm}); these are built from the two globally defined $Spin(6)$ spinors $\eta^1,\eta^2$ that enter in the 10d gravitino variation.

All the objects entering in (\ref{eq:massmatrix}) have been introduced previously, except the 4d \hbox{dilaton $\varphi$}, which is defined in terms of the 10d dilaton $\phi$ and the volume form $vol_6$ of $M_6$ by 
\be e^{-2\varphi}=\int e^{-2\phi}vol_6\;,\ee
and is used to define the Weyl rescaled 4d metric $g_{\mu\nu}^{(4)}$ entering in the 4d effective action:
\be\label{eq:g4}
g_{\mu\nu}^{(4)}=e^{-2\varphi}g_{\mu\nu}\;.
\ee
Recalling the definitions here above as well as eqs.$\:$(\ref{eq:normPhi}), (\ref{eq:K-standardForm}) and (\ref{eq:K+standardForm}), and assuming the 10d dilaton $\phi$ does not depend on the internal coordinates, we can write a chain of equalities which will be frequently used later on:
\be\label{eq:chain}
\int vol_6=\frac{1}{8}e^{-K_\pm}= e^{-2\varphi+2\phi}\;.
\ee

The $N=2$ gravitino mass matrix $S_{\mathcal A\mathcal B}$ contains the three Killing prepotentials $\mathcal P^x$. Indeed, its general form is:
\be \label{eq:MassMatrixAndPrepot} S_{\mathcal A \mathcal B}=\frac{i}{2}e^{\frac{K_V}{2}}(\sigma_x)_{\mathcal A}^{\;\;\mathcal C}\epsilon_{\mathcal B\mathcal C}\mathcal P^x\;=\;\frac{i}{2}e^{\frac{K_V}{2}}\left(\begin{array}{cc}\mathcal P^1-i\mathcal P^2 & -\mathcal P^3\\ -\mathcal P^3 & -(\mathcal P^1+i\mathcal P^2)\end{array}\right)\;,\ee
where $\epsilon_{\mathcal A\mathcal B}={0 \;1\choose -1\;0 }$ is the $SU(2)$ metric, $(\sigma_x)_{\mathcal A}^{\;\;\mathcal B}\,,x=1,2,3$ are the standard Pauli matrices and $K_V$ is the special K\"ahler potential for the scalars in the $N=2$ vector multiplets; for the type IIA compactifications on which we focus, $K_V\equiv K_+$.

Comparing (\ref{eq:massmatrix}) with (\ref{eq:MassMatrixAndPrepot}), and allowing for the formal $d_{H^{\mathrm{fl}}}\rightarrow \mathcal D$ extension, one deduces a geometric expression for the Killing prepotentials:  
\begin{eqnarray}
\nnb \mathcal P^1-i\mathcal P^2 & = &  2e^{\frac{K_-}{2}+\varphi}\int\langle \Phi_+ \,,\, \mathcal D \Phi_-  \rangle\qquad, \qquad \mathcal P^1+i\mathcal P^2 = 2e^{\frac{K_-}{2}+\varphi}\int\langle \Phi_+ \,,\, \mathcal D\bar \Phi_-  \rangle \\
\label{eq:KillingPrepGeneral} \mathcal P^3 & = &  -\frac{e^{2\varphi}}{\sqrt2}\int\langle \Phi_+ \,,\, G  \rangle \;.
\end{eqnarray}
Finally, using the expansions introduced in subsect.$\:$\ref{DiffConAndFluxes}, together with the ones for the pure spinors $\Phi_\pm$, eqs. (\ref{eq:expansPhi-}) and (\ref{eq:expansPhi+}), one obtains the $\mathcal P^x\,$ in terms of the quantities entering in the 4d effective action:
\begin{eqnarray}
\nonumber \mathcal P^1-i\mathcal P^2 &=& 2e^{\frac{K_-}{2}+\varphi}  V_-^T\mathbb S_- \mathbb Q V_+  = 2e^{\frac{K_-}{2}+\varphi} \big[(Z^Ie_{IA}- \mathcal G_I m_A^{I})X^A+(Z^I p^A_I- \mathcal G_I q^{IA})\mathcal{F}_A\big],\\ [2mm]
\label{eq:GeneralKillingPrepot} \mathcal P^1+i\mathcal P^2 &=& 2e^{\frac{K_-}{2}+\varphi}  \bar V_-^T\mathbb S_- \mathbb Q V_+  =         2e^{\frac{K_-}{2}+\varphi} \big[(\bar{Z}^I e_{IA}-\bar{\mathcal G}_I m_A^{\;\;I})X^A+(\bar Z^Ip^A_I- \bar{\mathcal G}_I q^{IA})\mathcal{F}_A\big],\\ [2mm]
\nonumber \mathcal P^3&=&    -\frac{e^{2\varphi}}{\sqrt 2}  V_G^T\mathbb S_+ V_+ =   -e^{2\varphi}\big[(e_{\mathrm{RR}A}+\xi^I e_{IA}-\tilde{\xi}_I m_A^{I})X^A+(m^A_{\mathrm{RR}} + \xi^I p^A_{I} -  \tilde\xi_I q^{IA} )\mathcal{F}_A \big],
\end{eqnarray}
where the symplectic vectors $V_\pm$ and $V_G$ are defined as
\be\label{eq:V+V-Vg}
V_+ =\left(\begin{array}{c} X^{A} \\ \mathcal F_{A}\end{array}\right)\:; \;\; V_- =\left(\begin{array}{c} Z^{I} \\ \mathcal G_{I}\end{array}\right) \:; \;\; V_G =\left(\begin{array}{c} G^{A} \\ -\tilde G_{A}\end{array}\right) =  \sqrt 2\left(\begin{array}{c} m_{\mathrm{RR}}^{A} \\ -e_{\mathrm{RR}A} \end{array}\right)   + \sqrt 2(\mathbb S_+)^{-1}\mathbb Q^T \mathbb S_-\left(\begin{array}{c} \xi^I \\ \tilde \xi_I \end{array}\right).
\ee

\subsection{$N=2$ supergravity picture and fermionic shifts}\label{4dsugraPicture}

In this subsection we discuss how the Killing prepotentials given above fit into the general formalism of 4d $N=2$ gauged supergravity. In particular, this will allow us to derive the form of the fermionic shifts, and express them in terms of these Killing prepotentials. The fermionic shifts (\ref{eq:N=2fermionicVar})-(\ref{eq:GaugMassMatrix}) will be the starting point to establish the supersymmetric vacuum conditions that will be studied in section \ref{VacuumCondit}. 

A consistent way of constructing an $N=2$ supergravity action containing the Killing prepotentials (\ref{eq:GeneralKillingPrepot}) has been given in ref.$\:$\cite{D'AuriaFerrTrig}, building on previous work \cite{GaugingHeisenberg}. The general framework is the one of gauged $N=2$ supergravity with (massive) tensor multiplets \cite{N=2withTensor1, N=2withTensor2}. Here we don't describe the complete 4d supergravity action, but just show how the Killing prepotentials emerge in this picture; then we deduce the related fermionic susy variations.

As above, we will choose a setting that corresponds to a type IIA compactification (the discussion for IIB would proceed in a perfectly mirror symmetric way).

The strategy adopted in \cite{D'AuriaFerrTrig} was to start from an (ungauged) $N=2$ supergravity of the kind obtained in type II compactifications on Calabi-Yau 3-folds, and then deform it by gauging the abelian isometries of the quaternionic metric associated with the kinetic terms for the hypermultiplets. A second step, allowing to introduce further interactions, was the dualization of a subset of the hyperscalars to antisymmetric 2-tensors.

The quaternionic manifold which is relevant for the theory under consideration is a \emph{special} one: its metric is determined via the so called c-map from the data of a special K\"ahler submanifold \cite{FerraraSabharwal}. In our case, this submanifold is the one describing the deformations of $\Phi_-\,$ (the $\Phi_+$-moduli $t^a\,,\;a=1,\ldots,b^+$, enter instead in the $N=2$ vector multiplets).

Let us first recall the principal features of this special quaternionic manifold. Its coordinates are the scalars $q^u = (\varphi, a, \xi^I, \tilde\xi_I, z^i)\,,\;u=1,\ldots,4(b^-+1)$, representing the bosonic components of the $N=2$ hypermultiplets. The quadruple $(\varphi, a, \xi^0, \tilde\xi_0)$ corresponds to the universal hypermultiplet, where $a$ is the axion coming from the dualization of the NS 2-form $B^{(4)}_{\mu\nu}$ extending along the 4d spacetime. All the other fields have already been introduced in the previous subsections; in particular, as above the complex scalars $z^i$ parameterize a special K\"ahler manifold $\mathscr M_-$. The c-map is realized by introducing the 1-forms \cite{FerraraSabharwal, D'AuriaFerrTrig}:
\begin{eqnarray}
\nnb u &=& ie^{\frac{K_-}{2}+\varphi}Z^I(d \tilde\xi_I - \mathcal M_{IJ}d\xi^J)\\
\nnb v &=& \frac{e^{2\varphi}}{2}\big[ d e^{-2\varphi}-i(d a+\tilde\xi_I d\xi^I-\xi^I d\tilde \xi_I  ) \big]\\
\nnb E &=& -\frac{i}{2} e^{\varphi-\frac{K_-}{2}}P_I(\im\mathcal G)^{-1\,IJ}(d \tilde\xi_J - \mathcal M_{JL}d\xi^L)  \\
\label{eq:FerrSabharw1forms}e &=& P_I d Z^I\;,
\end{eqnarray}
with $P_I=(P_0^{\;\underline{j}},P_i^{\;\underline{j}})=(-e_i^{\; \underline{j}}Z^i, e_i^{\; \underline{j}})\,$, where $e_i^{\; \underline{j}},\;(i,j=1,\ldots,b^-)$ are the vielbeine of the special K\"ahler manifold $\mathscr M_-$ (the underlined indices are the flat ones). The choice of special coordinates $Z^I=(1,z^i)$ is assumed\footnote{The matrix which in our conventions corresponds to $-2\im \mathcal G_{IJ}$ is called $N$ in \cite{FerraraSabharwal, D'AuriaFerrTrig}.}.
The quaternionic metric $h_{uv}$ is given by (we use $u$ and $v$ as quaternionic world indices, not to be confused with the 1-forms $u$ and $v$ introduced in (\ref{eq:FerrSabharw1forms})):
\begin{eqnarray}
\nnb h_{uv}d q^u d q^v &=&  \bar u\otimes u +  \bar v \otimes v  +  \bar E\otimes E + \bar e\otimes e\\
\label{eq:quaternionicMetric}&=& g^-_{i\bar\jmath}\,d z^id \bar z^{\bar\jmath} + (d\varphi)^2 + \frac{e^{4\varphi}}{4}\big(d a + (d V_\xi)^T \mathbb S_- V_\xi \big)^2 - \frac{e^{2\varphi}}{2}(d V_\xi)^T\tilde{\mathbb{M}}\,d V_\xi\;,
\end{eqnarray}
where $g^-_{i\bar\jmath}$ is the metric on $\mathscr M_-$, $\tilde{\mathbb{M}}$ corresponds to the matrix introduced in (\ref{eq:tildeM}), and \hbox{$V_\xi = (\xi^I ,\tilde\xi_I)^T$} is the symplectic vector containing the RR scalars.

Due to the fact that the holonomy of the quaternionic manifold is $SU(2)\times \mathcal H\,$, with $\mathcal H \subset Sp(2b^-+2)\,$ \cite{N=2review}, introducing flat indices $\mathcal A,\mathcal B=1,2\,$ and $\alpha,\beta=1,\ldots,2b^-+2$ running in the fundamental representations of $SU(2)$ and $Sp(2b^-+2)$ respectively, one can define the natural vielbeine $\mathcal U^{\mathcal A\alpha}_u$, satisfying the reality condition 
\be
\epsilon_{\mathcal A\mathcal B}\mathbb S^-_{\alpha\beta}\mathcal U^{\mathcal B\beta}=(\mathcal U^{\mathcal A\alpha})^*
\ee
and relating the metric $h_{uv}$ to the flat $SU(2)\cong Sp(2)$ and $Sp(2b^-+2)$ invariant metrics $\epsilon_{\mathcal A\mathcal B}$ and $\mathbb S^-_{\alpha\beta}\,$:
\be
h_{uv} = \mathcal U^{\mathcal A\alpha}_u \mathcal U_v^{\mathcal B\beta} \mathbb S^-_{\alpha\beta} \epsilon_{\mathcal A\mathcal B}\;.
\ee
For the metric (\ref{eq:quaternionicMetric}), we can choose the vielbein 1-forms:
\be
\label{eq:quaternVielbein}\mathcal U^{\mathcal A\alpha}  = \frac{1}{\sqrt 2}\left(\begin{array}{cccc}
\bar u & \bar e & -v & -E \\
\bar  v & \bar E & u & e
\end{array}\right)\;.
\ee
These will appear in the hyperini supersymmetry variations defined below.

The last ingredient we need is given by the connection 1-forms $\om^x,\;x=1,2,3\,$ for the $SU(2)$-bundle over the quaternionic manifold. In the present case these are given by \hbox{\cite{FerraraSabharwal, PolchinskiStrominger, Michelson}}:
\begin{eqnarray}
\nnb\om^1 &=& i(\bar u- u)\qquad,\qquad \om^2 \;=\; u+\bar u \\ [2mm]
\label{SU2Connection}\om^3 &=& \frac{i}{2}(v-\bar  v)+\frac{i}{2}\frac{Z^I\im\mathcal G_{IJ} d\bar Z^J - \bar Z^I \im\mathcal G_{IJ} d Z^J}{\bar Z^K\im\mathcal G_{KL} Z^L}\;.
\end{eqnarray}

A first deformation of the ungauged $N=2$ theory containing the quaternionic $\sigma$-model outlined above was obtained in \cite{D'AuriaFerrTrig} (see also \cite{GaugingHeisenberg}) by gauging the abelian isometries\footnote{The abelianity follows from the quadratic constraints written below eq.$\:$(\ref{eq:Sigma+-AsVectors}).} of the metric (\ref{eq:quaternionicMetric}) generated by the following choice of Killing vectors 
\be
\label{eq:ElectricKillingVectors} k_A = (-2e_{\mathrm{RR}A}-e_{IA}\xi^I + m_A^I\tilde\xi_I)\pd{\,}{a} + m_A^I\pd{\,}{\xi^I} + e_{IA}\pd{\,}{\tilde\xi_I}\;\,,\qquad A=0,1,\ldots, b^+\;,
\ee
where the `electric' charges $\,e_{\mathrm{RR}A} , e_{IA}, m_A^I$ are half of the parameters associated with the general set of fluxes described in subsect.$\:$\ref{DiffConAndFluxes}. In gauged supergravity to each such Killing vector is associated a set of three momentum maps $\mathcal P^x_A$, also called Killing prepotentials. These are given by the formula \cite{Michelson, D'AuriaFerrFre}
\be\label{eq:FormulaForKilling} \mathcal P^x_A=\om^x_u k^u_A\;,\ee
which is particularly simple due to the fact that in the present case the Lie derivative of the $SU(2)$ connection (\ref{SU2Connection}) along the vectors (\ref{eq:ElectricKillingVectors}) vanishes, and this causes the absence of further terms in (\ref{eq:FormulaForKilling}).
 
Plugging (\ref{SU2Connection}) and (\ref{eq:ElectricKillingVectors}) in (\ref{eq:FormulaForKilling}), one obtains
\begin{eqnarray}
\nnb\mathcal P^1_A &=& 2e^{\frac{K_-}{2}+\varphi}(\re Z^Ie_{IA}- \re\mathcal G_I m_A^{I})\qquad , \qquad \mathcal P^2_A = -2e^{\frac{K_-}{2}+\varphi}(\im Z^Ie_{IA}- \im\mathcal G_I m_A^{I})\\ [1mm]
\label{eq:electricP}\mathcal P^3_A &=& -e^{2\varphi}(e_{\mathrm{RR}A}+\xi^I e_{IA}-\tilde{\xi}_I m_A^{I})\;,
\end{eqnarray}
and we immediately see that the sums $\mathcal P^x_AX^A$ indeed correspond to the part of the $\mathcal P^x$ in eq.$\:$(\ref{eq:GeneralKillingPrepot}) containing the charges $\,e_{\mathrm{RR}A} , e_{IA}, m_A^I$.

In order to take into account the second half of flux parameters $\,m_{\mathrm{RR}}^A,p_I^A, q^{IA}$, the authors of ref.$\:$\cite{D'AuriaFerrTrig} performed a dualization of a subset of the scalars $\{\xi^I,\tilde\xi_I\}$, together with the axion $a$, to antisymmetric 2-tensors. Then the charges $\,m_{\mathrm{RR}}^A,p_I^A, q^{IA}$ could be introduced as mass terms for these tensors, in a way which is consistent with $N=2$ supersymmetry \cite{N=2withTensor1, N=2withTensor2}. Alternatively, using the `redundant' formalism described in \cite{DeWitSamtlebenTrig}, one could generate the same interactions by performing a gauging involving the magnetic gauge potentials and the quaternionic Killing vectors
\be
\label{eq:MagneticKillingVectors}\tilde k^A = (2m_{\mathrm{RR}}^A +p_I^A\xi^I - q^{IA}\tilde\xi_I )\pd{\,}{a} - q^{IA}\pd{\,}{\xi^I} - p_I^A\pd{\,}{\tilde\xi_I}\;,
\ee
and then integrating out the magnetic vector potentials, leaving in this way a theory with electric vectors and antisymmetric tensors (together with the other fields already present in the original action). In this sense, the flux parameters $m_{\mathrm{RR}}^A,p_I^A, q^{IA}$ can be interpreted as `magnetic' charges from the 4d $N=2$ viewpoint. We can then define the symplectic completion $\mathcal{\tilde P}^{xA}$ of the $\mathcal{ P}^{x}_A$ introduced above \cite{Michelson, DeWitSamtlebenTrig} as:
\be \mathcal{\tilde P}^{xA}=\om^x_u \tilde k^{uA}\;,\ee
yielding
\begin{eqnarray}
\nnb\mathcal{\tilde P}^{1A} &=& -2e^{\frac{K_-}{2}+\varphi}(\re Z^I p^A_I- \re{\mathcal G_I} q^{IA})\qquad,\qquad \mathcal{\tilde P}^{2A} = 2e^{\frac{K_-}{2}+\varphi}(\im Z^I p^A_I- \im \mathcal G_I q^{IA}) \\ [1mm]
\label{eq:magneticP}\mathcal{\tilde P}^{3A} &=& e^{2\varphi}(m^A_{\mathrm{RR}} + \xi^I p^A_{I} -  \tilde\xi_I q^{IA} ) \;.
\end{eqnarray}
It is worth remarking that the combinations of the $\xi^I, \tilde \xi_I$ entering in $\mathcal P^3_A$ and $\tilde{\mathcal P}^{3A}$ do not contain the scalars which have been dualized to antisymmetric tensors.

It is now easy to see that the symplectic invariant expressions
\be\label{eq:PxFromPxAandTildePxA}
\mathcal P^x= \mathcal P^x_AX^A - \mathcal{\tilde P}^{xA}\mathcal F_A\;
\ee
precisely reproduces the Killing prepotentials (\ref{eq:GeneralKillingPrepot}) provided by the compactification.

\vskip .5cm

\underline{\emph{$N=2$ fermionic shifts in the presence of electric and magnetic charges}}\\ [-2mm]

\noindent 

As discussed in \cite{N=2withTensor1, N=2withTensor2}, all the flux charges introduced above appear in the fermionic supersymmetry variations of the $N=2$ theory as generalized Fayet-Iliopoulos terms.

Besides the gravitini $\psi_{\mathcal A \mu}\,,\;\mathcal A=1,2$, the (positive chirality) fermions contained in the $N=2$ theory under consideration are the hyperini $\zeta_{\alpha}\,,\;\alpha=1,\ldots,2b^-+2$ and the gaugini $\lambda^{a\mathcal A}\,,\;a=1,\ldots,b^+$, associated with the hyper- and vector multiplets respectively. More precisely, the $\zeta_{\alpha}$ are the hyperini of the theory prior to the dualization of the axions: after the dualization, the $\zeta_{\alpha}$ belong to a scalar-tensor multiplet containing the undualized scalars as well as the antisymmetric 2-tensors \cite{N=2withTensor2}; however, for simplicity we will continue to call them hyperini. The $N=2$ fermionic transformation laws read:
\begin{eqnarray}
\nnb  \delta\psi_{\mathcal A\mu}  &= & \ldots + \nabla_\mu\ep_{\mathcal A} - S_{\mathcal A\mathcal B}\gamma^{(4)}_\mu\ep^{\mathcal B} \\
\label{eq:N=2fermionicVar} \delta\zeta_{\alpha} & = &  \ldots + N_{\alpha}^{\mathcal A}\ep_{\mathcal A} \\
\nnb \delta\lambda^{a\mathcal A} & = & \ldots + W^{a\mathcal A\mathcal B}\ep_{\mathcal B}\;,
\end{eqnarray}
The ``$\ldots$'' refer to terms which vanish on a bosonic, maximally symmetric spacetime and which therefore will not be relevant for the supersymmetric vacuum conditions we are going to analyse in the forthcoming section. The label $\,^{(4)}$ recalls that in the 4d effective theory we use the rescaled metric\footnote{The difference $\nabla^{(4)}_\mu\ep_+-\nabla_\mu\ep_+$ has been included into the dots since, being proportional to $\partial_\mu\varphi$, it vanishes when evaluated on a Poincar\'e invariant vacuum.} defined in (\ref{eq:g4}); hence, $\gamma_\mu^{(4)}=e^{-\varphi}\gamma_\mu$. Finally, $S_{\mathcal A\mathcal B}, N_{\alpha}^{\mathcal A}$ and $W^{a\mathcal A\mathcal B}$ are the mass matrices for the associated fermions. They contain the flux charges, and their expression is \cite{N=2withTensor2, Michelson}:
\begin{eqnarray}
\label{eq:GenFormS} S_{\mathcal A\mathcal B} & = & \frac{i}{2}e^{\frac{K_+}{2}}(\sigma_x)_{\mathcal A}^{\;\;\mathcal C}\epsilon_{\mathcal B\mathcal C}(\mathcal P^x_AX^A - \tilde{\mathcal P}^{xA}\mathcal F_A) \\ [2mm]
\label{eq:HypMassMatrix} N_\alpha^{\mathcal A} & = & 2e^{\frac{K_+}{2}}\mathcal U_{\;\;\alpha u}^{\mathcal A}(k_A^u\bar X^A - \tilde k^{uA}\bar{\mathcal F_A}) \\ [2mm]
\label{eq:GaugMassMatrix} W^{a{\mathcal A\mathcal B}}& = &  ie^{\frac{K_+}{2}} g_+^{a\bar b}(\sigma_x)_{\mathcal C}^{\;\;\mathcal B}\epsilon^{\mathcal C\mathcal A} ( \mathcal P^x_C  D_{\bar b} \bar X^C - \tilde{\mathcal P}^{xC} D_{\bar b} \bar{\mathcal F}_C )\;.
\end{eqnarray}
Notice that the vielbeine $\mathcal U_{\;\;\alpha u}^{\mathcal A}$ of the quaternionic manifold prior to the dualization of the axions appear in the hyperino mass matrix $N_\alpha^{\mathcal A}$.

Of course all the mass matrices vanish in the absence of fluxes. In this case we would have a continuum of $N=2$ supersymmetric vacuum configurations (with vanishing cosmological constant), all the scalar fields corresponding to massless moduli. In the presence of fluxes, the mass matrices (\ref{eq:GenFormS})-(\ref{eq:GaugMassMatrix}) are nontrivial, and determine a potential for the 4d supergravity action (see e.g. \cite{N=2review, N=2withTensor2}), in this way lifting a certain number of previously flat scalar directions. In subsect.$\;$\ref{VacuumCondInN=2} we will analyse the $N=1$ vacuum conditions which are established imposing the vanishing of the $N=2$ fermionic transformation laws (\ref{eq:N=2fermionicVar}) under a \emph{single} supersymmetry.


\section{$N=1$ vacuum conditions}\label{VacuumCondit}

\setcounter{equation}{0}

In this section we confront the 4d and 10d approaches to $N=1$ backgrounds. In subsect$\:$\ref{ExpPureSpEq} the equations characterizing the $N=1$ vacua at the 10d level are rewritten in a way which is suitable for the comparison with the conditions arising in the 4d approach. These are analysed in subsect.$\:$\ref{VacuumCondInN=2}, having as a starting point the $N=2$ theory described in sect.$\:$\ref{KKreduct}.

Starting from an $N=2$ theory, an $N=1$ vacuum can be obtained by spontaneous partial supersymmetry breaking. This is a concrete possibility when considering compactifications with fluxes, since the outcoming 4d supergravities possess a nontrivial potential due to the flux-generated gaugings. However, spontaneous partial susy breaking is non-generic: the old no-go theorem of \cite{2into1won'tgo} forbidding such phenomenon on Minkowski vacua can be circumvented only by the choice of a degenerate symplectic section in the vector multiplet sector, such that a prepotential doesn't exist \cite{MinimalHiggsBranch}. On the other hand, the no-go theorem does not constrain AdS vacua, which represent therefore an available possibility (see \cite{HousePalti} for an example in the context of IIA compactification).

Such an obstruction for $N=1$ solutions with vanishing vacuum energy is somehow reflected at the 10d level: it is well known that tadpole cancellation in a background with fluxes consisting of the product of Minkowski$_4$ with a compact $M_6$ manifold requires the presence of negative tension sources, such as orientifold planes. With an appropriate choice of the orientifold, the resulting 4d effective theory takes an $N=1$ form, and corresponds to a truncation of the previously $N=2$ action. At this point, the $N=1$ vacuum condition amounts just to an unbroken susy requirement\footnote{Here we are considering dimensional reductions of 10d supergravity on compact manifolds. Further possibilities are opened by allowing for a decompactification limit freezing a part of the moduli \cite{TaylorVafa}.}.

The two possibilities we have mentioned (spontaneous partial susy breaking and the \hbox{$N=2\to N=1$} truncation) are not unrelated, since the physics around an $N=1$ vacuum for energies well below the partial susy breaking scale has to be described by an $N=1$ theory (see \cite{LouisN=2->N=1} for a discussion), and in some cases such low energy theory can correspond to a truncation of the $N=2$ action. 

Here however we don't need to specify which is the mechanism leading to the $N=1$ vacua, and it will be sufficient to observe that a supersymmetric (bosonic) vacuum is characterized by the vanishing of the fermionic variations under the preserved supersymmetries. In particular, starting from an $N=2$ theory one has an (at least) $N=1$ vacuum if such a condition is satisfied by the variations associated with any chosen linear combination of the $N=2$ spinorial parameters $\ep_{\mathcal A}\,$, $\mathcal A=1,2$. This characterization applies also to $N=2\rightarrow N=1$ truncations, provided the linear combination of the two susy generators under which the vacuum is required to be invariant is the same as the one which is preserved at the level of the action.

We can therefore proceed introducing a two component vector $n_{\mathcal A}={\bar  a\choose b}\,$, where $a$ and $b$ are complex constants\footnote{The choice of writing $n_1=\bar a$ instead of $a$ is dictated by later convenience, see the forthcoming eq.$\:$(\ref{eq:splitForN=1}). Of course, the $a$ here has nothing to do with the axion considered in subsect.$\:$\ref{4dsugraPicture}.} satisfying $|a|^2+|b|^2=1$, and we select the preserved positive-chirality $N=1$ susy parameter $\ep$ by
\be\label{eq:DefEp+}
\ep=\bar n^{\mathcal A}\ep_{\mathcal A}\qquad\Longleftrightarrow\qquad \ep_{\mathcal A}=n_{\mathcal A}\ep\;,
\ee
where $\bar n^{\mathcal A}= {\bar a\choose b}^{\dag}$, and the two expressions (\ref{eq:DefEp+}) are equivalent since we put to zero the independent linear combination $b\ep_1-\bar a\ep_2$. The conjugated spinors $\ep^{\mathcal A}$ can be written as $\ep^{\mathcal A}=\ep^c n^{*\mathcal A}$, where $\ep^c\equiv\ep^*$ has negative 4d chirality and $n^{*\mathcal A}={ a\choose \bar b}\;$.

Recalling eq. (\ref{eq:10dSpinorsIIA}), we can write the 10d spinor parameters $\epsilon^{1,2}$ on the vacuum as
\begin{eqnarray}
\nnb\epsilon^1 &=&  \ep\otimes \bar a\eta^1_-  + \ep^c\otimes a\eta^1_+ \\
\label{eq:splitForN=1}\epsilon^2 &=& \ep\otimes b\eta^2_+  + \ep^c\otimes \bar b\eta^2_-\;.
\end{eqnarray}

Since we are interested in maximally symmetric spacetimes (Minkowski$_4$ or AdS$_4$), we can furthermore choose $\ep$ to satisfy the Killing spinor equation
\be
\label{eq:KillingEqOne}\nabla_\mu\ep = \frac{1}{2}\bar\mu\gamma_\mu\ep^c\;.
\ee
The complex parameter $\mu$ is related to the 4d spacetime cosmological constant by $\Lambda=-3|\mu|^2$.

\subsection{$N=1$ equations from the ten dimensional analysis}\label{ExpPureSpEq}

Before establishing the 4d $N=1$ vacuum conditions arising from the 4d effective action, let's see which is the outcome of the 10d analysis for $N=1$ backgrounds. 

A supersymmetric background configuration\footnote{At this stage, we cannot speak of a full solution of the 10d (classical) action, since the supersymmetry conditions alone do not imply all the equations of motion and Bianchi identities for the bosonic fields. Here we will not consider this issue (see e.g. \cite{KoerberTsimpis} for a very recent discussion).} of the 10d supergravity with four preserved supercharges is obtained by imposing the vanishing of the 10d fermionic transformations under the supersymmetry parameterized by the spinor ansatz (\ref{eq:splitForN=1}). Having this as a starting point, it has been shown in \cite{GMPT2} (and reviewed in \cite{GMPT3}) that $N=1$ backgrounds of type II theories have an internal manifold whose tangent plus cotangent bundle admits an $SU(3)\times SU(3)$ structure. The supersymmetry equations can then be rephrased in the framework of generalized complex geometry as differential conditions for the pair of $O(6,6)$ pure spinors associated with the \hbox{$SU(3)\times SU(3)$} structure. With reference to the decomposition (\ref{eq:splitForN=1}), such pure spinors can be written as the following bispinors:
\be
a\eta^1_+\otimes (b\eta^2_+)^{\dag} = \frac{a\bar b}{8} \Phi^0_+\qquad,\qquad a\eta^1_+\otimes (\bar b\eta^2_-)^{\dag} =\frac{ab}{8}\Phi^0_-\;,\\
\ee
where for $\Phi^0_\pm$ we have used def.(\ref{eq:defPhi0}). The complex parameters $a$ and $b$ could in general depend on the internal coordinates, and indeed this would be the case for supersymmetric solutions on warped backgrounds. However, here we are interested in a comparison with what results from the effective action approach. For this reason we restrict ourselves to a vanishing warp factor\footnote{Dimensional reduction on warped backgrounds is not fully understood yet. For progress in this sense see for instance \cite{GiddingsDeWolfe, GiddingsMaharana, deAlwis}.} and we assume both $a$ and $b$ to be constant. Moreover, we allow the 10d dilaton $\phi$ to depend on the external coordinates only. 

Finally, we have to pay attention to the 10d spinor conventions. Indeed, in \cite{GMPT2, GMPT3}, the type IIA `pure spinor equations' were derived assigning positive chirality to $\epsilon^1$ and negative chirality to $\epsilon^2$, while in (\ref{eq:splitForN=1}) we have done the opposite choice (following the conventions of \cite{GLW2}). We find that the type IIA supersymmetry equations for the ansatz (\ref{eq:splitForN=1}) are obtained from the ones given in \cite{GMPT3} upon implementing the following transformation:
\begin{eqnarray}
\nnb a\bar b\Phi^0_+ \;\to \;\bar a b\bar\Phi^0_+\quad &,&\quad ab\Phi^0_- \;\to \; - \overline{a b\Phi^0_-} \qquad,\qquad H \;\to \; -H\\
\label{eq:TransfPSE} \lambda(F) \;\to \; F\quad &,& \quad *F  \;\to \; -*\lambda(F)\;,
\end{eqnarray}
where the RR fluxes $F=F_0+F_2+F_4+F_6$ are just internal and precisely correspond to the ones introduced in (\ref{eq:F=eBG}), while the involution $\lambda$ is defined in (\ref{eq:lambda}). 

We remark that the type IIA pure spinor equations obtained in this way correspond precisely to the ones given in \cite{GMPT3} for type IIB, provided we exchange the $O(6,6)$ chirality of the pure spinors and of the RR field strengths and we conjugate the complex parameter $\mu$ given in (\ref{eq:KillingEqOne}) (this last transformation is harmless, because it does not modify the physical quantity associated with $\mu$, which is the 4d spacetime cosmological constant $\Lambda= -3|\mu|^2$). So type IIA with the ansatz (\ref{eq:splitForN=1}) and type IIB with a positive chirality choice for both $\epsilon^{1,2}$ lead to the same pure spinor equations.

Starting from the pure spinor equations of \cite{GMPT3}, performing the transformations (\ref{eq:TransfPSE}) and taking into account the assumptions on the warping and the dilaton, we arrive at:
\begin{eqnarray}
\label{eq:firstPureSpinorEqNoWarpDilaton}(d-H\wedge)(a\bar b\Phi^0_+) &=& -2\bar \mu \re(ab\Phi^0_-)\\
\label{eq:secondPureSpinorEqNoWarpDilaton}(d - H\wedge)(ab\Phi^0_-)&=& -3i\,\im(\mu a\bar b\Phi^0_+)\,+\, \frac{e^\phi}{2}\big( c_- F  +i*\lambda(F) \big)\;,
\end{eqnarray}
where $c_\pm = |a|^2 \pm|b|^2$. Consistently with our definition of $a$ and $b$, (see above eq.$\:$(\ref{eq:DefEp+})), we will fix $c_+=1$. Of course, any other choice for $c_+$ can be recovered by the redefinition $n^{\mathrm{old}}_{\mathcal A} = n^{\mathrm{new}}_{\mathcal A} /\sqrt{c_+}$.

We now rewrite (\ref{eq:firstPureSpinorEqNoWarpDilaton}) and (\ref{eq:secondPureSpinorEqNoWarpDilaton}) in a form more suitable for the forthcoming comparison with the effective theory approach. Separating the background flux and the exact pieces of the NS 3-form as $H=H^{\mathrm{fl}}+d B$, acting with $e^{-B}$ on the equations and recalling relation (\ref{eq:F=eBG}) as well as defs. (\ref{eq:dHfl}) for $d_{H^{\mathrm{fl}}}$ and (\ref{eq:def*B}) for the $B$-twisted Hodge operator $*_B$, we get:
\begin{eqnarray}
\nnb a \bar b\,d_{H^{\mathrm{fl}}}\Phi_+&=& -2\bar\mu  \re(ab\Phi_-)\\
ab\,d_{H^{\mathrm{fl}}}\Phi_-&=& -3i\im(\mu a\bar b \Phi_+) + \frac{e^\phi}{2}\big( c_-G + i*_B G \big)\;.\end{eqnarray}
$\Phi_\pm:=e^{-B}\Phi^0_\pm$ are the same pure spinors as appearing in the effective theory approach of sect.$\:$\ref{KKreduct}.

Even if the pure spinor equations were derived assuming the background to be fully geometric, it is formally possible to substitute the differential operator $d_{H^{\mathrm{fl}}}$ with the more general operator $\mathcal D$ defined in (\ref{eq:CurlyD}). This is suggested by what is done in the effective action approach, along the lines of \cite{GLW2} (see also \cite{MicuPaltiTas}). We therefore obtain the following generalized version of the pure spinor equations:
\begin{eqnarray}
\label{eq:1stPSEfinal}a \bar b\,\mathcal D\Phi_+&=& -2\bar\mu  \re(ab\Phi_-)\\
\label{eq:2ndPSEfinal} ab\,\mathcal D\Phi_-&=& -3i\im(\mu a\bar b \Phi_+) + \frac{e^\phi}{2}( c_-G + i*_B G )\;.\end{eqnarray}

For example, in the $SU(3)$ structure case, in which $\Phi_-=-i\Om$, we have $(\mathcal D\Om)_0 = R\llcorner\Om$, $(\mathcal D\Om)_2 = Q\cdot\Om\,$, $(\mathcal D\Om)_4 = d\Om\,$ and $(\mathcal D\Om)_6 = -H^{\mathrm{fl}}\wedge\Om$.

As already announced, our purpose is to compare these equations with the $N=1$ vacuum conditions arising from the 4d effective action. In order to do this, we need only the pure spinor modes corresponding to light scalars in 4d, and we can therefore use the expansions (\ref{eq:expansPhi-}), (\ref{eq:expansPhi+}) of $\Phi_\pm$ in terms of the basis $\Sigma^\pm$. Using the properties of the basis forms, it is also possible to perform the integral over the internal manifold, which we assume to be compact.

We obtain the version `in components' of the pure spinor equations by taking the integrated Mukai pairing of the first and the second pure spinor equations - eqs. (\ref{eq:1stPSEfinal}) and (\ref{eq:2ndPSEfinal}) - with the basis $\Sigma^\pm$ (seen as vectors of forms as in (\ref{eq:Sigma+-AsVectors})). Adopting the symplectic notation introduced in subsections \ref{AnsatzForms}-\ref{KillingPrep} (see in particular eqs.~(\ref{eq:SymplStrM-}), (\ref{eq:evenSymplecticStr}) for $\mathbb{S}_\pm$, eq.~(\ref{eq:SymplecticN}) for $\mathbb{N}$, eq.~(\ref{eq:ChargeMatrixQ}) for $\mathbb{Q}$, as well as eq.~(\ref{eq:V+V-Vg}) for $V_\pm$ and $V_G$), by a straightforward computation one can see that
\be\label{eq:1stEqInEffActLanguage}
\int\langle \,1^{\mathrm {st}}\textrm{ pure sp.eq.} \,,\, \Sigma^-\,\rangle\quad\quad\Longrightarrow\quad\quad
\framebox[1.17\width ][c]{$ a \bar b\mathbb Q V_+ = -2\bar\mu\re (abV_-)$}\;,\ee
\be\label{eq:2ndEqInEffActLanguage}\int\langle\,  2^{\mathrm {nd}}\textrm{ pure sp.eq.}\,,\, \Sigma^+ \,\rangle\;\:\Longrightarrow\;\:\framebox[1.10\width ][c]{$ab \,(\mathbb S_+)^{-1}\mathbb Q^T\mathbb S_- V_-= -3i\im (\mu a\bar b V_+) + c_-\frac{e^{\phi}}{2}V_G - i\frac{e^{\phi}}{2}\mathbb{N}V_G$}\,.\ee
\vskip .2cm
In this last derivation, it has been essential to dispose of eq.$\:$(\ref{eq:SymplecticN}), expressing the action of $*_B$ in terms of the special geometry data, in order to compute
\be
\int \langle *_B G , \Sigma^+ \rangle \,=\, -\mathbb N V_G\;.
\ee

\subsection{$N=1$ conditions from the effective action, and matching}\label{VacuumCondInN=2}

We now study the $N=1$ vacuum conditions arising from the effective action approach, showing that they precisely satisfy the integrated version of the pure spinor equations established here above.

At the end of subsect.$\:$\ref{4dsugraPicture} we wrote the form of the fermionic susy variations for the 4d $N=2$ effective theory corresponding to the type IIA compactification we considered. As it should be clear from the discussion at the beginning of this section, the 4d $N=1$ vacuum conditions amount to the vanishing of these fermionic variations under the single preserved supersymmetry, parameterized as in (\ref{eq:DefEp+}). From (\ref{eq:N=2fermionicVar}) we read:
\begin{eqnarray}
\label{eq:CondOnS}\langle \delta_{\ep}\psi_{\mathcal A\mu} \rangle=0 &\quad \Longleftrightarrow \quad & 2e^{-\varphi}S_{\mathcal A\mathcal B}n^{*\mathcal B} = n_\mathcal A\bar\mu \\ [2mm]
\label{eq:CondOnN}\langle \delta_{\ep}\zeta_{\alpha} \rangle=0 & \quad \Longleftrightarrow\quad  &  N_{\alpha}^{\mathcal A}n_{\mathcal A} = 0\\ [2mm]
\label{eq:CondOnW}\langle \delta_{\ep}\lambda^{a\mathcal A} \rangle=0 &\quad  \Longleftrightarrow\quad  & W^{a\mathcal A\mathcal B}n_{\mathcal B} = 0\;.
\end{eqnarray}
To get condition (\ref{eq:CondOnS}) we used (\ref{eq:KillingEqOne}) and $\gamma_\mu^{(4)}=e^{-\varphi}\gamma_\mu$. Eq.$\:$(\ref{eq:CondOnS}) relates the Killing prepotentials to the spacetime curvature parameter $\mu$. Recalling (\ref{eq:GenFormS}) and (\ref{eq:PxFromPxAandTildePxA}), its explicit form is
\be\label{eq:CondOnSExplicit}
ie^{\frac{K_+}{2}-\varphi}\left(\begin{array}{c}
a(\mathcal P^1-i\mathcal P^2) -\bar  b\mathcal P^3\\
- a\mathcal P^3 -\bar b(\mathcal P^1+i\mathcal P^2)
\end{array}\right) = \left(\begin{array}{c}
\bar a\bar \mu \\ b\bar \mu
\end{array}\right)\;.
\ee

Let's now analyse the implications following from the vanishing of the hyperini variation, eq. (\ref{eq:CondOnN}). Recalling (\ref{eq:HypMassMatrix}), this reads
\be
n_{\mathcal A}\mathcal U_{\;\;\alpha u}^{\mathcal A}(k_A^u\bar X^A - \tilde k^{uA}\bar{\mathcal F_A})=0\;.
\ee
Using (\ref{eq:quaternVielbein}) and (\ref{eq:FerrSabharw1forms}), substituting the expressions (\ref{eq:ElectricKillingVectors}), (\ref{eq:MagneticKillingVectors}) for $k_A$ and $\tilde k^A$, and recognizing the form (\ref{eq:GeneralKillingPrepot}) of the $\mathcal P^x$, we obtain the following set of conditions:
\begin{eqnarray}
\label{eq:HypVar1} a (\mathcal P^1-i\mathcal P^2) -2\bar b \mathcal P^3 &=& 0\\
\label{eq:HypVar2} 2 a \mathcal P^3 + \bar b(\mathcal P^1 + i\mathcal P^2) &=& 0\\
\label{eq:HypVar3} \bar b P_I (\im \mathcal G)^{-1\,IJ}\big[ (e_{JA}-\mathcal M_{JK}m_A^K)X^A + (p^A_J - \mathcal M_{JK}q^{KA})\mathcal F_A \big] &=& 0\\
\label{eq:HypVar4} a\bar P_I(\im \mathcal G)^{-1\,IJ}\big[ (e_{JA}-\overline{\mathcal M}_{JK}m_A^K)X^A + (p^A_J - \overline{\mathcal M}_{JK} q^{KA})\mathcal F_A \big] &=& 0.
\end{eqnarray}
The first two equations come from the vielbeine $\mathcal U_{\;\;\alpha}^{\mathcal A}$ corresponding to the 1-forms $u$ and $v$ given in (\ref{eq:FerrSabharw1forms}), while the last two are the conditions involving $E$ ($e$ doesn't contribute). The $P_I$ are the K\"ahlerian vielbeine defined below eq.$\:$(\ref{eq:FerrSabharw1forms}).

Comparing (\ref{eq:HypVar1}) and (\ref{eq:HypVar2}) with (\ref{eq:CondOnSExplicit}) we get
\be\label{eq:P3isMu}
i\frac{a}{2}(\mathcal P^1-i\mathcal P^2)= i\bar b\mathcal P^3=\bar a\bar \mu e^{\varphi - \frac{K_+}{2}}\qquad,\qquad   -\frac{i\bar b}{2}(\mathcal P^1 + i\mathcal P^2) =i a  \mathcal P^3= b\bar\mu e^{\varphi - \frac{K_+}{2}}\;,
\ee
which implies $(|a|^2-|b|^2)\bar \mu =0$; then if the vacuum is AdS, necessarily\footnote{$|a|=|b|$ is also necessary for a Minkowski background; however in this case the condition doesn't arise from the susy equations, but rather from the orientifold projection one is led to consider in order to cancel the tadpoles \cite{GMPT3}. Another way to arrive at the same conclusion is described in \cite{MartucciSmyth}.} $|a|=|b|$. Furthermore, notice that on a Minkowski vacuum ($\mu =0$) and for $a$ and $b$ being nonzero we have $\mathcal P^x=0$; therefore the gravitino mass matrix $S_{\mathcal A\mathcal B}$ vanishes (see (\ref{eq:MassMatrixAndPrepot})) and we cannot have spontaneous partial susy breaking in the $N=2$ theory. In order to obtain $N=1$ Minkowski vacua, an $N=2 \to N=1$ truncation of the action is required.

{F}rom now on we will assume $a\neq 0\,,\;b\neq 0$. The cases in which $a=0$ or $b=0$ could be studied separately; however, they are not relevant for the comparison with the pure spinor equations of the previous subsection, which were indeed established for nonvanishing $a$ and $b$.

Multiply eqs.$\:$(\ref{eq:HypVar3}) by $\frac{1}{2}e^{-K}(\im \mathcal G)^{-1\,LM}\bar P_M$ and (\ref{eq:HypVar4}) by $\frac{1}{2}e^{-K}(\im \mathcal G)^{-1\,LM}P_M$, then use the relations \cite{FerraraSabharwal}:
\begin{eqnarray}
\nnb \frac{1}{2}e^{-K_-}[(\im \mathcal G)^{-1}P^\dag P(\im \mathcal G)^{-1}]^{LJ} &=& (\im \mathcal G)^{-1\,LJ} + 2e^{K_-}Z^L\bar Z^J\\
& = & -(\im \mathcal M)^{-1\,LJ} -2e^{K_-}\bar Z^L Z^J
\end{eqnarray}
(see (\ref{eq:RelGM}) for the second equality). Recognizing expressions (\ref{eq:GeneralKillingPrepot}) for $\mathcal P^1\pm i\mathcal P^2$, we arrive at
\begin{eqnarray}
\nnb \bar b (\im\mathcal M)^{-1\,IJ}\big[(e_{JA}-\mathcal M_{JK}m_A^K)X^A + (p^A_J - \mathcal M_{JK}q^{KA})\mathcal F_A\big] + \bar Z^Ie^{\frac{K_-}{2}-\varphi}\bar b (\mathcal P^1-i\mathcal P^2) &=& 0\\
\nnb a(\im\mathcal M)^{-1\,IJ}\big[(e_{JA}-\overline{\mathcal M}_{JK}m_A^K)X^A + (p^A_J - \overline{\mathcal M}_{JK}q^{KA})\mathcal F_A\big] + Z^Ie^{ \frac{K_-}{2}-\varphi}a (\mathcal P^1+i\mathcal P^2) &=& 0\;.\\ \label{eq:vectorOf2eqs}
\end{eqnarray} \vskip -.2cm
\noindent Multiplying from the left (\ref{eq:vectorOf2eqs}) by ${a \choose -\bar b}^T$ and using (\ref{eq:P3isMu}) we conclude
\be\label{eq:FirstHalfFirstPSE}
a\bar  b (m_A^I X^A + q^{IA} \mathcal F_A)= -2\bar \mu \re(ab Z^I)\;,
\ee
where we have also used the fact that, because of the normalizations we adopted for the pure spinors $\Phi_\pm$, we have $e^{K_+}=e^{K_-}$.

A second independent linear combination of the two equations (\ref{eq:vectorOf2eqs}) can be obtained multiplying them by ${a \choose \bar b}^T$. Plugging (\ref{eq:FirstHalfFirstPSE}) in, using again (\ref{eq:P3isMu}) and recalling that \hbox{$\mathcal M_{IJ}Z^J=\mathcal G_I$}, we arrive at
\be\label{eq:SecondHalfFirstPSE}
a \bar b(e_{IA}X^A + p_I^A\mathcal F_A) = -2\bar \mu \re(ab\mathcal G_I)\;.
\ee

Employing the symplectic notation introduced in eqs. (\ref{eq:ChargeMatrixQ}), (\ref{eq:V+V-Vg}) , our conditions (\ref{eq:FirstHalfFirstPSE}) and (\ref{eq:SecondHalfFirstPSE}) can be summarized in a single equation for the symplectic vectors $V_\pm$:
\be\label{eq:ResultHyperiniVar}
\framebox[1.15\width ][c]{$  a \bar b\mathbb Q V_+ = -2\bar\mu\re (a bV_-)$}\;.\ee
As it is clear from a comparison with eq.$\:$(\ref{eq:1stEqInEffActLanguage}), the present condition precisely corresponds to the integrated first pure spinor equation.\\

The last condition to be analysed is the variation of the gaugini, eq.$\:$(\ref{eq:CondOnW}). Using (\ref{eq:GaugMassMatrix}) this reads
\be 
ie^{\frac{K_+}{2}} g_+^{a\bar b}D_{\bar b} \bar X^C( \mathcal P^x_C  - \mathcal N_{CE} \tilde{\mathcal P}^{xE}  )\sigma_x^{\mathcal A\mathcal B}n_{\mathcal B}  = 0\;,
\ee
where $\sigma_x^{\mathcal A\mathcal B}=(\sigma_x)_{\mathcal C}^{\;\;\mathcal B}\epsilon^{\mathcal C\mathcal A}$ and we have used (\ref{eq:DefMathcalN}) in order to factorize $D_{\bar b}\bar X^C$. Multiply this expression by $e^{\frac{K_+}{2}}D_a X^D$ in order to trade a lower case index with an upper case one; then using the special geometry relation
\be\label{eq:relgNforM+}
e^{K_+}D_aX^Dg_+^{a\bar b}D_{\bar b}\bar X^C=-\frac{1}{2}(\im \mathcal N)^{-1\,DC}-e^{K_+}\bar X^DX^C \ee 
(corresponding to the $\mathscr M_+$ version of (\ref{eq:usefulformula})) and recalling that $(\mathcal P^x_B - \mathcal N_{BC}\tilde{\mathcal P}^{x C})X^B =\mathcal P^x$ (see (\ref{eq:DefMathcalN}) and (\ref{eq:PxFromPxAandTildePxA})), we get
\be\label{eq:EqFromGauginiVar}
\sigma_x^{\mathcal A\mathcal B} n_{\mathcal B} \big[(\im \mathcal N)^{-1\,AB}(\mathcal P^x_B - \mathcal N_{BC}\tilde{\mathcal P}^{x C} ) + 2e^{K_+}\bar X^A \mathcal P^x\big]  = 0\;.
\ee
This is a vector of two equations ($\mathcal A=1,2$). Multiply it from the left by ${-\bar b\choose  a}^T=\bar n^{\mathcal C} \epsilon_{\mathcal C\mathcal A}$. Using (\ref{eq:CondOnSExplicit}) one sees that $\bar n^{\mathcal C} \epsilon_{\mathcal C\mathcal A}\sigma_x^{\mathcal A\mathcal B} n_{\mathcal B}\mathcal P^x =0$, therefore we are left with
\be\label{eq:LinComb1FromGaugini}
\big\{ 2\re[ab( \delta_x^{1}-i\delta_x^{2})]  + c_-\delta_x^{3} \big\}(\im\mathcal N)^{-1\,AB} (\mathcal P^x_B - \mathcal N_{BC}\tilde{\mathcal P}^{x C} ) = 0\;,
\ee
where we have introduced the parameter $c_-:= |a|^2-|b|^2$. Separating into imaginary and real parts we arrive respectively at:
\be
\re[ab( \tilde{\mathcal P}^{1 A}-i\tilde{\mathcal P}^{2 A})]  + \frac{c_-}{2}\tilde{\mathcal P}^{3 A} =0\qquad,\qquad  \re[ab(\mathcal P^1_A-i\mathcal P^2_A)]  + \frac{c_-}{2}\mathcal P^3_A  = 0\;.
\ee
Substituting the expressions (\ref{eq:electricP}), (\ref{eq:magneticP}) for $\mathcal P^x_A$ and $\tilde{\mathcal P}^{x A}$ and using (\ref{eq:chain}) as well as the definition of $G^A$ and $\tilde G_A$ below (\ref{eq:ExpRRstrengths}), we obtain the couple of equations
\be
\re(abZ^I)p_I^A - \re(ab\mathcal G_I) q^{IA} = \frac{c_-}{2}e^\phi G^A\quad , \quad \re(ab Z^I)e_{IA} - \re(ab \mathcal G_I) m^I_A  = \frac{c_-}{2}e^\phi \tilde G_A\;,
\ee
which can be assembled in a single equation for the symplectic vectors defined in (\ref{eq:V+V-Vg}):
\be\label{eq:GauginiSympl1}
(\mathbb S_+)^{-1}\mathbb Q^T\mathbb S_- \re(abV_-)=  \frac{c_-}{2}e^{\phi}V_G \;.
\ee

Multiplying the two equations (\ref{eq:EqFromGauginiVar}) by $ {\bar b\choose a}^T$ and using once again constraint (\ref{eq:P3isMu}), we get a second independent (recall that $a\neq 0\,,\;b\neq 0$) combination:
\be\label{eq:LinComb2FromGaugini}
\big\{ 2\,\im[a b(\delta_x^1-i\delta_x^2)] -i \delta_x^3 \big\}(\im\mathcal N)^{-1\,AB}(\mathcal P^x_B - \mathcal N_{BC}\tilde{\mathcal P}^{x C} ) = 12 e^{\frac{K_+}{2} + \varphi }\bar a b\bar \mu \bar X^A\;.
\ee
Following analogous steps to the ones which led us from (\ref{eq:LinComb1FromGaugini}) to (\ref{eq:GauginiSympl1}), we arrive at
\be\label{eq:GauginiSympl2}
(\mathbb S_+)^{-1}\mathbb Q^T\mathbb S_- \im(abV_-)= -3\im (\mu a\bar b V_+) - \frac{e^{\phi}}{2}\mathbb{N}\,V_G\;,
\ee
where the symplectic matrix $\mathbb N$ is given in (\ref{eq:SymplecticN}).

Conditions (\ref{eq:GauginiSympl1}), (\ref{eq:GauginiSympl2}) can be seen as the real and the imaginary parts of the single complex equation:
\be\label{eq:ResultGauginiVar}
\framebox[1.15\width ][c]{$ab (\mathbb S_+)^{-1}\mathbb Q^T\mathbb S_- V_-= -3i\im (\mu a\bar b V_+) + c_-\frac{e^{\phi}}{2}V_G - i\frac{e^{\phi}}{2}\mathbb{N}V_G\;$}\;.\ee
In this way we obtain a condition which exactly corresponds to the integrated second pure spinor equation, as it can be seen by comparison with eq.$\:$(\ref{eq:2ndEqInEffActLanguage}).
\vskip .3cm
Let us summarize the outcome of this section. At the 10d background level, we expanded the pure spinor equations on the basis of forms $\Sigma^\pm$ and we took the integral over the internal manifold, obtaining eqs.$\:$(\ref{eq:1stEqInEffActLanguage}) and (\ref{eq:2ndEqInEffActLanguage}). At the level of the 4d effective theory, we started from the vev of the fermionic variations under an arbitrary linear combination of the two $N=2$ supersymmetries, eqs.$\:$(\ref{eq:CondOnS})-(\ref{eq:CondOnW}), and we exploited the special geometry properties to rewrite the conditions in a more compact way. From the hyperini variation we obtained eq.$\:$(\ref{eq:ResultHyperiniVar}), corresponding to the integrated first pure spinor equation, while the gaugini transformation (together with constraint (\ref{eq:P3isMu})$\,$) yields eq.$\:$(\ref{eq:ResultGauginiVar}), which coincides with the integrated second pure spinor equation. The gravitini variation has been used to simplify the expressions, in particular to obtain constraint (\ref{eq:P3isMu}), which relates $\mathcal P^{1}\pm i \mathcal P^{2}$ or $\mathcal P^{3}$ to the spacetime curvature parameter $\mu$.

\section{Aspects of $N=2\rightarrow N=1$ theories}\label{N2toN1}
\setcounter{equation}{0}

In section \ref{VacuumCondit} we studied the conditions to have an $N=1$ vacuum starting from the $N=2$ effective supergravity defined by the compactification of type IIA on an $SU(3)\times SU(3)$ background. Physically, such solutions can be realized either by spontaneous partial supersymmetry breaking in the $N=2$ theory, or as supersymmetry-preserving solutions of an $N=1$ theory obtained as a consistent truncation of the $N=2$ action. In string theory, such $N=2\rightarrow N=1$ truncations can be realized including appropriate orientifold planes in the 10d background. Truncations can also be relevant for spontaneous partial susy breaking, in the sense that the $N=1$ theory describing the low energy physics around an $N=1$ vacuum which breaks $N=2$ spontaneously can in some special cases correspond to a truncation of the $N=2$ action. An example of this was provided in ref.$\,$\cite{HousePalti}, where the low energy $N=1$ effective action describing the fluctuations around the $N=1$ AdS$_4\times\,$half-flat vacuum solution of ref.$\,$\cite {LustTsimpis} was obtained by truncating an $N=2$ theory. This example is however special since the only hypermultiplet contained in the $N=2$ theory is the universal one.

In this section we study some aspects of the $N=1$ theory obtained as a generic truncation of the $N=2$ effective action described in section \ref{KKreduct}. In particular, we focus on the way the $N=1$ superpotential and D-terms are determined as linear combinations of the three $N=2$ Killing prepotentials $\mathcal P^x$. Then we write the F- and D- flatness conditions for $N=1$ vacua, establishing their relation with the $N=1$ conditions of the previous section. In this way we will we able to nicely reinterpret the matching with the pure spinor equations coming from the 10d analysis.

A thorough analysis of the conditions allowing to define a consistent \hbox{$N=2\to N=1$} truncation has been performed in \cite{Truncation1}, and extended in \cite{TruncationWithTensors} for the case in which tensor multiplets are also present.

\subsection{$N=1$ superpotential}

In subsect.$\:$\ref{KillingPrep} we briefly reviewed how refs.$\:$\cite{GLW1, GLW2} got the Killing prepotentials of the $N=2$ theory which is defined starting from a 10d background preserving eight supercharges. This strategy was further pursued in the same papers by restricting the background to preserve four supercharges rather than eight. In this way, as we will recall next, an expression for the $N=1$ superpotential $\mathcal W$ was obtained (see (\ref{eq:FirstExpreForeKW})).

The preserved $N=1$ spinor parameter can be chosen as in (\ref{eq:DefEp+}), and the correspondent linear combination of the $N=2$ gravitini defines the positive-chirality $N=1$ gravitino:  $\psi_\mu= \bar n^{\mathcal A}\psi_{\mathcal A\mu}$. Then, recalling the general form of the $N=2$ gravitini variation, eq.$\:$(\ref{eq:N=2GravVar}), one has:
\be\label{eq:N1GravVarFromN2}
\delta_{\ep}\psi_\mu=\bar n^{\mathcal A}\delta_{\ep}\psi_{\mathcal A\mu}=\nabla_\mu\ep -\bar n^{\mathcal A} S_{\mathcal A\mathcal B} n^{*\mathcal B} \gamma^{(4)}_\mu \ep^c\;.
\ee
On the other hand, the general form of the gravitino transformation in $N=1$ supergravity is
\be\label{eq:generalN=1gravVar}
\delta_{\ep}\psi_\mu=\nabla_\mu\ep -e^{\frac{K}{2}}\mathcal W\gamma_\mu^{(4)}\ep^c\;,
\ee
where the combination $e^{\frac{K}{2}}\mathcal W$ involving the $N=1$ K\"ahler potential $K$ and the superpotential $\mathcal W$ is related to the gravitino mass. 

Comparing (\ref{eq:N1GravVarFromN2}) and (\ref{eq:generalN=1gravVar}), one arrives at the identification \cite{GLW1, GLW2}:
\be
\label{eq:FirstExpreForeKW} e^{\frac{K}{2}}\mathcal W \;=\; \bar n^{\mathcal A} S_{\mathcal A\mathcal B} n^{*\mathcal B} \; = \; \frac{i}{2}e^{\frac{K_+}{2}} \big[ a^2(\mathcal P^1-i\mathcal P^2)-\bar  b^2(\mathcal P^1+i\mathcal P^2)-2 a \bar b \mathcal P^3   \big]\;,
\ee
where in the second equality eq.$\:$(\ref{eq:MassMatrixAndPrepot}) has been used.

At this point let us make a comment. The combination of the $N=2$ gravitini which is orthogonal to the one defining $\psi_\mu$ is $\tilde\psi_\mu := b \psi_{1\mu} - \bar a \psi_{2\mu}$. From the point of view of the $N=1$ theory, $\tilde\psi_\mu$ would be a component of a (possibly massive) spin $3/2$ multiplet. Such multiplets are usually not included in the standard supergravity action, and should therefore be truncated out of the spectrum. However, the truncation is consistent only if the variation of $\tilde \psi_\mu$ under the preserved supersymmetry vanishes identically: $\delta_{\ep}\tilde\psi_\mu \equiv 0$. Using the general form of the $N=2$ gravitini variation and mass matrix, eqs.$\:$(\ref{eq:N=2GravVar}) and (\ref{eq:MassMatrixAndPrepot}), this can be written as
\be
\label{eq:constrFromSecondGravitino} 
e^{\frac{K_+}{2}}\big[ab(\mathcal P^1-i\mathcal P^2)+\bar a \bar b(\mathcal P^1+i\mathcal P^2)+c_-\mathcal P^3 \big] \,= \, 0\;,
\ee
where as before $c_-=|a|^2-|b|^2$. Exploiting this constraint, we rewrite the combination $e^{\frac{K}{2}}\mathcal W$ in a slightly different form. Assuming $a\neq 0, b\neq 0$, multiplying (\ref{eq:constrFromSecondGravitino}) by $\frac{ic_-}{4\bar a b}$ and subtracting it from (\ref{eq:FirstExpreForeKW}), we get the more symmetric looking expression
\be\label{eq:2ndExprForeKW}
e^{\frac{K}{2}}\mathcal W\;=\; \frac{i}{4\bar a b}e^{\frac{K_+}{2}} \big[a b(\mathcal P^1-i\mathcal P^2)-\bar a \bar b(\mathcal P^1+i\mathcal P^2)-\mathcal P^3  \big]\;.
\ee
Notice that if $c_-=0\Leftrightarrow |a|^2=|b|^2=1/2$, then eq. (\ref{eq:FirstExpreForeKW}) already has this form. Substituting the geometric expressions (\ref{eq:KillingPrepGeneral}) for the three $\mathcal P^x$, we conclude that
\be
e^{\frac{K}{2}}\mathcal W \; = \;  \frac{i}{4\bar a b}e^{\frac{K_+}{2}+2\varphi} \Big[4ie^{\frac{K_-}{2}-\varphi}\int\langle \Phi_+ \,,\, \mathcal D \im(ab\Phi_-)  \rangle +   \frac{1}{\sqrt2}\int\langle \Phi_+ , G  \rangle\,  \Big]\;.
\ee
We identify the $N=1$ K\"ahler potential $K$ as \cite{GrimmLouisA,GrimmBen, HousePalti, VilladZwirner}:
\be\label{eq:N=1kahlerPot}
K\;=\;K_++ 4\varphi\;.
\ee
This yields the compact expression for the superpotential
\be
\label{eq:WfromN=2} \mathcal W = \frac{i}{4\bar ab}\int\langle \Phi_+ \,,\,\frac{1}{\sqrt 2} G^{\mathrm{fl}}+ \mathcal D \Pi_- \rangle\;,
\ee
where we have defined
\be\label{eq:defPi-} \Pi_-:=\frac{1}{\sqrt 2}A + i\im(C\Phi_-)\;.\ee
The so called compensator \cite{GrimmLouisA, GrimmBen}
\be\label{eq:DefC} C:= \sqrt 2 a be^{-\phi}=4 a b e^{\frac{K_-}{2}-\varphi}
\ee
(recall (\ref{eq:chain}) for the relation of the 10d dilaton $\phi$ with $K_-$ and $\varphi$) is a scalar trading the irrelevant rescaling freedom in $\Phi_-$ for the physical degree of freedom encoded in $\varphi$. In fact, the combination $C\Phi_-$ is invariant under (real) rescalings of $\Phi_-$.

In ref.$\:$\cite{GrimmBen} the form (\ref{eq:WfromN=2}) of the $N=1$ superpotential was derived in the context of type IIA compactifications in the presence of an $O6$ orientifold. Here we have a slightly different perspective, in that we are just requiring an $N=2\to N=1$ truncation, not necessarily induced by an orientifold. This is in principle more general: for instance, the orientifold requires $|a|=|b|$, while here we are not imposing $c_-=0$. It is not clear to us whether this really allows for more general constructions. An argument against this is that a 10d analysis indicates that $c_-$ should vanish for all compact $N=1$ solutions \cite{GMPT3, MartucciSmyth}. Restricting to $c_-=0$, anyway, does not necessarily mean considering an orientifold, and eq.$\:$(\ref{eq:WfromN=2}) should also give the correct superpotential of those $N=1$ low energy effective theories valid around $N=1$ AdS$_4$ vacua breaking $N=2$ spontaneously (at least for the cases in which these $N=1$ theories correspond to $N=2$ truncations). This seems to be confirmed by the fact that in the geometric $SU(3)$ structure case, the superpotential (\ref{eq:WfromN=2}) reduces to the one appearing in the example of ref.$\:$\cite{HousePalti} mentioned at the beginning of this section. Clearly, it would be interesting to find a concrete new example.

We obtain the form of the superpotential in terms of the flux charges and the 4d fields if we substitute into (\ref{eq:2ndExprForeKW}) the explicit expressions (\ref{eq:GeneralKillingPrepot}) of the $N=2$ Killing prepotentials:
\begin{eqnarray}
\nnb  \mathcal W&=&\frac{i}{4\bar ab}\Big\{ \big[ i\im(CZ^I) e_{IA} -i\im(C\mathcal G_I) m_A^I\big]X^A+ \big[ i\im(CZ^I) p^A_I -i\im(C\mathcal G_I) q^{IA}\big]\mathcal F_A  \\
\label{eq:explicitWinN=2fields} &+& (e_{\mathrm{RR}A}+ \xi^I e_{IA} - \tilde\xi_I m_{A}^{\;\;I}) X^A + (m^A_{\mathrm{RR}} + \xi^I p^A_{I} -  \tilde\xi_I q^{IA}) \mathcal F_A  \Big\}\;.
\end{eqnarray}
Eq. (\ref{eq:explicitWinN=2fields}) is still written in terms of the $N=2$ degrees of freedom, while we should restate it in $N=1$ variables.
Recall that, as discussed in subsect.$\:$\ref{4dsugraPicture}, in the $N=2$ theory a subset of the scalars $\xi^I, \tilde\xi_I$, together with the axion $a$, has been dualized to antisymmetric 2-tensors in order to allow the introduction of the magnetic charges $\,m_{\mathrm{RR}}^A,p_I^A, q^{IA}$. However, according to the remark below eq.$\:$(\ref{eq:magneticP}), the Killing prepotential $\mathcal P^3$ just contains the combinations of the $\xi^I, \tilde \xi_I$ which have not been dualized to antisymmetric tensors. Hence the same will be true for the expression (\ref{eq:explicitWinN=2fields}) of the superpotential. These scalars should be recombined with the other $N=2$ degrees of freedom $z^i,\varphi$ contained in (\ref{eq:explicitWinN=2fields}) in order to define appropriate holomorphic $N=1$ variables for the superpotential. Inspection shows that $\mathcal W$ depends holomorphically on the following combinations:
\be
\label{eq:defholomvar} U^I \,:=\,\xi^I + i\im(CZ^I)\qquad,\qquad \tilde U_I \,:=\,  \tilde \xi_I  +i\im(C\mathcal G_I).
\ee

Instead no redefinition is needed for the scalars $t^a$ coming from the $N=2$ vector multiplets since they appear in (\ref{eq:explicitWinN=2fields}) only through the holomorphic functions $X^A(t)$ and $\mathcal{F}_A(t)$.\\
From (\ref{eq:defPi-}) and (\ref{eq:expRRflAndPot}) we can see that $U^I$ and $\tilde U_I$ are precisely the coefficients of the expansion of $\Pi_-$ on the basis of odd forms:
\be\label{eq:Pi-InTermsOfU}
\Pi_-=U^I\alpha_I - \tilde U_I\beta^I\;.
\ee 
Therefore $\Pi_-$ defines the correct $N=1$ coordinates, and is the $N=1$ analog of $\Phi_-$ \cite{GrimmBen}. 

The form of the field redefinition (\ref{eq:defholomvar}) was already identified in \cite{GrimmLouisA, GrimmBen, HousePalti}. Here we have verified that it is appropriate for any $N=2\to N=1$ truncation, even in the presence of the general set of fluxes defined in subsect.$\:$\ref{DiffConAndFluxes}.

Substituting (\ref{eq:defholomvar}) into (\ref{eq:explicitWinN=2fields}), we have \cite{MicuPaltiTas}
\be\label{eq:ExplicitWinN=1fields}
\mathcal W=\frac{i}{4\bar a b}\Big[U^Ie_{IA}X^A - \tilde U_I m_A^{I}X^A + U^I p^A_I\mathcal{F}_A - \tilde U_I q^{IA}\mathcal{F}_A
 + X^Ae_{\mathrm{RR}A}+\mathcal{F}_A m^A_{\mathrm{RR}} \,  \Big]\;,
\ee
which now depends on holomorphic variables only. Notice that this form of the superpotential directly descends from (\ref{eq:WfromN=2}) if the expansion (\ref{eq:Pi-InTermsOfU}) is used.

\subsection{D-terms from $N=2 \;\rightarrow\;N=1$ truncations}\label{Dterms}

Having as a starting point the 4d $N=2$ supergravity defined by the $SU(3)\times SU(3)$ compactification of type IIA, we now derive the general form of the D-terms arising from an $N=2\to N=1$ truncation. As the superpotential, the D-terms are determined by a linear combination of the three $N=2$ Killing prepotentials. If the superpotential was obtained by looking at the gravitini variations, we will identify the D-terms by studying the gaugini transformations.

Before going into this, we need some more notions about $N=2\to N=1$ truncations. Unlike rigid supersymmetry, one cannot rewrite an $N=2$ supergravity in an $N=1$ form unless some restrictions are imposed. We have already discussed the necessity of truncating the spin 3/2 multiplet. Consistency with supersymmetry then imposes a series of constraints involving the other fields appearing in the action \cite{Truncation1, TruncationWithTensors}.

For the sake of writing an expression for the D-terms, we won't need to consider the whole set of constraints, rather we can restrict to the ones involving the $N=2$ vector multiplets. In particular, it is not necessary to deal with the more involved part of the story, namely the fact that (leaving aside the further complication due to the possible dualization to antisymmetric 2-tensors) the $N=2$ quaternionic manifold parameterized by the scalar components of the hypermultiplets has to reduce to a submanifold respecting the K\"ahler-Hodge structure required by $N=1$ supersymmetry. Some aspects of this will be needed in subsect.$\,$\ref{OrientifoldTrunc}, where we will study the F-flatness conditions in the case of an orientifold-induced truncation.

An $N=2$ vector multiplet is composed of one vector, one complex scalar and two Weyl fermions (the gaugini), and splits in an $N=1$ vector multiplet and an $N=1$ chiral multiplet. The consistent truncation acts in such a way that out of $n_V$ $N=2$ vector multiplets (for us $n_V=b^+\equiv$ dim$\mathscr M_+$), the resulting $N=1$ theory inherits just $n_{Ch}\leq n_V$ chiral multiplets and $\widehat n_V= n_V-n_{Ch}$ vector multiplets. In more detail, splitting the indices as $A = (\check{A},\widehat A)\;$, with $A=0,\ldots,n_V\;$, $\check{A}=0,\ldots,n_{Ch}\;$ and $\widehat A=1,\ldots,\widehat n_V=n_V-n_{Ch}$, we have the following conditions \cite{Truncation1}:
\be\label{eq:TruncCond1}
A_\mu^{\check A}=0\quad,\quad X^{\widehat A}=0\ee
Notice that $A_\mu^0$ is always truncated. If we use special coordinates $t^a=X^a/X^0$ for $\mathscr M_+$, then the submanifold inherited by the $N=1$ theory is parametrized by the $t^{\check a}$. Other conditions are:
\begin{eqnarray}
\nnb \mathcal F_{\widehat A}=0\quad &;&\quad \mathcal N_{\check A\widehat B}=0\\
\label{eq:TruncCond2} g^+_{\check a \widehat{\bar b}}= 0\quad &;&\quad D_{\check a}X^{\widehat B}= D_{\widehat a}X^{\check B}=0\;.
\end{eqnarray}

We now deduce the expression of the $N=1$ $D$-terms by studying the gaugino variations, adapting an analogous derivation performed in \cite{Truncation1}. In \cite{Truncation1} this was done for the choice $\ep=\ep_1\,,\;\ep_2 =0$ of the susy parameters, while here we allow for an arbitrary linear combination $\ep_{\mathcal A}=n_{\mathcal A}\ep$, and moreover we set everything in the context of flux compactifications.

When splitting each $N=2$ vector multiplet in two $N=1$ supermultiplets, a linear combination of the two gaugini $\lambda^{a\mathcal A}\,,\;\mathcal A=1,2$ pairs up with the vector $A_\mu^a$ and becomes the gaugino of the $N=1$ vector multiplet, while the orthogonal combination enters in the chiral multiplet together with the scalar $t^a$. In order to recognize which combination of the gaugini belonging to a given $N=2$ vector multiplet corresponds to the $N=1$ chiral fermion and which other should be identified with the $N=1$ gaugino, it is sufficient to study the $N=2$ gaugini variation under the one preserved supersymmetry. Indeed, the chiral fermion has to transform into the scalar, while the $N=1$ gaugino goes into the vector field strength. The general form (ignoring three fermions terms) of the (positive-chirality) gaugini variation for the $N=2$ theory we are considering is\footnote{The derivative of the $t^a$ is not covariantized since in the $N=2$ effective action obtained from flux compactifications as described in this paper one does not have gaugings of the special K\"ahler isometries.} \cite{N=2review}
\be\label{eq:GeneralN=2gaugVar}
\delta\lambda^{a\mathcal A} \;=\; \partial_\mu t^a \gamma^{\mu}\ep^{\mathcal A} - G^{(-)a}_{\mu\nu}\gamma^{\mu\nu}\epsilon^{\mathcal A\mathcal B}\ep_{\mathcal B}  + W^{a\mathcal A\mathcal B}\ep_{\mathcal B}\;.
\ee
While the gaugino mass matrix $W^{a\mathcal A\mathcal B}$ is defined in (\ref{eq:GaugMassMatrix}), we won't need the precise definition of $G_{\mu\nu}^{(-)a}$, corresponding to the anti self-dual part of the ``dressed field strength'' for the vectors inside the $N=2$ vector multiplets.

With our definition (\ref{eq:DefEp+}) of the $N=1$ susy parameter $\ep$, we see that the relevant linear combination for the $N=1$ gaugino is:
\be\label{eq:tildelambda} \tilde{\lambda}^{a}\equiv \bar{n}^{\mathcal A}\epsilon_{\mathcal A\mathcal B}\lambda^{a \mathcal B}\,,\ee
Indeed this projects (\ref{eq:GeneralN=2gaugVar}) on the term containing the field strength, excluding the term containing the scalar $t^a$:
\be\label{eq:deltatildelambda}
\delta_{\ep}\tilde{\lambda}^{a} \;=\;  G^{(-)a}_{\mu\nu}\gamma^{\mu\nu} \ep  + \bar{n}^{\mathcal A}\epsilon_{\mathcal A\mathcal B} W^{a\mathcal B\mathcal C}n_{\mathcal C}\ep\;.
\ee
The projection on the term containing $\partial_\mu t^a$ is instead obtained by considering
\be
\rho^a \equiv n_{\mathcal A}\lambda^{a\mathcal A}\;,\ee
so that
\be
\delta_{\ep}\rho^a\; =\; \partial_\mu t^a \gamma^\mu\ep^c + n_{\mathcal A}W^{a\mathcal A\mathcal B}n_{\mathcal B}\ep\;.
\ee

Two steps are still needed in order to get the identification of the $N=1$ gaugini. First, we should recall that conditions (\ref{eq:TruncCond1}) imply that (with the special coordinates choice \hbox{$t^a=X^a/X^0$}) from a given $N=2$ vector multiplet we retain either the $N=1$ vector multiplet or the chiral multiplet. In particular, requiring $A_\mu^{\check A}=0$ requires $\tilde \lambda^{\check a}=0 $ too. So we are left with the $\tilde{\lambda}^{\widehat a}$ only. Second, by looking at the variations of the surviving vectors $\delta_{\ep}A_\mu^{\widehat A}$, and comparing with the generic susy transformation of an $N=1$ vector, one realizes that the correct identification for the $N=1$ gaugini $\lambda^{\widehat A}$ is \cite{Truncation1}:
\be\label{eq:N=1gaugFromTrunc}
\lambda^{\widehat A}=-2 e^{\frac{K_+}{2}}D_{\widehat b} X^{\widehat A}\, \tilde{\lambda}^{\widehat{b}}\;.
\ee
Similar arguments lead us to put $\rho^{\widehat a}=0$ and to identify the $n_{Ch}$ $N=1$ chiral fermions with the $\rho^{\check a}$.

Having now the expression (\ref{eq:N=1gaugFromTrunc}) for the $N=1$ gaugini arising from the $N=2\to N=1$ truncation, we can compare their supersymmetry variation with the general form of the gaugini variation in 4d $N=1$ supergravity, which reads (up to three fermions terms):
\be\label{eq:N=1gauginoVar}
\delta \lambda^{\widehat A}=F^{(-)\widehat A}_{\mu\nu}\gamma^{\mu\nu}\ep + iD^{\widehat A}\ep\;,
\ee
where $F_{\mu\nu}^{(-)\widehat A}$ is the (anti self-dual) $N=1$ field strengths and $D^{\widehat A}$ are the D-terms, whose generic form is:
\be\label{eq:GeneralFormDterms}
D^{\widehat A}= -2(\im f_{\widehat A\widehat B})^{-1}\mathscr P_{\widehat B}\;,
\ee
where $\mathscr P_{\widehat B}$ is the Killing prepotential of the $N=1$ theory depending on the scalars in the chiral multiplets and $f_{\widehat A\widehat B}$ is the vector kinetic matrix, which is holomorphic in the $N=1$ scalars.

Comparison of (\ref{eq:N=1gauginoVar}) with $\delta_\ep\lambda^{\widehat A}= -2 e^{\frac{K_+}{2}}D_{\widehat b} X^{\widehat A}\delta_\ep\tilde{\lambda}^{\widehat {b}}$, $\;\delta_\ep\tilde{\lambda}^{a}$ being given in (\ref{eq:deltatildelambda}), with the further information that $-2 e^{\frac{K_+}{2}}D_{\widehat b} X^{\widehat A} G^{(-)\widehat b}_{\mu\nu}$ reduces to $F^{(-)\widehat A}_{\mu\nu}$ \cite{Truncation1}, provides the identification
\begin{eqnarray}
\nonumber D^{\widehat A}&=& 2ie^{\frac{K_+}{2}} D_{\widehat c} X^{\widehat A} \bar{n}^{\mathcal C}\epsilon_{\mathcal C\mathcal A}W^{\widehat c\mathcal A\mathcal B}n_{\mathcal B}\\ [2mm]
&=&  -2e^{K_+}D_{\widehat c}X^{\widehat A}g_+^{\widehat c\widehat{\bar d}}D_{\widehat{\bar d}}\bar X^{\widehat B}\, \big(\bar{n}^{\mathcal C}(\sigma_x)_{\mathcal C}^{\;\;\mathcal B}n_{\mathcal B}\big) \big(\mathcal P^x_{\widehat B}-\mathcal N_{\widehat B\widehat C}\tilde{\mathcal P}^{x\widehat C}\big) \;.
\end{eqnarray}
We have also used (\ref{eq:DefMathcalN}) in order to factorize $D_{\bar b} \bar X^C$ in the expression (\ref{eq:GaugMassMatrix}) for $W^{ a\mathcal A\mathcal B}$. Recalling the special geometry formula (\ref{eq:relgNforM+}) and the fact that $X^{\widehat A}=0$, we obtain
\be
\label{eq:almostDterms}  D^{\widehat A} = (\im\mathcal{N})^{-1\, \widehat A \widehat B} \Big\{ 2\re\big[a b(\mathcal P^1_{\widehat B}-i\mathcal P^2_{\widehat B})\big] - \mathcal N_{\widehat B\widehat C}2\re\big[ab (\tilde{\mathcal P}^{1\widehat C} - i\tilde{\mathcal P}^{2\widehat C})\big] +  c_- (\mathcal P^3_{\widehat B}  -\mathcal N_{\widehat B\widehat C}\tilde{\mathcal P}^{3\widehat C}) \Big\}\;.\ee
In \cite{Truncation1} it is shown that $\overline{\mathcal N}_{\widehat A\widehat B}$ is holomorphic on the reduced manifold\footnote{When the $N=2$ prepotential exists, this can be seen from the $\mathscr M_+$ analogous of (\ref{eq:relMandG}): one checks that $\overline{\mathcal N}_{\widehat A\widehat B} = \mathcal F_{\widehat A\widehat B}$, which is holomorphic in the $t^a$.}, and by comparison with (\ref{eq:GeneralFormDterms}) it can then be identified with the holomorphic kinetic matrix $f_{\widehat A\widehat B}$ of the $N=1$ theory. 

Substituting the expressions (\ref{eq:GeneralKillingPrepot}) for the Killing prepotentials and using the definition (\ref{eq:DefC}) of $C$, we finally obtain our expression for the D-terms:
\begin{eqnarray}
\nnb D^{\widehat A} &=& \sqrt 2 e^{2\varphi}(\im\mathcal{N})^{-1\, \widehat A \widehat B} \Big\{ \re(CZ^{I})e_{ I\widehat B} -\re(C\mathcal G_{I})m_{\widehat B}^{ I} - \mathcal N_{\widehat B\widehat C}\big[\re(CZ^{I})p^{\widehat C}_{I}-\re(C\mathcal G_{ I})q^{ I\widehat C} \big]\\  
\label{eq:Dterms}&-& \frac{c_-}{2} (\tilde G_{\widehat B}  -\mathcal N_{\widehat B\widehat C}G^{\widehat C}) \Big\}\;.
\end{eqnarray}
Since the $N=2\to N=1$ truncation reduces also the hypersector, it is understood that the index $I$ runs now over the surviving fields only.

Notice that, due to the fact that the graviphoton $A_\mu^0$ is always projected out by the \hbox{$N=2\rightarrow  N=1$} truncation, the charges $e_{I0}, m_{0}^I$ do not appear in the expression for the \hbox{D-terms}. In the specific context of $SU(3)$ structure compactifications, these charges are associated with the NS 3-form flux: $H^{\mathrm{fl}}=m_0^I\alpha_I-e_{I0}\beta^I$, which therefore contributes to the superpotential only.

Furthermore, we can check that the D-terms vanish when considering a (geometric) Calabi-Yau orientifold with general RR fluxes \cite{GrimmLouisA}; this is because in the Calabi-Yau case all the basis forms are closed, i.e. $e_{Ia} = m_a^I=0$ (recall the ansatz (\ref{eq:DiffCondSU3})), while the RR fluxes contained in $G^A, \tilde G_A$ (see below eq. (\ref{eq:ExpRRstrengths})) don't contribute because the orientifold condition imposes $|a|=|b|\Leftrightarrow c_-=0$.

More generally, we observe that the $N=1$ theory does not have D-terms if $c_- = 0$ and $\mathcal D \re(ab \Phi_-)=0$. This is a `generalized half-flatness' condition for the manifold $M_6$ \cite{GrimmBen, GMPT3}. In the $SU(3)$ structure case this becomes $d\re(iab\Om)=0$, which, together with  the constraint $d(J\wedge J)=0$ (being always satisfied when adopting the ansatz (\ref{eq:DiffCondSU3}) for the expansion forms), characterizes a half-flat manifold.

Finally, let us compare the D-flatness condition with the results of section$\:$\ref{VacuumCondit}. The combination of the Killing prepotentials defining the D-terms corresponds exactly to the one appearing in the vacuum condition (\ref{eq:LinComb1FromGaugini}); indeed, the D-terms are defined precisely by the same combination of the $N=2$ gaugino variations which has been taken to write (\ref{eq:LinComb1FromGaugini}). The only difference is that here a part of the degrees of freedom has been eliminated by the $N=2\to N=1$ truncation. According to the computation we did below eq.$\:$(\ref{eq:LinComb1FromGaugini}), we conclude that the D-flatness equation for $N=1$ supersymmetric solutions corresponds to the real part of the second pure spinor equation, eq.$\:$(\ref{eq:2ndEqInEffActLanguage}), once this last is expanded in terms of the $N=1$ degrees of freedom.

\subsection{Supersymmetric vacuum conditions for $O6$-induced truncations}\label{OrientifoldTrunc}

The $N=1$ vacuum conditions we have analysed in section \ref{VacuumCondInN=2} are also valid for the case of $N=2\to N=1$ truncations. Of course, the truncation reduces the number of degrees of freedom and, since it has to be consistent with the preserved supersymmetry, part of the constraints presented in sect.$\:$\ref{VacuumCondInN=2} will be automatically satisfied. For instance, as we have discussed above eq.$\:$(\ref{eq:constrFromSecondGravitino}), truncating the $N=2$ gravitini combination corresponding to $\tilde\psi_\mu$ goes together with $\delta_\ep\tilde\psi_\mu = 0$, and this has to be imposed already at the level of the action.

Here we want to reinterpret the conditions of section \ref{VacuumCondInN=2} in the language of $N=1$ supergravity. We will also re-establish the correspondence with the pure spinor equations, this time expanded in terms of the $N=1$ degrees of freedom.

We have already seen in the previous subsection how the vanishing of the $\ep$-generated susy variation of the $N=2$ gaugini combination (\ref{eq:tildelambda}) corresponds in the truncated theory to the D-flatness condition, which therefore yields the real part of the second pure spinor equation.

The $\langle\delta_{\ep}\psi_\mu\rangle = 0$ condition concerning the $N=1$ gravitino is also readily treated using (\ref{eq:generalN=1gravVar}) and (\ref{eq:KillingEqOne}), yielding a relation between the spacetime curvature parameter $\mu$ and the vev of $e^{\frac{K}{2}}\mathcal W$ (the gravitino mass):
\be\label{eq:RelMuW}
\bar\mu= 2\langle e^{\frac{K}{2}-\varphi} \mathcal W \rangle\;.
\ee

In order to write the F-flatness conditions associated with the chiral multiplets, one needs a more detailed knowledge of the $N=2\to N=1$ truncation, and in particular of the way the $N=2$ hypermultiplet sector is reduced to $N=1$ chiral multiplets (or better, since antisymmetric 2-tensors are in principle present, how the $N=2$ scalar-tensor multiplet reduces to $N=1$ chiral and linear multiplets). For this reason we restrict ourselves to the explicit example of truncation provided by the inclusion of an $O6$ orientifold in the IIA background. The resulting 4d $N=1$ action was derived in \cite{GrimmLouisA} for Calabi-Yau compactifications, while the generalization to $SU(3)$ and $SU(3)\times SU(3)$ structures has been discussed in \cite{GrimmBen}.

In the following we summarize just the features that will be needed in order to compute the supersymmetric vacuum conditions.

The BPS condition associated with the $O6$ orientifold gives $a=\bar b{e^{i\theta}}$, where $\theta$ is an arbitrary phase. This implies $c_-=0\,$ and $2ab= e^{i\theta}$.

Beside constraints (\ref{eq:TruncCond1}), (\ref{eq:TruncCond2}) concerning the $N=2$ vector multiplet sector, even/odd parity of the internal forms under the orientifold projection imposes the following constraints on the $N=2$ hypermultiplet sector\footnote{In this section the real and the imaginary parts of $C\Phi_-$ (and of its coefficients) are exchanged with respect to \cite{GrimmLouisA, GrimmBen}. This harmless difference can be traced back to the fact that in the $SU(3)$ structure (or Calabi-Yau) case our $\Phi_- = Z^I\alpha_I-\mathcal G_I\beta^I$ reduces to $i\Om$ instead of $\Om\,$.} (prior to the dualization of the axions):
\be
\xi^{\widehat I}=\tilde\xi_{\check I}=\im(CZ^{\widehat I})=\im(C\mathcal G_{\check I})=\re(CZ^{\check I})=\re(C\mathcal G_{\widehat I})=0,
\ee
where the index $I=0,1,\ldots, b^-$ has been split as $I=(\check I, \widehat I\,)\;$.

The $N=1$ scalar degrees of freedom are then encoded in
\be
\Phi_+=X^{\check A}\om_{\check A} - \mathcal F_{\check A}\tilde\om^{\check A}\qquad,\qquad \Pi_-=U^{\check I}\alpha_{\check I} - \tilde U_{\widehat I}\beta^{\widehat I}\;.
\ee
For the case in which no axions are dualized, the (real) dimension of the scalar manifold parameterized by $U^{\check I}, \tilde U_{\widehat I}$ is $2b^- + 2$, equal to half the dimension of the $N=2$ original quaternionic manifold (notice that all the $U^{\check I}, \tilde U_{\widehat I}$ fields are dynamical, since the unphysical $Z^0$ is `compensated' by the 4d dilaton $\varphi$ contained in $C$).

The K\"ahler potential (\ref{eq:N=1kahlerPot}) of the $N=1$ theory reads
\be\label{eq:KahlPotForO6}
K= -\ln i( \bar X^{\check A}\mathcal F_{\check A} -X^{\check A}\bar{\mathcal F}_{\check A})+4\varphi\;.
\ee
Its dependence on the $N=1$ chiral scalars $U^{\check I}$, $\tilde U_{\widehat I}$ is implicit in $\varphi$. Indeed, using the definition (\ref{eq:DefC}) of $C$ and recalling that $i\int\langle\Phi_- ,\bar\Phi_-\rangle = e^{-K_-}$, one shows immediately the relation between $\varphi$ and $C\Phi_-$:
\begin{eqnarray}\label{eq:phi4Phi-}
 e^{-2\varphi}=\frac{i}{4}\int \langle C\Phi_-, \overline{C\Phi_-}\,\rangle &=& \frac{1}{2}\int \langle\, \re(C\Phi_-)\,,\, \im(C\Phi_-)\,\rangle \\
\nnb &=& \frac{1}{2}\big[ \im(CZ^{\check I})\re(C\mathcal G_{\check I})  \,-\,  \re(CZ^{\widehat I})\im(C\mathcal G_{\widehat I}) \big]\;.
\end{eqnarray}
From the first line of (\ref{eq:phi4Phi-}), it follows \cite{GrimmBen} that $e^{-2\varphi}$ entering in $e^{-K}$ takes the form of a Hitchin functional. The real and imaginary parts of the pure spinor $C\Phi_-$ are related through the Hitchin map, which can also be expressed as $\re(C\Phi_-)=*_B\im(C\Phi_-)\;$. Hence $\re(C\mathcal G_{\check I})$ and $\re(CZ^{\widehat I})$ are functions of $\im(CZ^{\check I})$ and $\im(C\mathcal G_{\widehat I})$.

Recalling that $\Pi_-= \frac{1}{\sqrt 2}A + i\im(C\Phi_-) = U^{\check I}\alpha_{\check I} - \tilde U_{\widehat I} \beta^{\widehat I}$, we can see that $e^{-2\varphi}$ depends only on the imaginary parts of $U^{\check I}$ and $\tilde U_{\widehat I}$. Shifts of the RR scalars corresponding to the real parts of $U^{\check I}$ and $\tilde U_{\widehat I}$ are therefore isometries of the K\"ahler metric.\\ [-3mm]

As an aside, we remark that the above also describes the example of the $N=2\to N=1$ truncation exhibited in \cite{HousePalti}, even if no orientifold was introduced there. As already said, this example is however special, since it starts from compactifications on half-flat manifolds leading to $N=2$ theories without hypermultiplets, except the universal one (so $\mathrm{dim}\mathscr M_-=0$); in the $N=1$ truncation only $U^0=\xi^0+i\im(CZ^0)$ is kept.\\

\underline{\emph{F-flatness in the $U^{\check I}$ and $\tilde U_{\widehat I}$ directions}}\\ [-2mm]

\noindent The F-flatness condition associated with the chiral multiplets coming from the $N=2$ hypersector could be studied demanding the vanishing of the chiral fermion susy transformations, and then exploiting the results of section \ref{VacuumCondInN=2}. Equivalently, we choose to evaluate the K\"ahler covariant derivatives of the superpotential with respect to the chiral scalars $U^{\check I}, \tilde U_{\widehat I}$, and impose
\be\label{eq:FflatU}
0= D_{U^{\check I}}\mathcal W \equiv (\partial_{U^{\check I}} + \partial_{U^{\check I}} K)\mathcal W\qquad,\qquad 0 = D_{\tilde U_{\widehat I}}\mathcal W \equiv (\partial_{\tilde U_{\widehat I}} + \partial_{\tilde U_{\widehat I}}K)\mathcal W\;.\ee

\noindent From (\ref{eq:ExplicitWinN=1fields}) we immediately find the partial derivatives of the superpotential:
\be\label{eq:derUKsuperpot}\partial_{U^{\check I}} \mathcal W =  \frac{i}{4\bar a b} ( e_{{\check I}{\check A}}X^{\check A}  +p^{\check A}_{\check I}\mathcal{F}_{\check A} )\qquad,\qquad \partial_{\tilde U_{\widehat I}} \mathcal W =  -\frac{i}{4 \bar a b} ( m_{\check A}^{{\widehat I}}X^{\check A} +q^{{\widehat I}{\check A}}\mathcal{F}_{\check A} ) \;.
\ee
The derivatives of the K\"ahler potential (\ref{eq:KahlPotForO6}) are less trivial. Since $K$ depends implicitly on $\im U^{\check I}=\im(CZ^{\check I})$ and $\im \tilde U_{\widehat I}=\im(C\mathcal G_{\widehat I})$ through $\varphi=\varphi(\im(CZ^{\check I}),\im(C\mathcal G_{\widehat I}))$, we have
\be\label{eq:PartialK1}
\partial_{U^{\check I}}K = 4\partial_{U^{\check I}}\varphi \;=\; -2i\partial_{\im (CZ^{\check I})}\varphi\qquad,\qquad\partial_{\tilde U_{\widehat I}}K = 4\partial_{\tilde U_{\widehat I}}\varphi \;= \; -2i\partial_{\im(C\mathcal G_{\widehat I})}\varphi\;.
\ee
In order to evaluate this, we use the following property for the variation of a Hitchin functional
\be\label{eq:VariatKahlPot}
\delta e^{-2\varphi} \equiv \frac{i}{4} \delta \int \langle C\Phi_-, \overline{C\Phi_-}\, \rangle = \int \langle \re(C\Phi_-), \delta\im(C\Phi_-)\, \rangle\;,
\ee
which can be derived considering the decomposition under representations of $SU(3)\times SU(3)$ and recalling the fact that the Mukai pairing picks just the singlet. In terms of the moduli of $\,\im( C\Phi_-) = \im(CZ^{\check I})\alpha_{\check I} - \im(C\mathcal G_{\widehat I})\beta^{\widehat I}$, (\ref{eq:VariatKahlPot}) is rewritten as\footnote{We also checked this explicitly by computing and inverting the jacobian for the change of variables $\big(\,e^{-\varphi}, \im(a\bar bZ^{\check\imath}), \re(a \bar bZ^{\widehat\imath})\,\big)\;\longrightarrow\;\big(\,\im(CZ^{\check I}), \im(C\mathcal G_{\widehat I})\,\big)$, where the unphysical $Z^0$ has not been included in the old variables. The result confirms (\ref{eq:dere2phi}).}
\begin{eqnarray}
\nnb\pd{e^{-2\varphi}}{\im (CZ^{\check I})} &=& \int \langle \re(C\Phi_-), \alpha_{\check I}\, \rangle \;=\;  \re(C\mathcal G_{\check I})\\
\label{eq:dere2phi}\pd{e^{-2\varphi}}{\im (C\mathcal G_{\widehat I})}&=& -\int \langle \re(C\Phi_-), \beta^{\widehat I}\, \rangle \;=\; - \re(CZ^{\widehat I})\;.
\end{eqnarray}
We conclude that
\be
\partial_{U^{\check I}}K = ie^{2\varphi}\re(C\mathcal G_{\check I})\qquad,\qquad \partial_{\tilde U_{\widehat I}}K = -ie^{2\varphi}\re(C Z^{\widehat I})\;.
\ee
Recalling the definition (\ref{eq:DefC}) of $C$, the fact that with our choice for the normalization of the pure spinors $e^{K_-} =e^{K_+}$ and eqs.$\:$(\ref{eq:N=1kahlerPot}), (\ref{eq:RelMuW}), we obtain
\be
(\partial_{U^{\check I}}K) \mathcal W= 2i \bar \mu\re(a b\mathcal G_{\check I})\qquad,\qquad (\partial_{\tilde U_{\widehat I}}K)\mathcal W=- 2i \bar \mu\re(a bZ^{\widehat I})\;.
\ee
It is now straightforward to see that the two sets of conditions (\ref{eq:FflatU}) precisely
give \be\label{eq:FflatnessUVSympl}\framebox[1.15\width ][c]{$ a \bar b\mathbb Q V_+ = -2\bar\mu\re (a bV_-)$}\;.\ee
Here $V_\pm$ contain only the truncated fields: they are the remnants of the $N=2$ symplectic sections. The charge matrix is also reduced accordingly. In agreement with our discussion of subsect.$\:$\ref{ExpPureSpEq}, eq.$\:$(\ref{eq:FflatnessUVSympl}) corresponds to the first pure spinor equation, expanded in the $N=1$ degrees of freedom and integrated over the internal manifold.\\

\underline{\emph{F-flatness in the $t^{\check a}$ directions}}\\ [-2mm]

\noindent In order to write the F-flatness condition associated with the $N=1$ chiral multiplets $(t^{\check a}, \rho^{\check a})$ descending from the $N=2$ vector multiplet sector ($\rho^{\check a}$ are the chiral fermions), we will build on the results of subsect.$\,$\ref{VacuumCondInN=2}. Imposing $\langle\delta_{\ep}\rho^{\check a}\rangle=0$ is clearly the same thing as requiring $D_{\check a} \mathcal W \equiv (\partial_{t^{\check a}}+\partial_{t^{\check a}}K)\mathcal W=0$. Indeed, the form of the variations of the chiral fermions dictated by $N=1$ supergravity is
\be\label{eq:ChiralFermVar}
\delta_{\ep}\rho^{\check a} = \partial_\mu t^{\check a} + 2e^{\frac{K}{2}}g_+^{\check a\check{\bar b}}D_{\check{\bar b}}\overline{\mathcal W}\;.
\ee
The chiral fermions $\rho^{\check a}$ have been identified in subsect.$\,$\ref{Dterms} with the $N=2$ gaugini combination $ n_{\mathcal A}\lambda^{\check a\mathcal A}$. Therefore we have the F-flatness condition $0=\langle\delta_{\ep}\rho^{\check a}\rangle =n_{\mathcal A}\langle\delta_{\ep}\lambda^{\check a\mathcal A}\rangle$, where in $\delta_{\ep}\lambda^{\check a\mathcal A}$ one should consider only the non-truncated degrees of freedom. Since $c_-=0\,$, $\,n_{\mathcal A}\langle\delta_{\ep}\lambda^{\check a\mathcal A}\rangle=0$ is equivalent to $\bar b \langle\delta_\ep\lambda^{\check a 1}\rangle + a \langle\delta_\ep\lambda^{\check a2}\rangle  =0$, and this corresponds to eq. (\ref{eq:LinComb2FromGaugini}). At this point the computation becomes identical to the one in subsect.$\,$\ref{ExpPureSpEq}, and we conclude that $\langle\delta_{\ep}\rho^{\check a}\rangle=0$ leads to
\be\label{eq:FflatnesstSympl}\framebox[1.15\width ][c]{$(\mathbb{S}_+)^{-1}\mathbb Q^T\mathbb{S}_- \im(abV_-)= -3\im (\mu a\bar b  V_+) -\frac{e^{\phi}}{2}\mathbb{N}V_G$}\;.\ee
Here again the symplectic vectors $V_\pm$ and $V_G$ contain just the components surviving the $N=2\to N=1$ truncation. Eq. (\ref{eq:FflatnesstSympl}) corresponds to the imaginary part of the integrated second pure spinor equation, expanded in the $N=1$ degrees of freedom and integrated over the internal manifold. 

In subsect.$\,$\ref{ExpPureSpEq}, in order to arrive at (\ref{eq:GauginiSympl2}) we needed the constraint (\ref{eq:P3isMu}). In the present $N=1$ setting this constraint can be rederived as follows. Assuming that the vacuum satisfies (\ref{eq:FflatnessUVSympl}), from (\ref{eq:explicitWinN=2fields}) we see that the vev of the superpotential is:
\be
4i\bar a b \langle\mathcal W\rangle =  \frac{2i\bar \mu}{ a\bar b } \big[\re(a b\mathcal G_{\check I})\im(CZ^{\check I})-\re(a bZ^{\widehat I})\im(C\mathcal G_{\widehat I})\big] - \frac{1}{\sqrt 2}( \tilde G_{\check A}X^{\check A} +  G^{\check A}\mathcal F_{\check A}  ) \;.
\ee
Now multiply both sides by $e^{2\varphi}$: recalling (\ref{eq:DefC}) and (\ref{eq:phi4Phi-}) the first term on the RHS gives $4\sqrt8 i\bar \mu\bar a be^\phi$, while by eq. (\ref{eq:GeneralKillingPrepot}) the last term corresponds to $\mathcal P^3$; for the LHS, use (\ref{eq:RelMuW}), (\ref{eq:N=1kahlerPot}) and $e^{-\frac{K_+}{2}+\varphi}=\sqrt 8e^\phi$ (see (\ref{eq:chain})). We get the relation:
\be\label{eq:constraint}
2\sqrt 8 \bar \mu\bar a be^{\phi}= i\mathcal P^3\;,\ee
which is (\ref{eq:P3isMu}) expressed in this $N=1$ context.
\vskip .2cm
Let us summarize the correspondence between the supersymmetric vacuum conditions arising in the $N=1$ effective action and the pure spinor equations resulting from the 10d approach. In order to perform the comparison, the pure spinor equations have to be expanded on the basis $\Sigma^\pm$, truncated to the $N=1$ degrees of freedom only, and then integrated over the compact 6d manifold. The D-flatness constraint matches the real part of the second pure spinor equation, while the F-flatness condition for the chiral multiplets coming from the $N=2$ vector multiplets corresponds to its imaginary part. F-flatness with respect to the chiral multiplets descending from the $N=2$ hypersector provides instead the first pure spinor equation.

Even though we have performed the analysis of the present subsection for the orientifold case, it is pretty clear that it should be applicable more generally to any $N=2\to N=1$ truncation.

\section{Conclusions}\label{conclusions}

The main purpose of this paper was to confront the 4d and 10d approaches to $N=1$ vacua of type II theories. We considered the $N=2$ and $N=1$ 4d effective actions obtained by off-shell flux compactifications on $SU(3)\times SU(3)$ backgrounds. We established the $N=1$ vacuum conditions, and we showed they satisfy an integrated version of the $N=1$ constraints in 10d, written in the generalized geometry formulation of \cite{GMPT2, GMPT3}.

We remark that we have verified the correspondence in the presence of a very large set of fluxes, composed by the RR ones as well as by all the charges generated by the action of the extended differential $\mathcal D$ mapping the even basis forms to the odd ones and vice versa. These include the NS and the geometric fluxes, as well as a complementary set of charges which turn out to be associated with nongeometric backgrounds, as argued in \cite{GLW2}.

Although we have explicitly performed the comparison in a type IIA setting, we expect the matching be the same for type IIB. Indeed, both the 10d pure spinor equations and the Killing prepotentials leading to the 4d $N=1$ vacuum conditions display a very mirror symmetric aspect: to pass from IIA to IIB and back again, basically one just has to exchange the pure spinors $\Phi_+\leftrightarrow \Phi_-$ and the RR fluxes $F^{\mathrm{even}}\leftrightarrow F^{\mathrm{odd}}$.

Another subject we discussed is how, when considering $SU(3)\times SU(3)$ backgrounds, the generalized geometry formalism allows to recover some of the standard features of the $N=2$ effective actions obtained from familiar string compactifications. In particular, building on the decomposition of the pure spinor variations under the $SU(3)\times SU(3)$ structure, we verified that the metric describing the fluctuations of the internal metric and $B$-field as inherited from the 10d supergravity indeed is reproduced by the special K\"ahler metrics derived from the logarithm of the Hitchin functionals for even/odd pure spinors.

The decomposition in $SU(3)\times SU(3)$ representations was also the tool used to analyse the action of the $*_B$ operator on the basis of forms. In this way we showed how to obtain the symplectic matrix $\mathbb M$ contributing to define (through a generalization of the c-map for Calabi-Yau manifolds) the $N=2$ quaternionic sigma-model. 

However, one should recall that the possibility to obtain an actual 4d effective theory is subject to the existence of a finite basis of forms on the internal manifold selecting the light degrees of freedom in 4d; these forms are assumed to respect a quite restrictive series of constraints. Clearly, it would be of much interest to prove the concreteness of such an ansatz by providing some explicit examples in which all the requirements are satisfied.

\vskip 1cm

{\large \bf Acknowledgments}\\ [2mm]
We are greatly indebted to Mariana Gra{\~n}a, Michela Petrini and Alessandro Tomasiello for many helpful explications and correspondence. This paper, and in particular the idea to reproduce the pure spinor equations from a 4d off-shell approach, grew out of numerous discussions of DC with Mariana Gra{\~n}a and Michela Petrini. We also wish to thank Ruben Minasian, Massimo Porrati and Fabio Zwirner for discussions as well as Jan Louis for correspondence. 

DC would like to thank the String Theory group at the University of Roma ``Tor Vergata'', and in particular Francesco Fucito, for the warm hospitality and discussions. This work is supported in part by the EU grants MRTN-CT-2004-005104 and MRTN-CT-2004-512194 as well as (DC) by the ``Programme Vinci 2006 de l'Universit\'e Franco-Italienne''.

\vskip 1.3cm

\appendix
\section{Notation and conventions}\label{conventions}
\setcounter{equation}{0}

\subsection{Indices}
\vskip .2cm

\begin{center}
\begin{tabular}{|c|c|c|}
\hline
letters & 	range					& labeling  \\
\hline
$\mu,\nu,\ldots$  & $0,\ldots,3$ & 4d spacetime coords. \\
$m,n,\ldots$ &  $1,\dots,6\;$ & 6d compact coords. \\
$\Lambda, \Sigma, \ldots$ & $1,\ldots,12$ & vector repr. of $O(6,6)$, i.e. $T\oplus T^*$ coords. \\
$A,B,\ldots$ & $0,1,\ldots,b^+$ & projective coords. for $\mathscr M_+$\\
$a,b,\ldots$ & $1,\ldots,b^+$ & coordinates for $\mathscr M_+$ \\
$I,J,\ldots$ & $0,1,\ldots, b^-$ &  projective coords. for $\mathscr M_-$ \\
$i,j,\ldots$ & $1,\ldots, b^-$ &   coordinates for $\mathscr M_-$   \\
$u,v,\ldots$ & $1,\ldots,4(b^-+1)$ & quaternionic coordinates \\
$\mathcal A,\mathcal B,\ldots$ & $1,2\;\;$ & fundamental repr. of $SU(2)$\\
$\alpha,\beta,\ldots$ & $1,\ldots,2(b^-+1)\;$ & fundamental repr. of $Sp(2b^-+2,\mathbb{R})$ \\
\hline
\end{tabular}
\end{center}

\subsection{Clifford algebra and spinor conventions}
The spacetime metric has mostly $+$ signature: $(-,+,+,\ldots\,)\,$.
We choose a Majorana representation for the \emph{Cliff}(3,1) and \emph{Cliff}(6) gamma matrices. The \emph{Cliff}(3,1) gamma matrices $\gamma^\mu$ are all real; they are hermitian, except $\gamma^0$ which is antihermitian. The \emph{Cliff}(6) gamma matrices $\gamma^m$ are all purely imaginary and hermitian. The 4d and 6d chirality matrices are respectively:
\be
\gamma_5=\frac{i}{4!}\epsilon_{\mu\nu\rho\sigma}\gamma^{\mu\nu\rho\sigma}=i\gamma^0\gamma^1\gamma^2\gamma^3\qquad,\qquad \gamma=-\frac{i}{6!}\epsilon_{mnpqrs}\gamma^{mnpqrs}\;,
\ee
so that both $\gamma_5$ and $\gamma$ are purely imaginary and hermitian. The 10d chirality matrix is $\Gamma_{11}=\gamma_5\otimes\gamma$, and is real and hermitian.

Note that most of the 4d supergravity literature adopts a $(+---)$ signature convention for the 4d metric, and this leads to a difference in the gamma matrices consisting in a factor of $i$. Such a difference is reflected in the supergravity formulas appearing in the main text.

We define the charge conjugation in such a way that for any $Spin(3,1)$ spinor $\ep$, its charge conjugate $\ep^c$ is just given by the complex conjugate, $\ep^c=\ep^*$. If $\ep$ is a Weyl spinor with positive chirality ($\gamma_5\ep=\ep$), then its charge conjugate $\ep^c \equiv \ep^*$ is again a Weyl spinor, with negative chirality, and vice versa. Similarly, if $\eta_+$ is a $Spin(6)$ spinor with positive chirality ($\gamma\eta_+=\eta_+$), then $\eta_-\equiv \eta_+^*$ has negative chirality.

\section{Mukai pairing and Clifford map}\label{MukaiAndClifford}

\setcounter{equation}{0}

We summarize here some relations involving the Mukai pairing and the Clifford map (defined in (\ref{eq:DefMukai}) and (\ref{eq:CliffMap}) respectively) which are useful in the generalized geometry computations. We adopt the conventions of \cite{GMPT3}, where a part of the formulas collected here can be found. We also added an explicit computation concerning the relation between pure spinors and generalized almost complex structures.\\

First note that the six dimensional Hodge star $*$ is defined as:
\be
\label{eq:defHodge*} *\,e^{a_1\ldots a_k}=\frac{1}{(6-k)!}\epsilon_{a_{k+1}\ldots a_6}^{\phantom{a_{k+1}\ldots a_6}a_1\ldots a_k}\,e^{a_{k+1}\ldots a_6}\;,
\ee
giving a quite unusual supplementary minus sign on odd forms.
\vskip .3cm
The following are properties of the Mukai pairing, holding for $A, C\in\wedge^\bullet T^*$ and $B\in\wedge^2T^*$. We recall that the Mukai pairing is antisymmetric in six dimensions.
\begin{eqnarray}
\lambda(e^BA)&=&e^{-B}\lambda(A)\;,\\
\label{eq:Bdrops}\langle e^{-B} A, e^{-B}C \rangle &=& \langle  A, C \rangle\;,\\
\label{eq:*passes}\langle A,*C \rangle &=& \langle C,*A \rangle\;,\\
\label{eq:lambdaPasses}\langle A_\pm,\lambda(C_\pm) \rangle &=& \pm\langle C_\pm,\lambda(A_\pm) \rangle\\
\label{eq:GammaPass} \langle A , \Gamma^\Lambda C\rangle &=& \langle C, \Gamma^\Lambda A  \rangle\;,
\end{eqnarray}
where the \emph{Cliff}$(6,6)$ gamma matrices $\Gamma^\Lambda$ correspond to $d x^m\wedge$ or $\iota_{\partial_m}$ as in the main text.
\vskip .3cm
Under the Clifford map (\ref{eq:CliffMap}), the Mukai pairing translates as:
\be\label{eq:MukaiUnderClifford}
\langle A_k , C_{6-k} \rangle\; =\; \frac{(-)^{k}}{8}\tr(\!\slas{*A_{k}}\!\slas{\;\;C_{6-k}})vol_6 \;=\; \frac{i}{8}\tr(\gamma \:\slash\!\!\!\! A_k ^{\;T}\!\slas{\;\;C_{6-k}})vol_6\;,
\ee
where $vol_6$ is the volume form of $M_6$ and the trace is taken over the spinor indices of the \emph{Cliff}$(6)$ gamma matrices. For the second equality, we have used
\be
\slas{*A_k} = i\gamma \!\slas{\:\lambda(A_k)} = (-)^k i\gamma\, /\!\!\!\!A_k^T \;,
\ee
obtained from
\be\label{eq:action*lambda}
\begin{picture}(10,11)(-15,1)
\put(0,0){\line(3,1){30}}
\end{picture}*\lambda(A) =i\gamma \slash \!\!\!\!A
\ee
together with $\slas{\:\lambda(A_k)} = (-)^k /\!\!\!\! A_k^T$.
\vskip .3cm
The \emph{Cliff}(6,6) action on even/odd forms $C_\pm \in \wedge^\bullet T^*$ translates under the Clifford map as:
\be
\begin{picture}(10,10)(-10,4)
\put(0,0){\line(4,1){50}}
\end{picture}
dx^m\wedge C_\pm
 \;= \;\frac{1}{2}[\gamma^m ,\, /\!\!\!\!C_\pm\,]_\pm \qquad,\qquad
\begin{picture}(13,10)(-10,3)
\put(0,0){\line(4,1){35}}
\end{picture} i_{\partial_m} C_\pm\; = \;\frac{1}{2}[\gamma_m ,\, /\!\!\!\!C_\pm\,]_\mp \;,
\ee
where $[\:,\,]_\pm$ stands for anticommutator/commutator. In the main text we use the action of the antisymmetrized product of two \emph{Cliff}$(6,6)$ gamma matrices:
\be\Gamma^{\Lambda\Sigma} = \big(\, dx^m\wedge dx^n\wedge\;,\; \frac{1}{2} [dx^m\wedge, \iota_{\partial_n}] \;,\; \frac{1}{2} [\iota_{\partial_m} , dx^n\wedge]\;,\;\iota_{\partial_m}\iota_{\partial_n}\, \big)\;.\ee
Under the Clifford map this becomes:
\begin{eqnarray}
\nnb\begin{picture}(10,10)(-10,4)
\put(0,0){\line(6,1){80}}
\end{picture}
dx^m\wedge dx^n\wedge C_\pm
&= &\frac{1}{4}\big [ \gamma^{mn}\, /\!\!\!\!C_\pm\, \pm \, \gamma^m\, /\!\!\!\!C_\pm \gamma^n\, \mp \, \gamma^n \,/\!\!\!\!C_\pm \gamma^m\,- \, /\!\!\!\!C_\pm \gamma^{nm}  \big ]\\ [1mm]
\nnb\begin{picture}(10,10)(-10,4)
\put(0,0){\line(6,1){75}}
\end{picture}
\frac{1}{2} [dx^m\wedge, \iota_{\partial_n}] C_\pm
&= &\frac{1}{4}\big [ \gamma^m_{\;\;\,\,n}\, /\!\!\!\!C_\pm\, \mp \, \gamma^m\, /\!\!\!\!C_\pm \gamma_n\, \mp \, \gamma_n \,/\!\!\!\!C_\pm \gamma^m\,+ \, /\!\!\!\!C_\pm \gamma_{n}^{\;m}  \big ]\\ [1mm]
\nnb\begin{picture}(10,10)(-10,4)
\put(0,0){\line(6,1){75}}
\end{picture}
\frac{1}{2}[\iota_{\partial_m}, dx^n\wedge] C_\pm
&= &\frac{1}{4}\big [ \gamma_m^{\;\;\,n}\, /\!\!\!\!C_\pm\, \pm \, \gamma_m\, /\!\!\!\!C_\pm \gamma^n\, \pm \, \gamma^n \,/\!\!\!\!C_\pm \gamma_m\,+ \, /\!\!\!\!C_\pm \gamma^{n}_{\;\;m}  \big ]\\ [1mm]
\label{eq:2GammaUnderCliff}\begin{picture}(10,10)(-10,4)
\put(0,0){\line(5,1){45}}
\end{picture}
\iota_{\partial_m}\iota_{\partial_n} C_\pm
&= &\frac{1}{4}\big [ \gamma_{mn}\, /\!\!\!\!C_\pm\, \mp \, \gamma_m\, /\!\!\!\!C_\pm \gamma_n\, \pm \, \gamma_n \,/\!\!\!\!C_\pm \gamma_m\,- \, /\!\!\!\!C_\pm \gamma_{nm}  \big ]\;.
\end{eqnarray}
When $C_\pm =\Phi^0_\pm$ corresponds to one of the pure spinors defining the $SU(3)\times SU(3)$ structure, then it is not difficult to see how each term appearing in (\ref{eq:2GammaUnderCliff}) transforms under \hbox{$SU(3)\times SU(3)$}, and therefore we can locate its position in the diamond (\ref{eq:Udiamond}). Indeed, comparing with the explicit basis given in (\ref{eq:BasisDiamond}), we see that for instance $\gamma^m\Phi^0_+\gamma^n\in U_{\bf{\bar 3}, \bf{3}}$, while $\Phi^0_+\gamma^{mn}$ contains a term proportional to $\Phi^0_+\in U_{\bf{1}, \bf{\bar 1}}$ and a term belonging to $U_{\bf{1}, \bf{\bar 3}}\,$.
\vskip .3cm
As an example of how to use this technology, we can check the correspondence of the generalized almost complex structures defined from the pure spinors (\ref{eq:DefPhi_pm}) via the formula (\ref{eq:RelJPhi}) with the matrices $\mathcal J_\pm$ given in eq. (\ref{eq:CurlyJ+-}). Start from the case of vanishing $B$, and write (\ref{eq:RelJPhi}) for $\Phi^0_\pm$. Recalling eq.$\,$(\ref{eq:normPhi}) and the basis (\ref{eq:Gamma66}) for the \emph{Cliff}(6,6) gamma matrices we have:
\be\label{eq:JfromPhiStep1}
\mathcal J^{\;\Lambda}_{\pm\;\Sigma} = \frac{1}{2vol_6} \left(\begin{array}{cc}\langle \re\Phi^0_\pm \,,\, \frac{1}{2}[dx^m\wedge ,\iota_{\partial_n}] \re \Phi^0_\pm \rangle & \langle \re\Phi^0_\pm \,,\, dx^m\wedge dx^n\wedge\re \Phi^0_\pm \rangle \\ [2mm] 
\langle \re\Phi^0_\pm \,,\, \iota_{\partial_m}\iota_{\partial_n}\re \Phi^0_\pm \rangle & \langle \re\Phi^0_\pm \,,\, \frac{1}{2}[\iota_{\partial_m}, dx^n\wedge ]\re \Phi^0_\pm \rangle \end{array}\right)\;.
\ee
We can now evaluate this in the bispinor picture, using eqs. (\ref{eq:MukaiUnderClifford}) and (\ref{eq:2GammaUnderCliff}). For instance, for the lower block on the left, we have:
\begin{eqnarray}
\nnb\frac{1}{2 vol_6}\langle \re\Phi^0_\pm , \iota_{\partial_m}\iota_{\partial_n}\re \Phi^0_\pm \rangle &=& \frac{i}{8^2} \tr\big[\gamma (\slas{\re \Phi^0_\pm})^T  ( \gamma_{mn}\, \slas{\re\Phi^0_\pm}  + \slas{\re\Phi^0_\pm} \gamma_{mn}  )\big ]\\
 &=&  \frac{i}{2}(\eta^{1\dag}_+ \gamma_{mn} \eta^1_+ + \eta^{2\dag}_\pm \gamma_{mn} \eta^2_\pm) =  \frac{1}{2}(J_1 \pm J_2)\;.
\end{eqnarray}
In the first equality we have written only the nonzero terms, while to get the second line we substituted (\ref{eq:defPhi0}) and used $\pm i \eta^{1,2\dag}_\pm \gamma_{mn}\eta^{1,2}_\pm= J_{1,2}$. The evaluation of the other blocks is analogous, and we obtain eq.$\:$(\ref{eq:CurlyJ+-}) with $B=0$. When considering pure spinors with nonvanishing $B$, in (\ref{eq:JfromPhiStep1}) we have $\Phi_\pm=e^{-B}\Phi^0_\pm$ instead of $\Phi^0_\pm$. We wish to make $e^{-B}$ pass through the $dx^m\wedge $ and $\iota_{\partial_m}$ and then use (\ref{eq:Bdrops}). While the $dx^m\wedge$ commute with $e^{-B}$, for the contractions we have $\iota_{\partial_m} e^{-B} = e^{-B}(\iota_{\partial_m}- B_{mn}dx^n\wedge)$. Taking this into account we recover the two matrices ${1\; 0 \choose -B\; 1}$ and ${1\;\; 0 \choose B\; 1}$ of eq. (\ref{eq:CurlyJ+-}).

\section{Special K\"ahler geometry formulas}\label{SpecialGeometry}
\setcounter{equation}{0}

In this appendix we collect some properties of local special K\"ahler geometry which are used in the main text. Thorough discussions of this subject can be found, for instance, in refs. \cite{N=2review, WhatIsSpecialKaehler?, Freed}. Here we present the formulas in the notation referring to the special K\"ahler manifold $\mathscr M_-$ introduced in subsect.$\:$\ref{AnsatzForms}; modulo switching the notation, it is understood they are also valid for $\mathscr M_+$.

Recall that a local special K\"ahler manifold $\mathscr M_-$ of complex dimension $b^-$ is a Hodge-K\"ahler manifold (with line bundle $\mathcal L$) with the further structure of a holomorphic flat \hbox{$Sp(2b^-+2,\mathbb R)$} vector bundle $\mathcal S$ over it. We denote the holomorphic section of the $\mathcal S\otimes \mathcal L $ bundle by
\be V_-=\left(\begin{array}{c}Z^I \\ \mathcal G_J\end{array}\right)\;,\qquad I,J=0,\ldots,b^-\;.\ee
The K\"ahler potential has to be expressed in terms of $V_-$ as:
\be\label{eq:KahlPotAppendix}
K_-=-\ln (-iV_-^T\mathbb S_-\bar V_-) \,=\,-\ln i(\bar Z^I \mathcal G_I-\bar{\mathcal G}_J Z^J)\;,
\ee
where $\mathbb S_-$ is the $Sp(2b^-+2)$ metric. In the compactification context of subsect.$\:$\ref{AnsatzForms}, the symplectic structure is provided by the Mukai pairing as in eq.$\:$(\ref{eq:SymplStrM-}), and the holomorphic section is encoded in $\Phi_-=Z^I\alpha_I-\mathcal G_I\beta^I\,$, so that $K_-=-\ln i\int\langle \Phi_- , \bar\Phi_-\rangle\,$.

The following relations define the \emph{period matrix} $\mathcal M_{IJ}\,$:
\be
\mathcal G_I=\mathcal M_{IJ}Z^J\qquad,\qquad D_i\mathcal G_J=\overline{\mathcal M}_{JK}D_iZ^K\;,
\ee
where the K\"ahler covariant derivative acting on the holomorphic section is $D_i=\partial_i+\partial_iK_-$.

Whenever a prepotential $\mathcal G$ can be introduced, we have $\mathcal G_I = \partial_I \mathcal G\,$ and
\be\label{eq:homGI}
\mathcal G_I = \mathcal G_{IJ}Z^J\qquad,\qquad \textrm{where \;} \mathcal G_{IJ} := \partial_I\partial_J \mathcal G\;. 
\ee
In this case the period matrix $\mathcal M_{IJ}$ can be expressed as
\be\label{eq:relMandG}
\mathcal{M}_{IJ}= \overline{\mathcal G}_{IJ} + 2i\frac{(\im \mathcal G_{IK})Z^K(\im \mathcal G_{J L}) Z^L}{Z^M(\im \mathcal G_{MN}) Z^N}\;.
\ee
Other identities that can be shown are:
\be\label{eq:RelGK}
Z^I\im \mathcal G_{IJ}\bar Z^J=-\frac{1}{2}e^{-K_-} \quad\quad \textrm{(following directly from (\ref{eq:KahlPotAppendix}) and (\ref{eq:homGI}))}
\ee
\be
\label{eq:usefulformula} D_kZ^I g_-^{k\bar{l}}  D_{\bar{l}}\bar{Z}^J = -\frac{1}{2}e^{-K_-}(\im\mathcal{M})^{-1\:IJ} - \bar{Z}^I Z^J\;.
\ee
Finally, using (\ref{eq:relMandG}) and (\ref{eq:RelGK}), one can see that
\begin{eqnarray}
\nnb(\im \mathcal M)^{-1\,IJ} & = & -(\im \mathcal G)^{-1\,IJ} -2e^{K_-}(Z^I\bar Z^J + \bar Z^IZ^J)\\ [1mm]
\label{eq:RelGM} \,[\re\mathcal M (\im\mathcal M)^{-1}]_I^{\;\;J} & = & - [\re \mathcal G(\im \mathcal G)^{-1}]_I^{\;\;J} -2e^{K_-}( \mathcal G_I\bar Z^J+\bar{ \mathcal G}_IZ^J) \\ [1mm]
\nnb [\im\mathcal M+\re\mathcal M(\im\mathcal M)^{-1}\re\mathcal M]_{IJ} & = &   -[\im \mathcal G+\re \mathcal G(\im \mathcal G)^{-1}\re \mathcal G]_{IJ} -2e^{K_-}( \mathcal G_I\bar{ \mathcal G}_J +\bar{ \mathcal G}_I \mathcal G_J)\;.
\end{eqnarray}



\begin{thebibliography}{99}


\bibitem{FluxReviews} Recent reviews are:\\
M.~Grana, {\it Flux compactifications in string theory: A comprehensive review}, Phys.\ Rept.\  {\bf 423} (2006) 91 [arXiv:hep-th/0509003];\\
M.~R.~Douglas and S.~Kachru,
  {\it Flux compactification},
  Rev.\ Mod.\ Phys.\  {\bf 79} (2007) 733
  [arXiv:hep-th/0610102].\\
R.~Blumenhagen, B.~Kors, D.~Lust and S.~Stieberger,
  {\it Four-dimensional String Compactifications with D-Branes, Orientifolds and
  Fluxes},
  arXiv:hep-th/0610327.

\bibitem{GurLouisMicuWaldr}
  S.~Gurrieri, J.~Louis, A.~Micu and D.~Waldram,
  {\it Mirror symmetry in generalized Calabi-Yau compactifications},
  Nucl.\ Phys.\  B {\bf 654} (2003) 61
  [arXiv:hep-th/0211102].

\bibitem{GurrMicuIIBhalfFlat}
S.~Gurrieri and A.~Micu,
  {\it Type IIB theory on half-flat manifolds},
  Class.\ Quant.\ Grav.\  {\bf 20} (2003) 2181
  [arXiv:hep-th/0212278].

\bibitem{GLW1} M.~Grana, J.~Louis and D.~Waldram,
  {\it Hitchin functionals in $N = 2$ supergravity},
  JHEP {\bf 0601}, 008 (2006)
  [arXiv:hep-th/0505264].

\bibitem{TomasMirrorSymFl}
  A.~Tomasiello,
  {\it Topological mirror symmetry with fluxes},
  JHEP {\bf 0506} (2005) 067
  [arXiv:hep-th/0502148].

\bibitem{HousePalti} T.~House and E.~Palti, {\it Effective action of (massive) IIA on manifolds with SU(3) structure}, Phys.\ Rev.\  D {\bf 72} (2005) 026004 [arXiv:hep-th/0505177].


\bibitem{MinasianKashani}
	A.~K.~Kashani-Poor and R.~Minasian,
  {\it Towards reduction of type II theories on SU(3) structure manifolds},
  JHEP {\bf 0703}, 109 (2007)
  [arXiv:hep-th/0611106].


\bibitem{LouisMicuCYwithFlux}
J.~Louis and A.~Micu,
  {\it Type II theories compactified on Calabi-Yau threefolds in the presence  of
  background fluxes},
  Nucl.\ Phys.\  B {\bf 635} (2002) 395
  [arXiv:hep-th/0202168].


\bibitem{DallAgataFluxRev}
  G.~Dall'Agata,
  {\it String compactifications with fluxes},
  Class.\ Quant.\ Grav.\  {\bf 21} (2004) S1479.


\bibitem{GaugingHeisenberg}
  R.~D'Auria, S.~Ferrara, M.~Trigiante and S.~Vaula,
  {\it Gauging the Heisenberg algebra of special quaternionic manifolds},
  Phys.\ Lett.\  B {\bf 610} (2005) 147
  [arXiv:hep-th/0410290];
  {\it Scalar potential for the gauged Heisenberg algebra and a non-polynomial
  antisymmetric tensor theory},
  Phys.\ Lett.\  B {\bf 610} (2005) 270
  [arXiv:hep-th/0412063].


\bibitem{D'AuriaFerrTrig}
R.~D'Auria, S.~Ferrara and M.~Trigiante,
  {\it On the supergravity formulation of mirror symmetry in generalized
  Calabi-Yau manifolds},
  arXiv:hep-th/0701247.

\bibitem{HitchinGenCY}
N.~Hitchin,
  {\it Generalized Calabi-Yau manifolds},
  Quart.\ J.\ Math.\ Oxford Ser.\  {\bf 54} (2003) 281
  [arXiv:math/0209099].

\bibitem{GualtieriThesis}
M.~Gualtieri,
  {\it Generalized complex geometry},
  Oxford University DPhil Thesis (2004)
  [arXiv:math.dg/0401221].

\bibitem{GLW2} 
M.~Grana, J.~Louis and D.~Waldram,
  {\it $SU(3) \times SU(3)$ compactification and mirror duals of magnetic fluxes},
  JHEP {\bf 0704}, 101 (2007)
  [arXiv:hep-th/0612237].


\bibitem{GrimmBen} 
  I.~Benmachiche and T.~W.~Grimm,
  {\it Generalized $N = 1$ orientifold compactifications and the Hitchin functionals},
  Nucl.\ Phys.\  B {\bf 748} (2006) 200
  [arXiv:hep-th/0602241].



\bibitem{HullGeoForNonGeo}
C.~M.~Hull,
  {\it A geometry for non-geometric string backgrounds},
  JHEP {\bf 0510} (2005) 065
  [arXiv:hep-th/0406102].

\bibitem{SheltonTaylorWecht}
  J.~Shelton, W.~Taylor and B.~Wecht,
  {\it Nongeometric flux compactifications},
  JHEP {\bf 0510} (2005) 085
  [arXiv:hep-th/0508133]; {\it Generalized flux vacua},
  JHEP {\bf 0702} (2007) 095
  [arXiv:hep-th/0607015].

\bibitem{GrangeShafNameki}
P.~Grange and S.~Schafer-Nameki,
  {\it T-duality with H-flux: Non-commutativity, T-folds and $G \times G$ structure},
  Nucl.\ Phys.\  B {\bf 770} (2007) 123
  [arXiv:hep-th/0609084].

\bibitem{Ellwood}
I.~T.~Ellwood,
  {\it NS-NS fluxes in Hitchin's generalized geometry},
  arXiv:hep-th/0612100.


\bibitem{MicuPaltiTas}
A.~Micu, E.~Palti and G.~Tasinato,
  {\it Towards Minkowski vacua in type II string compactifications},
  JHEP {\bf 0703}, 104 (2007)
  [arXiv:hep-th/0701173].


\bibitem{HitchinFunctionals}
 N.~J.~Hitchin,
  {\it The geometry of three-forms in six and seven dimensions},
  arXiv:math.dg/0010054; {\it Stable forms and special metrics}, 
  in M.~Fernandez and J.~A.~Wolf (eds.),
  ``Global differential geometry: the mathematical legacy of Alfred Gray'', 
  Contemp. Math. {\bf 288}, 2001 [arXiv:math/0107101].


\bibitem{GMPT1}
  M.~Grana, R.~Minasian, M.~Petrini and A.~Tomasiello,
  {\it Supersymmetric backgrounds from generalized Calabi-Yau manifolds},
  JHEP {\bf 0408} (2004) 046
  [arXiv:hep-th/0406137].


\bibitem{GMPT2}
  M.~Grana, R.~Minasian, M.~Petrini and A.~Tomasiello,
  {\it Generalized structures of $N=1$ vacua},
  JHEP {\bf 0511}, 020 (2005)
  [arXiv:hep-th/0505212].

\bibitem{GMPT3}
M.~Grana, R.~Minasian, M.~Petrini and A.~Tomasiello,
  {\it A scan for new $N=1$ vacua on twisted tori},
  JHEP {\bf 0705} (2007) 031
  [arXiv:hep-th/0609124].


\bibitem{JeschekWitt1} 
C.~Jeschek and F.~Witt,
  {\it Generalised $G(2)$-structures and type IIB superstrings},
  JHEP {\bf 0503} (2005) 053
  [arXiv:hep-th/0412280].


\bibitem{WittThesis}
F.~Witt,
  {\it Special metric structures and closed forms}, 
  Oxford University DPhil Thesis (2004)
  [arXiv:math/0502443].


\bibitem{Tomasiello} A.~Tomasiello, {\it Reformulating Supersymmetry with a Generalized Dolbeault Operator},
  arXiv:0704.2613 [hep-th].
  

\bibitem{GenHodgeDec} M.~Gualtieri, {\it Generalized geometry and the Hodge decomposition}, arXiv:math.dg/0409093.


\bibitem{CandelasOssa}
  P.~Candelas and X.~de la Ossa,
  {\it Moduli space of Calabi-Yau manifolds},
  Nucl.\ Phys.\  B {\bf 355} (1991) 455.

\bibitem{Suzuki}
  H.~Suzuki,
  {\it Calabi-Yau compactification of type IIB string and a mass formula of the
  extreme black holes},
  Mod.\ Phys.\ Lett.\  A {\bf 11} (1996) 623
  [arXiv:hep-th/9508001].

\bibitem{CeresoleEtAl1} 
  A.~Ceresole, R.~D'Auria and S.~Ferrara,
  {\it The Symplectic Structure of N=2 Supergravity and its Central Extension},
  Nucl.\ Phys.\ Proc.\ Suppl.\  {\bf 46}, 67 (1996)
  [arXiv:hep-th/9509160].


\bibitem{ChiossiSalamon}
  S.~Chiossi and S.~Salamon,
  {\it The intrinsic torsion of $SU(3)$ and $G_2$ structures}, 
  in ``Differential geometry'', Valencia, (2001)
  [arXiv:math.dg/0202282].



\bibitem{N=2withTensor1} 
G.~Dall'Agata, R.~D'Auria, L.~Sommovigo and S.~Vaula,
  {\it $D = 4$, $N = 2$ gauged supergravity in the presence of tensor multiplets},
  Nucl.\ Phys.\  B {\bf 682}, 243 (2004)
  \hbox{[arXiv:hep-th/0312210]}.


\bibitem{N=2withTensor2}
R.~D'Auria, L.~Sommovigo and S.~Vaula,
  {\it $N = 2$ supergravity Lagrangian coupled to tensor multiplets with  electric
  and magnetic fluxes},
  JHEP {\bf 0411}, 028 (2004)
  \hbox{[arXiv:hep-th/0409097]}.


\bibitem{N=2review}
L.~Andrianopoli, M.~Bertolini, A.~Ceresole, R.~D'Auria, S.~Ferrara, P.~Fre and T.~Magri,
  {\it $N = 2$ supergravity and $N = 2$ super Yang-Mills theory on general scalar
  manifolds: Symplectic covariance, gaugings and the momentum map},
  J.\ Geom.\ Phys.\  {\bf 23}, 111 (1997)
  [arXiv:hep-th/9605032];\\
  L.~Andrianopoli, M.~Bertolini, A.~Ceresole, R.~D'Auria, S.~Ferrara and P.~Fre',
  {\it General Matter Coupled $N=2$ Supergravity},
  Nucl.\ Phys.\  B {\bf 476}, 397 (1996)
  \hbox{[arXiv:hep-th/9603004]}.


  
\bibitem{FerraraSabharwal}
S.~Ferrara and S.~Sabharwal,
  {\it Quaternionic Manifolds for Type II Superstring Vacua of Calabi-Yau
  Spaces},
  Nucl.\ Phys.\  B {\bf 332}, 317 (1990).

\bibitem{PolchinskiStrominger}
J.~Polchinski and A.~Strominger,
  {\it New Vacua for Type II String Theory},
  Phys.\ Lett.\  B {\bf 388} (1996) 736
  [arXiv:hep-th/9510227].

\bibitem{Michelson}
J.~Michelson,
  {\it Compactifications of type IIB strings to four dimensions with  non-trivial
  classical potential},
  Nucl.\ Phys.\  B {\bf 495} (1997) 127
  [arXiv:hep-th/9610151].

\bibitem{D'AuriaFerrFre}
R.~D'Auria, S.~Ferrara and P.~Fre,
  {\it Special and quaternionic isometries: general couplings in N=2 supergravity
  and the scalar potential},
  Nucl.\ Phys.\  B {\bf 359} (1991) 705.

\bibitem{DeWitSamtlebenTrig}
B.~de Wit, H.~Samtleben and M.~Trigiante,
  {\it Magnetic charges in local field theory},
  JHEP {\bf 0509} (2005) 016
  [arXiv:hep-th/0507289].



\bibitem{2into1won'tgo} S.~Cecotti, L.~Girardello and M.~Porrati,
  {\it Two Into One Won't Go},
  Phys.\ Lett.\  B {\bf 145} (1984) 61.

\bibitem{MinimalHiggsBranch}
  S.~Ferrara, L.~Girardello and M.~Porrati,
  {\it Minimal Higgs Branch for the Breaking of Half of the Supersymmetries in N=2 Supergravity},
  Phys.\ Lett.\  B {\bf 366} (1996) 155
  \hbox{[arXiv:hep-th/9510074]}.

\bibitem{TaylorVafa} 
	T.~R.~Taylor and C.~Vafa,
  {\it RR flux on Calabi-Yau and partial supersymmetry breaking},
  Phys.\ Lett.\  B {\bf 474} (2000) 130
  [arXiv:hep-th/9912152].


\bibitem{LouisN=2->N=1}
J.~Louis,
  {\it Aspects of spontaneous $N = 2 \rightarrow N = 1$ breaking in supergravity},
  arXiv:hep-th/0203138.


\bibitem{KoerberTsimpis}
P.~Koerber and D.~Tsimpis, 
	{\it Supersymmetric sources, integrability and generalized-structure compactifications}, 
	arXiv:0706.1244 [hep-th].

\bibitem{GiddingsDeWolfe}
  O.~DeWolfe and S.~B.~Giddings,
  {\it Scales and hierarchies in warped compactifications and brane worlds},
  Phys.\ Rev.\  D {\bf 67} (2003) 066008
  [arXiv:hep-th/0208123].

\bibitem{GiddingsMaharana}
	S.~B.~Giddings and A.~Maharana,
  {\it Dynamics of warped compactifications and the shape of the warped
  landscape},
  Phys.\ Rev.\  D {\bf 73} (2006) 126003
  [arXiv:hep-th/0507158].


\bibitem{deAlwis}
S.~P.~de Alwis,
  {\it On potentials from fluxes},
  Phys.\ Rev.\  D {\bf 68} (2003) 126001
  [arXiv:hep-th/0307084]; {\it Brane worlds in 5D and warped compactifications in IIB},
  Phys.\ Lett.\  B {\bf 603} (2004) 230
  [arXiv:hep-th/0407126].


\bibitem{MartucciSmyth}
  L.~Martucci and P.~Smyth,
  {\it Supersymmetric D-branes and calibrations on general $N = 1$ backgrounds},
  JHEP {\bf 0511} (2005) 048
  [arXiv:hep-th/0507099].


\bibitem{LustTsimpis}
 D.~Lust and D.~Tsimpis,
  {\it Supersymmetric $AdS_4$ compactifications of IIA supergravity},
  JHEP {\bf 0502}, 027 (2005)
  [arXiv:hep-th/0412250].



\bibitem{Truncation1}
L.~Andrianopoli, R.~D'Auria and S.~Ferrara,
  {\it Supersymmetry reduction of N-extended supergravities in four  dimensions},
  JHEP {\bf 0203}, 025 (2002)
  [arXiv:hep-th/0110277];\\
{\it Consistent reduction of $N = 2 \to N = 1$ four dimensional supergravity
  coupled to matter},
  Nucl.\ Phys.\  B {\bf 628}, 387 (2002)
  [arXiv:hep-th/0112192].

\bibitem{TruncationWithTensors}
R.~D'Auria, S.~Ferrara, M.~Trigiante and S.~Vaula,
  {\it $N = 1$ reductions of $N = 2$ supergravity in the presence of tensor
  multiplets},
  JHEP {\bf 0503}, 052 (2005)
  [arXiv:hep-th/0502219].


\bibitem{GrimmLouisA} 
T.~W.~Grimm and J.~Louis,
  {\it The effective action of type IIA Calabi-Yau orientifolds},
  Nucl.\ Phys.\  B {\bf 718}, 153 (2005)
  [arXiv:hep-th/0412277];\\
T.~W.~Grimm,
  {\it The effective action of type II Calabi-Yau orientifolds},
  Fortsch.\ Phys.\  {\bf 53}, 1179 (2005)
  [arXiv:hep-th/0507153].

  

\bibitem{VilladZwirner}
  G.~Villadoro and F.~Zwirner,
  {\it $N = 1$ effective potential from dual type-IIA $D6/O6$ orientifolds with
  general fluxes},
  JHEP {\bf 0506} (2005) 047
  [arXiv:hep-th/0503169].

  
\bibitem{WhatIsSpecialKaehler?}
B.~Craps, F.~Roose, W.~Troost and A.~Van Proeyen,
  {\it What is special K\"ahler geometry?},
  Nucl.\ Phys.\  B {\bf 503} (1997) 565
  [arXiv:hep-th/9703082].


\bibitem{Freed}
D.~S.~Freed,
  {\it Special Kaehler manifolds},
  Commun.\ Math.\ Phys.\  {\bf 203} (1999) 31
  \hbox{[arXiv:hep-th/9712042]}.

\bibitem{KoerberMartucci}
P.~Koerber and L.~Martucci,
  {\it From ten to four and back again: how to generalize the geometry},
  arXiv:0707.1038 [hep-th].


\end{thebibliography}
\end{document}